\documentclass{nature_plusfigure}
\usepackage{graphicx}
\usepackage{subcaption}
\usepackage{aas_macros,amsmath,amssymb,xcolor,url}
\usepackage{multirow}
\usepackage{longtable,hhline}
\usepackage{upgreek}
\usepackage{hyperref}
\usepackage{graphicx}
\usepackage{lineno}

\usepackage[none]{hyphenat}
\usepackage{xspace}

\usepackage[symbol]{footmisc}
\usepackage{cleveref}

\usepackage{multibib}
\newcites{meth}{Methods References}

\makeatletter
\newcommand{\maincite}[1]{\begingroup\@fileswfalse\cite{#1}\endgroup}
\makeatother

\newcounter{suppnote}
\renewcommand{\thesuppnote}{\arabic{suppnote}}
\newcommand{\suppnote}[2]{%
  \refstepcounter{suppnote}%
  \subsection*{Supplementary Note \thesuppnote\ | #1}%
  \label{#2}%
}
\newcommand{\suppref}[1]{%
  \hyperref[#1]{Supplementary Note~\ref*{#1}}%
}
\newcommand{\suppdiscussion}[2]{%
  \phantomsection%
  \subsection*{Supplementary Discussion | #1}%
  \label{#2}%
}
\newcommand{\suppdiscussionref}[1]{%
  \hyperref[#1]{Supplementary Discussion}%
}

\title{A temporally clustered gravitational wave pair and a Type IIb supernova}

\author{Sheng Yang$^{1}$\footnote{sheng.yang@hnas.ac.cn}, Ting-Wan Chen$^{2}$, Weichangfeng Guo$^{3}$, Brian D.~Metzger$^{4,5}$, Yuhao Zhang$^{6,7,8}$, Amar Aryan$^{2}$, Juntao Bai$^{1}$, Thomas de Boer$^{9}$, Mattia Bulla$^{10,11,12}$, Enrico Cappellaro$^{13}$, Sunil Choudhary$^{3,1}$, He Gao$^{14,15}$, Mansi M. Kasliwal$^{16}$, Darshan Kumar$^{1}$, Cheng-Han Lai$^{2}$, Chien-Cheng Lin$^{9}$, Liang-Duan Liu$^{6,7,8}$, Gregory S. H. Paek$^{9}$, Jesper Sollerman$^{17}$, Richard J. Wainscoat$^{9}$, Zhi-Qiang You$^{1}$, Yun-Wei Yu$^{6,7,8}$, Jianping Yuan$^{18}$, Xing-Jiang Zhu$^{19,14}$, Zong-Hong Zhu$^{15,20}$\footnote{zhuzh@bnu.edu.cn}}
\begin{document}

\maketitle

\begin{small}
\begin{affiliations}
\label{sec:affiliations}

\item Institute for Gravitational Wave Astronomy, Henan Academy of Sciences, Zhengzhou 450046, Henan, China
\item Graduate Institute of Astronomy, National Central University, 300 Jhongda Road, 32001 Jhongli, Taiwan
\item Department of Physics, University of Western Australia, Crawley WA 6009, Australia
\item Department of Physics and Columbia Astrophysics Laboratory, Columbia University, New York, NY 10027, USA
\item Center for Computational Astrophysics, Flatiron Institute, 162 5th Ave, New York, NY 10010, USA
\item Institute of Astrophysics, Central China Normal University, Wuhan 430079, China
\item Laboratory for Compact Object Astrophysics and Astronomical Technology, Central China Normal University, Wuhan 430079, China
\item Education Research and Application Center, National Astronomical Data Center, Wuhan 430079, China
\item Institute for Astronomy, University of Hawai`i, 2680 Woodlawn Drive, Honolulu, HI 96822, USA
\item Department of Physics and Earth Science, University of Ferrara, via Saragat 1, I-44122 Ferrara, Italy
\item INFN, Sezione di Ferrara, via Saragat 1, I-44122 Ferrara, Italy 
\item INAF, Osservatorio Astronomico d’Abruzzo, via Mentore Maggini snc, 64100 Teramo, Italy 
\item Istituto Nazionale di Astrofisica, Osservatorio Astronomico di Padova, vicolo dell’Osservatorio 5, 35122 Padova, Italy
\item Institute for Frontier in Astronomy and Astrophysics, Beijing Normal University, Beijing 102206, China
\item School of Physics and Astronomy, Beijing Normal University, Beijing 100875, China
\item Division of Physics, Mathematics and Astronomy, California Institute of Technology, Pasadena, CA 91125, USA
\item Department of Astronomy, Stockholm University, Sweden
\item Xinjiang Astronomical Observatory, Chinese Academy of Sciences, 150 Science 1-Street, Urumqi, Xinjiang 830011, People’s Republic of China
\item  Department of Physics, Faculty of Arts and Sciences, Beijing Normal University, Zhuhai 519087, China
\item School of Physics and Technology, Wuhan University, Wuhan 430072, China  

\end{affiliations}
\end{small}

\begin{abstract}

Compact object mergers are generally expected long after the stellar collapse that forms their remnants. Here we report two marginal gravitational wave (GW) candidates, S190910d and S190910h, separated by 7.1~hr, whose three-dimensional localizations overlap at the position and distance of the Type~IIb supernova SN~2019rsq. The GW pair lies in the extreme tail of the joint distribution of temporal separation and localization overlap, with a pairwise null probability of $7.6\times10^{-5}$ under our Monte Carlo null test. Independent re-analyses recover both as signal-like candidates. Under a common-origin hypothesis, S190910h remains consistent with a low-mass compact binary with source-frame masses near $2.5+1.3\,M_\odot$, whereas S190910d favours a strongly asymmetric $\sim34+3\,M_\odot$ system whose observed amplitude is compatible with the supernova distance. SN~2019rsq is a rapidly evolving stripped-envelope Type~IIb explosion whose light curve is consistent with stellar collapse occurring near the GW activity. Although the supernova coincidence alone is not statistically rare, its timing and physical properties provide a stellar collapse anchor for the unusual GW pair. A fragmenting collapsar disk provides a plausible framework that can accommodate the component masses, hour-scale delays and supernova properties, raising the possibility that multiple compact object mergers can accompany a single stellar collapse episode.

\end{abstract}

\bigskip

Core-collapse SNe (CCSNe) and compact binary mergers are usually regarded as subsequent, temporally separated stages of stellar evolution. Core collapse leaves behind a neutron star or black hole, and only after a long delay ($\gtrsim 10^5$ yr\cite{Tauris2017DNS,Andrews2019DNS,Broekgaarden2021BHNS}), in a suitable binary configuration, can such remnants later merge as a binary neutron star (BNS) or neutron star-black hole (NSBH) system. BNS and NSBH mergers are known to produce gravitational waves (GWs), short gamma ray bursts (GRBs)\cite{2012Sci...337..932G} and r-process powered kilonovae\cite{1998ApJ...507L..59L,2010MNRAS.406.2650M,2013ApJ...774...25K,2020LRR....23....1M}, as exemplified by the joint detection of GW170817, GRB~170817A, and AT~2017gfo\cite{2017PASA...34...69A,2017Natur.551...64A,2017Sci...358.1556C,2017ApJ...848L..17C,Diaz_2017,2017Sci...358.1570D,Valenti_2017,2017Sci...358.1559K,2017Natur.551...67P,2017Sci...358.1574S,2017Natur.551...75S,Tanvir_2017,2017PASJ...69..101U}.
The prompt outcome of massive star collapse is instead a CCSN and, in rare cases, a long GRB accompanied by a broad-lined Type~Ic SN, as typified by GRB~980425 and SN~1998bw\cite{2001ApJ...562..670W,1999ApJ...516..788W}. 

Recent observations, however, have begun to blur this conventional separation. 
Several long GRBs, including GRB~211211A\cite{2022Natur.612..228T,2022Natur.612..223R}, GRB~230307A\cite{2024Natur.626..737L} and GRB~191019A\cite{2025ApJ...979..159S}, have been discussed in connection with kilonova or kilonova-like emission and a compact-object merger origin, whereas the sub-second GRB~200826A\cite{Ahumada2021GRB200826A} was accompanied by a supernova and securely linked to massive-star collapse. Most recently, GRB~160425A\cite{Li2026GRB160425A} exhibited two distinct prompt emission episodes separated by only a few minutes, one consistently short-like and merger-like and the other long-like and collapsar-like. Together, these events demonstrate that the traditional division of GRBs by duration does not map uniquely onto their physical origin, and more broadly that observational signatures associated with stellar collapse and compact-object mergers can overlap within the same transient phenomenology.
Recent claims of ``superkilonova'' systems have further raised the possibility that compact object merger activity may accompany stellar collapse itself. 
In particular, two Type~IIb supernovae have now been discussed in spatial and temporal coincidence with low-mass GW candidates: SN~2025ulz, proposed in association with S250818k\cite{2025ApJ...995L..59K,2025ApJ...995L..27G,Franz_2025}, and SN~2025adtq, identified during follow-up of S251112cm\cite{Hall2026S251112cm}. Although neither association is yet conclusive, these cases motivate a systematic search for related configurations. Subsequent radio observations of SN~2025ulz revealed non-thermal emission consistent with fast ejecta interacting with circumstellar material and also compatible with an off-axis jet, leaving the broader superkilonova interpretation viable\cite{Tanner2026}. Together, these developments motivate the possibility that stellar collapse and compact object merger activity may, in rare cases, unfold within the same short-lived astrophysical episode.
One physical framework capable of producing such a sequence is fragmentation of the accretion disk formed during the collapse of a rapidly rotating massive star, in which gravitational instability can produce multiple compact remnants that subsequently merge on short timescales\cite{2007ApJ...658.1173P,2024ApJ...971L..34M,2025ApJ...991L..22C}.
A closely related collapsar-disk origin has also recently been proposed for the strongly asymmetric GW event GW190814\cite{Baibhav2026}.

\begin{figure}[ht!]
  \centering
  \includegraphics[width=0.85\columnwidth]{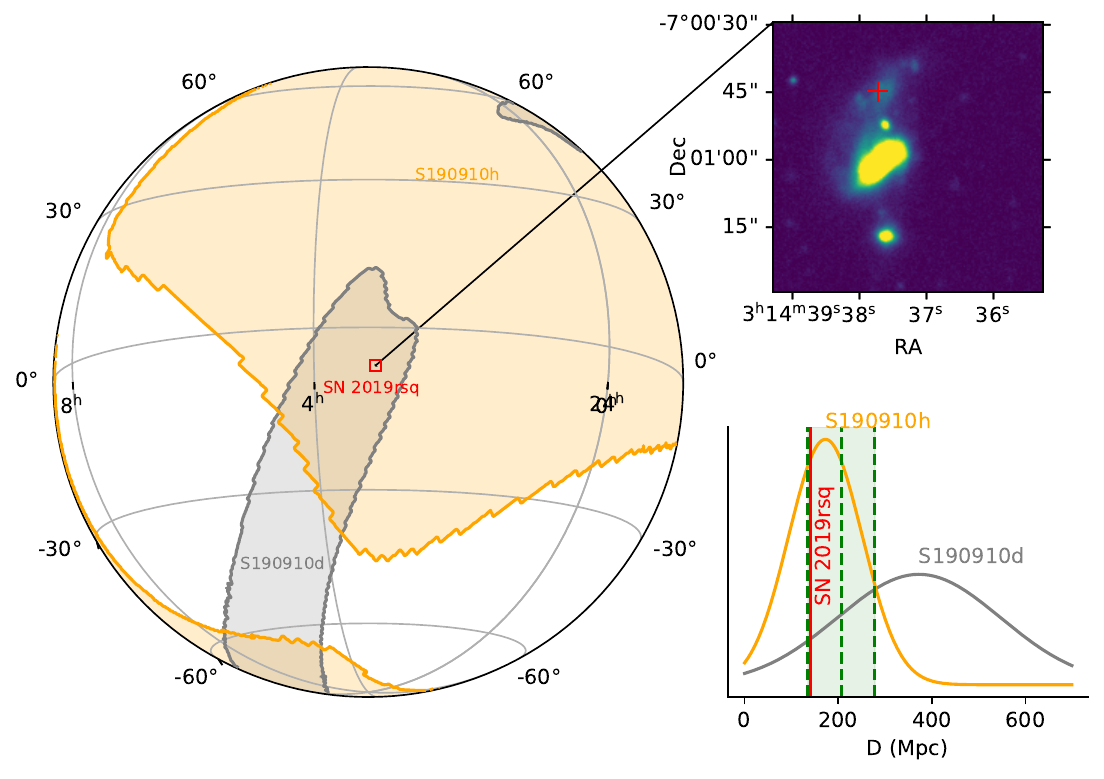}
  \caption{\textbf{Three dimensional localization overlap of S190910d, S190910h, and SN~2019rsq.}
  \textbf{Left panel:} Sky localizations of S190910d (grey) and S190910h (orange) shown as 90\% credible regions. The overlapping region contains the position of SN 2019rsq, marked by the red box.
  \textbf{right upper Panel:} Host galaxy environment of SN~2019rsq. The location of the supernova is indicated by the red plus. The image was obtained from the DESI archival service.
  \textbf{Lower right panel:} Line-of-sight distance information evaluated at the SN coordinates from the public 3D GW localization products. For visualization, the grey and orange curves show the Gaussian likelihood kernels associated with the per-pixel radial-ansatz parameters $(\mu,\sigma)$ for S190910d $(372.37,174.57)$~Mpc and S190910h $(172.60,78.55)$~Mpc. After removal of the common volumetric $D_L^2$ prior, their product gives an approximate common-distance likelihood centred at $D_{\rm comb}=206.23\pm71.63$~Mpc (green; Methods, Section~\ref{sec:SI2}). The red vertical line marks the redshift distance of SN~2019rsq, $D_{\rm SN}=140.6$~Mpc, which lies within the broad support of this consistency estimate. This construction is not a formal joint parameter-estimation posterior.} 
  \label{fig:localization}
\end{figure}

Motivated by this possibility, we carried out a systematic archival search for spatially and temporally compatible associations between public GW triggers and spectroscopically confirmed stripped-envelope (SE) supernovae (see Methods, Section~\ref{sec:SI1}). Using temporal proximity and two-dimensional sky localization as the discovery criteria, the search yielded two apparent GW--SN matches: S190910d/SN~2019rsq and S190910h/SN~2019rsq. Because both matches involve the same Type~IIb supernova, they identify a single candidate triplet in which the two GW triggers are separated by only 7.1~hr. Both candidates were classified in the public low-latency analyses as likely involving at least one neutron star: S190910d was assigned a 98\% probability of being NSBH-like, whereas S190910h was assigned a 61\% probability of being BNS-like, with ${\rm HasNS}>99\%$ for both. The publicly released LIGO--Virgo localizations of the two candidates also show substantial three-dimensional overlap, and their line-of-sight distance information is compatible with a common distance that includes the redshift-inferred distance of SN~2019rsq (Fig.~\ref{fig:localization}).

\begin{figure}[ht!]
  \centering
  \includegraphics[width=0.65\columnwidth]{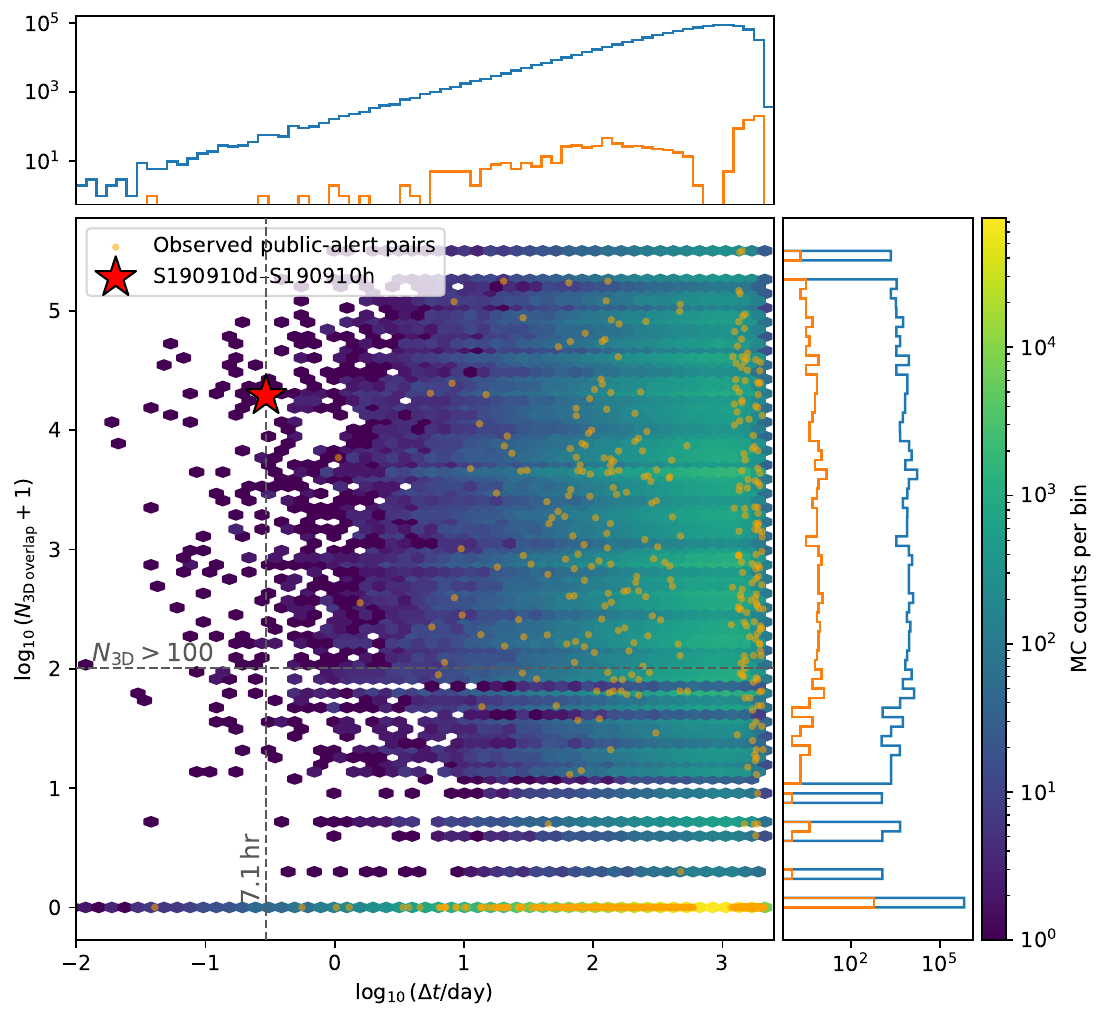}
  \caption{\textbf{Rarity of close-in-time NS/mass-gap GW pairs with consistent 3D localizations.} The main panel shows GW trigger pairs in the plane of temporal separation, $\log_{10}(\Delta t/\mathrm{day})$, versus 3D localization overlap strength, $\log_{10}(N_{\mathrm{3D}}+1)$, where $N_{\mathrm{3D}}$ is the number of HEALPix pixels (NSIDE$=256$; $\approx 0.052~\mathrm{deg}^2$ per pixel) that overlap in sky position and satisfy the adopted $\mu\pm\sigma$ radial-kernel overlap criterion. The coloured hexagonal density map shows the Monte Carlo sample obtained by randomizing GW trigger times while preserving the GW sky localizations and radial-distance information. Orange points mark observed GW pairs from the selected O3--O4b NS-involved or mass-gap public-alert sample, and the red star marks S190910d--S190910h. The vertical dashed line marks the observed temporal separation of the pair ($\Delta t \simeq 7.1$ hr), while the horizontal dashed line marks the threshold $N_{\mathrm{3D}}>100$ adopted in the coincidence-rate calculation, corresponding to a distance-consistent overlap area of approximately $5.2~\mathrm{deg}^2$. Marginal distributions for $\Delta t$ (top) and $N_{\mathrm{3D}}$ (right) are shown for both populations. Restricting attention to pairs within 5 days, the selected O3--O4b sample contains 10 close-in-time GW pairs, three of which show non-zero 3D overlap in the public localization products. In addition to S190910d--S190910h ($N_{\rm 3D}=19{,}368$), these are S241230ev--S250101k ($N_{\rm 3D}=5{,}872$, $\Delta t\simeq1.07$ d) and S230731an--S230802aq ($N_{\rm 3D}=356$, $\Delta t\simeq1.57$ d). Both comparison systems are BBH-dominated in their public low-latency classifications. S190910d--S190910h therefore stands out by combining the shortest timescale among the substantially 3D overlapping systems with the largest 3D overlap.}
  \label{fig:overlap}
\end{figure}

Close-in-time public GW candidates do occur, but few combine hour-scale temporal proximity with substantial three-dimensional localization overlap. To quantify the rarity of S190910d--S190910h independently of its association with SN~2019rsq, we performed Monte Carlo coincidence tests under the null hypothesis that the public GW triggers are unrelated (see Methods, Section~\ref{sec:SI2}). Using the NS-involved or mass-gap parent sample and our 3D search criteria, the pairwise null probability for an unrelated GW pair to occur within 7.1~hr and satisfy the adopted localization-overlap criterion is $7.6\times10^{-5}$. Because both triggers are near the detection threshold, their inferred luminosity distances are particularly susceptible to the distance--inclination degeneracy and to waveform and prior assumptions. We therefore repeated the calculation using only temporal proximity and two-dimensional sky overlap, obtaining $1.41\times10^{-4}$ at the observed 7.1~hr separation. 
The comparable rarity after removing the distance information shows that the clustering is not driven solely by the GW distance inference, but is already unusual in time and projected sky position.
In the joint plane of temporal separation and 3D localization overlap, S190910d--S190910h occupies the sparsely populated tail of the null distribution (Fig.~\ref{fig:overlap}), combining an unusually short separation with substantial localization overlap.

Both S190910d and S190910h are near-threshold events. We therefore re-analysed both candidates using the Summed Parallel Infinite Impulse Response (\textsc{SPIIR}) matched-filter framework\cite{spiir_7} (see Methods, Section~\ref{sec:SI3}). Both recoveries show template-consistent matched-filter morphologies and rank above the loudest background trigger in the corresponding data segment, supporting a signal-like interpretation. S190910d is also independently recovered by the IAS O3a search as GW190910\_012619, with an astrophysical probability of $p_{\rm astro}=0.58$\cite{Olsen2022IAS,IASO3aData}. These checks support retaining S190910d and S190910h as a physically interesting clustered pair of GW candidates.

The re-analyses also sharpen the source interpretation. S190910h is recovered by a low-mass BNS-like template with a detector-frame chirp mass of $\mathcal{M}^{\rm det}\simeq1.53\,M_\odot$. For S190910d, dedicated precessing parameter estimation robustly favours a strongly asymmetric mass configuration, with $q\simeq0.086$, insensitive to extending the lower prior boundary from $q=0.05$ to $q=0.01$. The source-frame component-mass posterior favours representative values near $m_1^{\rm src}\simeq34\,M_\odot$ and $m_2^{\rm src}\simeq2.9\,M_\odot$, closely matching the source-frame mass region independently favoured by the IAS analysis of the same trigger\cite{Olsen2022IAS,IASO3aData}. Thus, S190910h occupies a low-mass compact binary regime, whereas S190910d is consistent with a strongly asymmetric black-hole--compact object system with a few-solar-mass secondary.

We next tested whether these configurations remain viable under a common-origin hypothesis with SN~2019rsq. Conditioning S190910h on the supernova position and a finite distance prior centred on the SN distance yields a stable low-mass solution near source-frame component masses $m_1^{\rm src}\simeq2.5\,M_\odot$ and $m_2^{\rm src}\simeq1.3\,M_\odot$. Applying the same fully conditioned strategy to S190910d produces a substantially less informative posterior, so we separate its intrinsic and extrinsic consistency tests. The source-frame mass scale is constrained by the prior-robust precessing analysis, which favours representative values near $m_1^{\rm src}\simeq34\,M_\odot$ and $m_2^{\rm src}\simeq2.9\,M_\odot$, while compatibility with the supernova distance is tested independently through geometric projection. Placing this representative configuration at $D_{\rm L}=140.6$~Mpc and converting the source-frame masses to their corresponding detector-frame values for waveform generation yields a network SNR no greater than the observed $\rho_{\rm net}\simeq8.6$ for $6.6\%$ of isotropically distributed orientations. Thus, geometric projection alone can reproduce the observed signal strength for a source at the supernova distance. S190910h, by contrast, remains compatible with the SN hypothesis directly in the conditioned parameter estimation and does not require the same separate amplitude-consistency test.
The 7.1-hr separation between the two GW triggers also substantially changes the detector geometry: at the SN position, Earth rotation rotates the L1 polarization-response vector $(F_+,F_\times)$ by approximately $98^\circ$. The two mergers can therefore be observed through markedly different polarization projections even if they arise from the same system and have similar underlying orbital geometries. Together, the independent recoveries, source-parameter inference, geometric projection and time-dependent detector response show that the common-source interpretation is self-consistent in signal morphology, intrinsic mass scale and observed amplitude.

Within the overlapping localization region lies SN~2019rsq, a stripped-envelope supernova initially classified as broad-lined Type~Ic and subsequently established as Type~IIb following the emergence of hydrogen and helium features at later phases. Our independent photometric and spectroscopic re-analysis confirms this evolution: the early spectra show broad features and high expansion velocities, whereas the later spectra reveal clearer hydrogen and helium signatures characteristic of a Type~IIb event (see Methods, Section~\ref{sec:SI4}). Its optical peak occurred approximately 24~days after the earlier GW trigger, S190910d.

\begin{figure}[ht!]
  \centering
  \includegraphics[width=0.85\columnwidth]{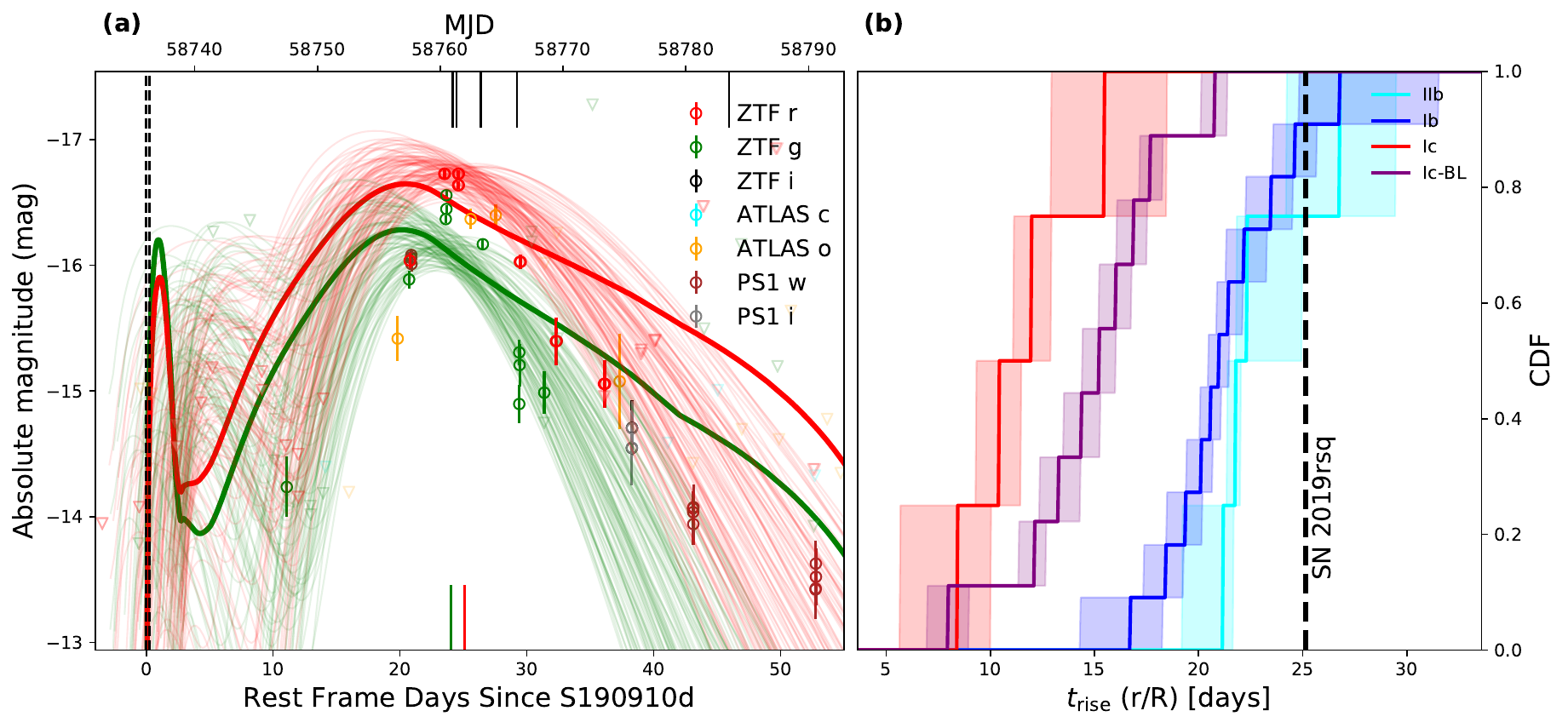}
  \caption{\textbf{Temporal match between SN~2019rsq and the GW events S190910d and S190910h.} \textbf{a: Multi-band light curves of SN~2019rsq with model comparisons.} Observed ZTF, ATLAS, and PS1 photometry is shown as a function of time. The lower and upper x-axes give the rest-frame phase relative to S190910d and the Modified Julian Date (MJD), respectively, while the left y-axes show absolute magnitudes. All magnitudes are corrected for Galactic extinction. Solid curves show a representative \texttt{MESA}+\texttt{SNEC} hydrodynamic model reproducing the overall light curve evolution. Faint curves show \texttt{TransFit} realizations obtained by sampling the physical parameter space and retaining solutions that reproduce the photometry while placing the explosion epoch within approximately 3~days of S190910d; the 100 best-fitting accepted realizations are shown. Upper black vertical ticks mark the five epochs of optical spectroscopy of SN~2019rsq, spanning observations near maximum through the later Keck/LRIS spectrum (see Extended Data Fig.~\ref{fig:spec1}). Lower green and red ticks indicate the epochs of peak brightness in the $g$ and $r$ bands, respectively. Black dashed vertical lines mark S190910d and S190910h. \textbf{b: Optical rise-time distributions of stripped-envelope SNe.} As a complementary empirical consistency check, cumulative distributions of $r/R$-band rise times for SNe~IIb, Ib, Ic, and Ic-BL are constructed from the literature compilation of \cite{2015A&A...574A..60T}. Shaded regions show the 16th--84th percentile ranges obtained by propagating the reported measurement uncertainties through Monte Carlo sampling. The black dashed vertical line marks the rise time of SN~2019rsq obtained when S190910d is adopted as the explosion epoch. Under this assumption, the inferred rise time lies within the observed stripped envelope SN distribution, providing a consistency check on the proposed temporal association rather than an independent determination of the explosion epoch.}
  \label{fig:lc2019rsq}
\end{figure}

The available photometry does not sample an identifiable shock-breakout or early shock-cooling phase and therefore does not determine the explosion epoch tightly on its own. We instead tested whether an explosion occurring near the GW activity is compatible with the observed light curve. Diffusion-based and radiation-hydrodynamic calculations show that physically reasonable Type~IIb ejecta structures can reproduce the photometric evolution under this condition. Conversely, adopting S190910d as the explosion epoch implies a rise time of approximately 24~days, well within the observed Type~IIb distribution (Fig.~\ref{fig:lc2019rsq}). The optical evolution is therefore consistent with stellar collapse occurring during the same episode as the GW activity.

At the same time, SN~2019rsq exhibits a relatively narrow, rapidly evolving light curve compared with classical Type~IIb events and belongs to the small population of FBOT-like Type~IIb supernovae identified by \cite{ho2023}. Such behaviour has been associated with compact progenitors, strongly stripped hydrogen envelopes and binary interaction. Our broader comparison with well-observed Type~IIb and related SE supernovae shows notable similarities between SN~2019rsq and SN~2025ulz in their main-peak light curve evolution, colour behaviour and spectral progression. Energetic Type~IIb and transitional Ib/IIb events such as SN~2003bg and SN~2020cpg further show that residual hydrogen and Type~IIb-like spectra can coexist with energetic explosions and comparatively massive stripped progenitors\cite{Mazzali2009,Teffs2022}. Taken together, these comparisons place SN~2019rsq in an unusual but observationally established regime of fast, strongly stripped and energetic stellar explosions that is compatible with the massive-star collapse picture considered below.

SN~2019rsq also occurs in a disturbed, interacting morphology host galaxy, projected onto a tidal-tail-like structure with ongoing star formation, consistent with a young massive-star environment (see Methods, Section~\ref{sec:SI5}). The supernova coincidence itself is not highly significant in isolation: the estimated probabilities for a stripped-envelope or Type~IIb supernova to occur by chance within the adopted GW overlap volume and temporal window are approximately $0.69$ and $0.35$, respectively (see Methods, Section~\ref{sec:SI2}). 
These estimates treat the broader stripped-envelope and Type~IIb populations as a whole and do not condition on the fast, FBOT-like properties of SN~2019rsq. Quantitatively, the strongest anomaly therefore remains the rare temporally clustered and three-dimensionally consistent GW pair; SN~2019rsq provides the spatial, distance, temporal and stellar collapse anchor that makes a common physical interpretation possible.

\begin{figure}[ht]
    \centering
    \includegraphics[page=1,width=0.48\textwidth,]{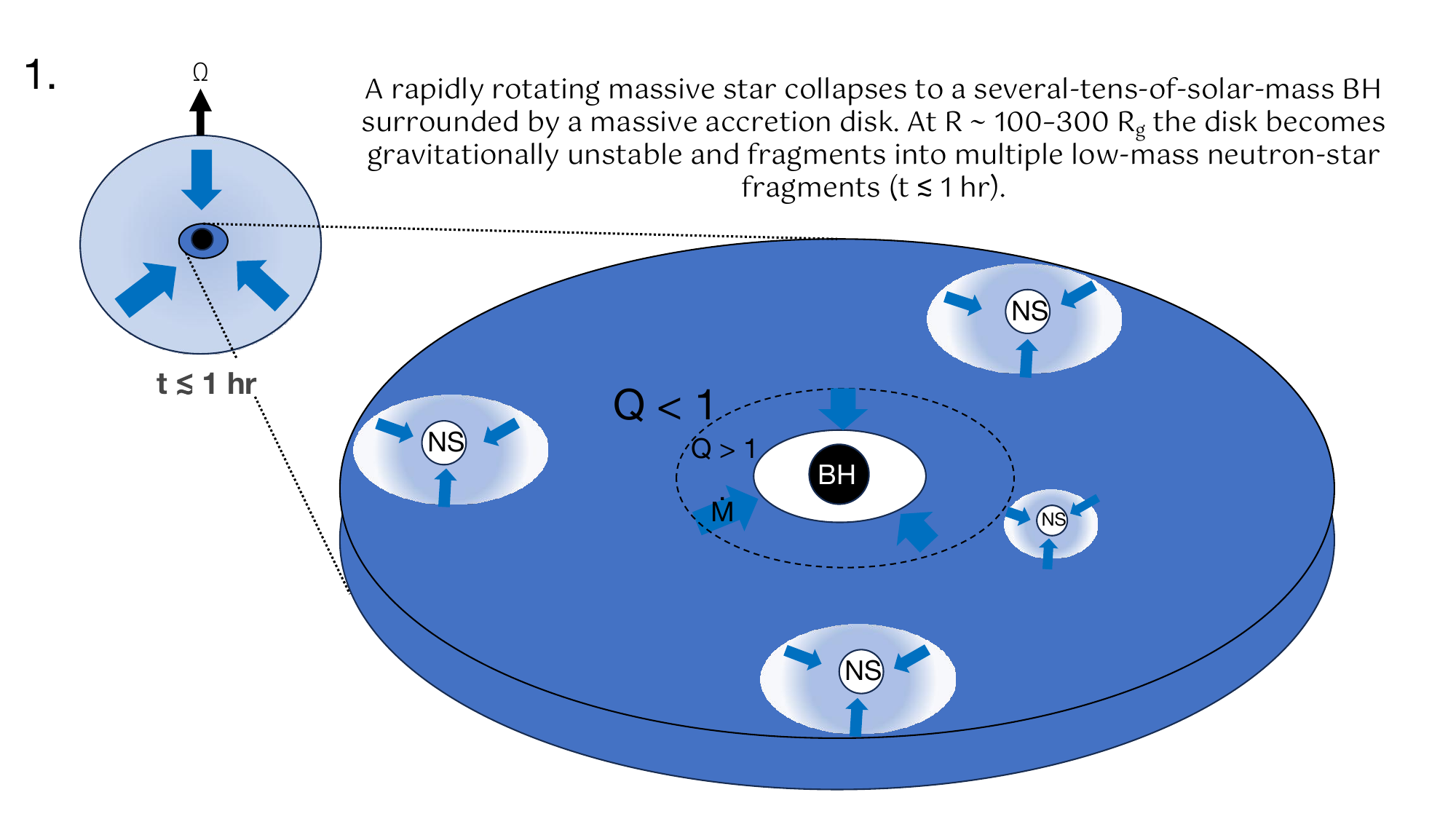}\hfill
    \includegraphics[page=2,width=0.48\textwidth,]{Main_figures/cartoon.pdf}\vspace{2mm}
    \includegraphics[page=3,width=0.48\textwidth,]{Main_figures/cartoon.pdf}\hfill
    \includegraphics[page=4,width=0.48\textwidth,]{Main_figures/cartoon.pdf}
    \caption{\textbf{Schematic illustration of the proposed fragmenting-disk scenario.}
    \textbf{a,} Core collapse of a rapidly rotating massive star forms a black hole of several tens of solar masses surrounded by a gravitationally unstable accretion disk that fragments at radii of order $100$--$300R_g$.
    \textbf{b,} An accretion-grown compact fragment or earlier fragment-merger remnant reaches a mass of a few solar masses and inspirals into the central black hole, producing S190910d, while the outer stellar layers expand as SN~2019rsq. Other lower-mass fragments remain in the outer disk and may pair into binaries.
    \textbf{c,} A low-mass compact binary formed in the disk merges $\simeq7.1$~hr later, producing S190910h, while the supernova ejecta continue to expand toward the optical peak approximately 24~days after S190910d.
    \textbf{d,} Subsequent fragment interactions and few-body scattering may produce additional compact-object mergers or delayed central-engine activity on timescales extending from days to weeks or months.}
    \label{fig:scenario}
\end{figure}

In this respect, a plausible physical framework for the observed sequence is provided by disk fragmentation during the collapse of a rapidly rotating massive star (see Methods, Section~\ref{sec:SI6}; Fig.~\ref{fig:scenario}). Core collapse can form a massive black hole surrounded by a dense, neutron-rich accretion disk that becomes gravitationally unstable and fragments on short dynamical timescales at radii of order $100$--$300\,R_g$\cite{2007ApJ...658.1173P,2024ApJ...971L..34M,2025ApJ...991L..22C}. The initially formed fragments are expected to be relatively low mass, but subsequent accretion, migration and fragment--fragment interactions can grow some of them into more massive compact remnants.

One such accretion-grown fragment or fragment-merger remnant could merge with the central black hole to produce S190910d, whose inferred source-frame masses are approximately $34+3\,M_\odot$. A closely related collapsar-disk interpretation has recently been proposed for GW190814, with its strongly asymmetric $\sim23+2.6\,M_\odot$ source-frame mass configuration\cite{Baibhav2026}. Elsewhere in the same disk, lower-mass fragments can pair into compact binaries and merge through gravitational radiation, providing a natural route to S190910h. Direct inspiral times in the fragmentation region span hours to days, naturally encompassing the observed $7.1$~hr separation between the two GW candidates. The resulting sequence, i.e., a strongly asymmetric black-hole--compact object merger followed hours later by a lower-mass compact binary merger while the stripped-envelope supernova ejecta expand, is therefore consistent with the hierarchy of outcomes and timescales expected in fragmenting collapsar disks. 

The optical modelling provides a complementary consistency check on this picture. The rapid evolution of SN~2019rsq favours a relatively small observable ejecta mass, but does not tightly constrain the total mass participating in the collapse, because a substantial fraction of the star may instead remain in the central black hole, accretion disk or fallback material. Recent simulations of rapidly rotating massive-star collapse show that progenitors of several tens of solar masses can form massive black-hole--disk systems while ejecting only a few solar masses of material\cite{Fujibayashi2024,CrosatoMenegazzi2025}. The low ejecta mass implied by the optical light curve is therefore not intrinsically inconsistent with the several-tens-of-solar-mass black hole favoured by the GW analysis.

A further requirement is sufficient angular momentum to form an extended accretion disk rather than allowing the stellar material to fall nearly radially into the central black hole. Recent stellar-evolution calculations show that rapidly rotating, stripped massive stars can retain substantial and structured specific angular momentum up to collapse, sufficient to form centrifugally supported accretion disks around the newly formed compact object\cite{Renzo2026}. We therefore regard the progenitor rotation as a separate consistency requirement, while the characteristic fragmentation radii of $R_{\rm form}\sim100$--$300\,R_g$ and the associated hour-scale merger times are set by dedicated collapsar-disk calculations. Taken together, the mass partition, angular-momentum budget and disk dynamical timescales occupy a common physically allowed regime consistent with the observed supernova and GW sequence.

Importantly, the same framework does not require every subsequent compact object interaction to occur promptly. In a multi-fragment disk, migration and few-body interactions can redistribute orbital energy and angular momentum, leaving surviving fragments or merger remnants on wider or eccentric orbits and extending subsequent merger delays from days to weeks or months\cite{Baibhav2026}. The absence of a confidently identified third GW trigger immediately following S190910h therefore does not strongly constrain the scenario. We nevertheless examined the detector coverage during this interval and found substantial variations in directional sensitivity and periods of incomplete multi-detector operation (see Methods, Section~\ref{sec:SI7}).

A long-duration burst, GRB~190915A, occurred approximately five days later and is positionally compatible with the broader GW--SN region, although its broad localization and longer delay preclude treating it as evidence for the association. If physically related, it could instead represent delayed or reactivated high-energy activity from the evolving fragmenting-disk system (\suppref{supp:grb190915a}). The radio observations provide an independent constraint on relativistic outflow activity (see Methods, Section~\ref{sec:radio}). At the distance of SN~2019rsq, a long-lived relativistic afterglow comparable to GRB~030329 would have remained readily detectable in the available ASKAP and VLASS imaging and is therefore disfavoured. Weaker, off-axis, baryon-loaded or choked outflows, or explosions into a sufficiently low-density environment, remain allowed. Thus, the multi-wavelength data disfavour a powerful, long-lived, radio-bright relativistic jet, but do not require the absence of central-engine activity.

Regardless of whether the SN~2019rsq system ultimately proves to be a rare coincidence or a genuine example of multiple GW signals produced during stellar collapse, it identifies temporally clustered GW activity as a promising target for multi-messenger searches. The recently proposed SN~2025ulz/S250818k association, together with the later SN~2025adtq/S251112cm comparison case, motivates testing whether this phenomenology recurs (\suppdiscussionref{supp:comparison}). Rapidly evolving Type~IIb supernovae accompanied by low-mass or strongly asymmetric GW candidates may define a previously under-explored phenomenological regime. Future GW observing runs, coupled with wide-field optical surveys and rapid spectroscopic classification, should determine whether such systems represent a genuine population or a small number of chance alignments.

\clearpage
\noindent\textbf{Figures.} \\
Figure~1: Three-dimensional localization overlap of S190910d, S190910h, and SN~2019rsq.\\
Figure~2: Rarity of close-in-time NS-involved GW pairs with consistent 3D localizations. \\
Figure~3: Temporal match between SN~2019rsq and the GW events S190910d and S190910h. \\
Figure~4: Schematic illustration of the proposed scenario. \\

\clearpage

\noindent {\bf References}
\bibliographystyle{naturemag}
\bibliography{references}

\newpage

\begin{methods}

\section{Datasets and Matching Procedure}
\label{sec:SI1}
\subsection{Supernova sample}
\label{sec:SI11}

We restricted the transient search to stripped-envelope (SE) supernovae because collapsar-disk fragmentation requires a compact, strongly stripped, rapidly rotating progenitor\maincite{2024ApJ...971L..34M}, and because the post-cutoff Type~IIb systems SN~2025ulz\maincite{2025ApJ...995L..59K,2025ApJ...995L..27G,Franz_2025} and SN~2025adtq\maincite{Hall2026S251112cm} provide external observational motivation. 

We constructed the SE-SN sample using the public Zwicky Transient Facility (ZTF) Bright Transient Survey (BTS) catalogue\footnote{https://sites.astro.caltech.edu/ztf/bts/bts.php}, queried on 2025 October 31. We selected events with spectroscopic classifications consistent with SE explosions, including Types~IIb, Ib, Ic and closely related subtypes such as Ic-BL, and required the source information needed for the spatial and distance matching described below.

For the primary GW--SN search, we restricted the supernova sample to the interval relevant to the O3--O4b GW sample. Because the matching procedure allows the main optical peak to occur up to 30 days after a GW trigger, the SN interval extends from the beginning of O3 on 2019 April 1 through 2025 February 27, 30 days beyond the end of O4b. The resulting primary sample contains 676 spectroscopically classified SE SNe, including 122 Type~IIb SNe. Events outside this interval are used only for phenomenological comparison and do not enter the formal search statistics.

For each SN, the measured redshift was converted to a luminosity distance assuming a flat $\Lambda$CDM cosmology with $H_0=70~{\rm km\,s^{-1}\,Mpc^{-1}}$ and $\Omega_{\rm m}=0.3$. As an independent transient-side completeness check, we repeated the archival search using the broader Transient Name Server (TNS) database\footnote{https://www.wis-tns.org/}; the method and limitations of this check are described in \suppref{supp:samples}.

\subsection{GW event sample}\label{sec:SI12}

Motivated by the low-mass compact remnants predicted in fragmenting collapsar disks (Section~\ref{sec:SI6}), we targeted public GW alerts with a non-negligible low-latency probability of belonging to the BNS, NSBH or mass-gap source categories.

We assembled the GW public-alert sample from the LIGO--Virgo--KAGRA (LVK) public-alert archive hosted by GraceDB\footnote{https://gracedb.ligo.org/}, queried through the GraceDB API\footnote{https://ligo-gracedb.readthedocs.io/} on 2025 October 31. For the primary statistical analysis, we fixed the sample cutoff at the end of O4b on 2025 January 28 and included public alerts from O3 through O4b. Within this interval GraceDB contains 247 non-retracted public GW superevents. We retained superevents whose preliminary low-latency classification assigned at least a 1\% probability to a BNS, NSBH or mass-gap source category, yielding 43 selected superevents. The 1\% threshold is an inclusive search criterion rather than a statement that the low-latency classification of any individual near-threshold candidate is precise.

For each selected candidate, we cross-checked its subsequent catalogue status against the cumulative LVK transient catalogue through GWTC-2\citemeth{2021PhRvX..11b1053A}, GWTC-3\citemeth{2023PhRvX..13d1039A} and GWTC-4\citemeth{2025arXiv250818082T}. This catalogue-status check does not alter the primary public-alert sample used for the archival search. Non-retracted public candidates satisfying the predefined source-class criterion were retained irrespective of subsequent catalogue status. This intentionally inclusive strategy retains S190910d and S190910h, whose signal-like nature is examined independently in Section~\ref{sec:SI3}.

We retrieved the public BAYESTAR localization products and event metadata directly from GraceDB. For consistency across the public-alert sample, these localization products were used for the archival matching and population-level GW-pair calculations; later offline LALInference products were used only for the target-specific robustness check described in Section~\ref{sec:SI22}. Of the 247 non-retracted superevents, 246 have localization products containing usable line-of-sight distance information. The remaining event, S200114f, has a two-dimensional localization only and is retained only in calculations that do not require distance information. The event selections and the code used to reproduce them will be released through the analysis repository described in the Code Availability statement.

\subsection{Spatial matching}

For each GW trigger, we read the HEALPix\citemeth{2005ApJ...622..759G} localization probability map $p(\Omega)$ using the \texttt{ligo.skymap} utilities\footnote{https://lscsoft.docs.ligo.org/ligo.skymap/}. All maps were resampled to a common resolution of \texttt{NSIDE}$=256$ using HEALPix up/down-grading\footnote{https://healpy.readthedocs.io/en/latest/} and renormalized after resampling.

We defined the 90\% credible sky region by ranking pixels in descending posterior probability and selecting the minimal set enclosing 90\% of the total sky probability. Each SN sky position was converted to a HEALPix pixel at the same resolution. A GW--SN angular match was recorded when the SN pixel lay within the GW 90\% credible region.

Candidate associations were selected using this angular criterion together with the temporal criterion below. We then evaluated the radial-distance information at the SN position for the matched candidates.

\subsection{Temporal matching}

To identify plausible GW--SN associations, we required the main optical SN peak to occur within a physically motivated interval after the GW trigger, under the hypothesis that the GW activity occurs near the epoch of stellar collapse. Observed stripped-envelope SNe span a broad range of rise times: mean rest-frame $r/R$-band rise times are $22.9\pm0.8$, $21.3\pm0.4$, $11.5\pm0.5$ and $14.7\pm0.2$~days for SNe~IIb, Ib, Ic and Ic-BL, respectively\maincite{2015A&A...574A..60T}, while Type~IIb events can extend to approximately 26~days\citemeth{Prentice2019}. Early shock-breakout or shock-cooling emission, when present, occurs on substantially shorter timescales and is distinct from the main optical peak considered here (for example, SN~1993J\citemeth{1994ApJ...429..300W}).

We define the source-rest-frame delay $\Delta t_{\rm rest} = \frac{t_{\rm SN}^{\rm peak}-t_{\rm GW}}{1+z_{\rm SN}}$, and adopt the matching criterion $10~{\rm d}<\Delta t_{\rm rest}<30~{\rm d}$. This interval encompasses the observed diversity of the main optical rise of stripped-envelope SNe while remaining restrictive against chance temporal coincidences. The same rest-frame convention is used in Fig.~\ref{fig:lc2019rsq}. For SN~2019rsq specifically, we additionally compare its inferred rise time with the narrower Type~IIb distribution and use the corresponding Type~IIb timescale in the subclass-specific chance-coincidence estimate of Section~\ref{sec:SI2}.

\subsection{Identification of the SN~2019rsq/S190910d--S190910h configuration}

Applying the angular and temporal criteria above yields two apparent GW--SN matches in the archival search: S190910d/SN~2019rsq and S190910h/SN~2019rsq. Because both matches involve the same optical transient, the search identifies a single candidate triplet comprising one SE SN and two GW triggers separated by only 7.1~hr.

The two GW localizations also overlap substantially in three dimensions, and their line-of-sight distance information at the position of SN~2019rsq is compatible with a common distance that includes the SN redshift distance (Fig.~\ref{fig:localization}). The optical peak occurs approximately 24 rest-frame days after S190910d (Fig.~\ref{fig:lc2019rsq}). The statistical rarity of the GW pair is quantified in Section~\ref{sec:SI2}; the two near-threshold candidates are independently re-analysed in Section~\ref{sec:SI3}. A broader TNS search and the post-cutoff SN~2025ulz/SN~2025adtq systems provide additional completeness and external checks (\suppref{supp:samples}).

\section{Statistical Significance and Chance Coincidence Tests}
\label{sec:SI2}

\subsection{Observed close-in-time GW pairs}
\label{sec:SI21}

Among the 43 selected public GW superevents, 10 pairs are separated by less than 5~days and only three retain non-zero 3D overlap after incorporating the line-of-sight distance information. S190910d--S190910h combines the shortest separation (7.1~hr) with the largest distance-compatible overlap, $N_{\rm 3D}=19{,}368$ pixels at \texttt{NSIDE}$=256$ ($\simeq1{,}016~{\rm deg^2}$), and is the only such pair associated with an SE SN selected by the angular and temporal archival search (Fig.~\ref{fig:overlap}). The other overlapping pairs and their classifications are listed in \suppref{supp:stats}.

\subsection{Common-distance consistency of the two GW candidates}
\label{sec:SI22}

To test distance compatibility at the SN position and define the approximate volume used below, we use the radial ansatz of the public three-dimensional localization products for S190910d and S190910h. Along a fixed line of sight it can be written schematically as $p(D_L\mid\Omega,d)\propto D_L^2\exp[-(D_L-\mu)^2/(2\sigma^2)]$, where the $D_L^2$ factor represents the standard volumetric distance prior. Because the two public products use the same distance-prior form, we remove this common factor for the present consistency calculation and combine the Gaussian likelihood kernels $\exp[-(D_L-\mu_i)^2/(2\sigma_i^2)]$. Their product has $\sigma_{\rm comb}^2 = \left( \sigma_1^{-2}+\sigma_2^{-2} \right)^{-1}$ and $\mu_{\rm comb}=\sigma_{\rm comb}^2(\mu_1/\sigma_1^2+\mu_2/\sigma_2^2)$. At the SN position, $\mu_{\rm S190910d}=372.37$~Mpc, $\sigma_{\rm S190910d}=174.57$~Mpc and $\mu_{\rm S190910h}=172.60$~Mpc, $\sigma_{\rm S190910h}=78.55$~Mpc, giving an approximate common-distance likelihood centred at $206.23\pm71.63$~Mpc. The SN luminosity distance, $D_{\rm SN}=140.6$~Mpc, lies within its broad support (Fig.~\ref{fig:localization}). This construction is a consistency estimate and is used to define the effective distance range below; it is not a formal joint parameter-estimation posterior.

As a target-specific robustness check, we also evaluated the later offline LALInference three-dimensional localization products for S190910d and S190910h at the position of SN~2019rsq. After removal of the same $D_L^2$ volumetric distance prior, the corresponding line-of-sight likelihood kernels have $(\mu,\sigma)=(366.34,228.61)$~Mpc for S190910d and $(17.50,132.27)$~Mpc for S190910h. The SN distance remains consistent with each kernel at approximately the $1\sigma$ level, and the two likelihood centres differ by only $1.32\sigma$ relative to their combined widths. These later products are used only as a target-specific robustness check and were not used to define the archival discovery selection.

\subsection{Chance probability of an unrelated GW pair}
\label{sec:SI23}

We quantified the rarity of the observed configuration with Monte Carlo simulations under the null hypothesis that the public GW candidates are mutually unrelated in time. Using the selected GW sample described in Section~\ref{sec:SI1}, we generated $10^6$ pairwise realizations over the O3--O4b calendar baseline, $T_{\rm span}\simeq2130$~days (2019 April 1 to 2025 January 28). In each realization, two distinct superevents were drawn from the sample and assigned independent occurrence times uniformly over this baseline while retaining their localization products.

For each realization, $N_{\rm 2D}$ is the number of common HEALPix pixels between the two 90\% credible sky regions, and $N_{\rm 3D}$ is the subset of these pixels for which the corresponding adopted $\mu\pm\sigma$ radial-kernel intervals overlap, $|\mu_1-\mu_2|<\sigma_1+\sigma_2$.
A coincident 3D pair was counted when its temporal separation was smaller than the adopted threshold and both $N_{\rm 2D}>100$ and $N_{\rm 3D}>100$. At \texttt{NSIDE}=256, 100 pixels correspond to approximately $5.2~{\rm deg}^2$; this threshold is used to reject trivial boundary-level overlaps rather than to impose a physical angular scale.

For the 43-superevent selected sample, the empirical probability of an unrelated GW pair occurring within the observed 7.1~hr separation and satisfying the adopted 3D criteria is $P_{\rm 2GW}=7.6\times10^{-5}$. Because the line-of-sight distance inference of near-threshold candidates can be sensitive to waveform and prior assumptions, we repeated the calculation using only temporal proximity and two-dimensional sky overlap. Under the same $N_{\rm 2D}>100$ criterion, $P_{\rm 2D}=1.41\times10^{-4}$ at 7.1~hr. The clustering is therefore already unusual in time plus projected sky position, while the public distance information strengthens the three-dimensional consistency. Repeating the analysis with the full public-alert sample and varying the temporal threshold from 6 to 24~hr gives probabilities of the same order (\suppref{supp:stats}). 

\subsection{Chance probability of the supernova coincidence}
\label{sec:SI24}

Given the occurrence of S190910d and S190910h as a temporally clustered GW pair with substantial 3D overlap, we next estimated the probability that an unrelated SE or Type~IIb SN would fall within the corresponding spacetime volume. 

The expected number of unrelated SNe is $N_{\rm exp}=R_{\rm SN}V_{\rm eff}\Delta t_{\rm eff}$, where $R_{\rm SN}$ is the volumetric SN rate, $V_{\rm eff}$ is the effective 3D GW overlap volume, and $\Delta t_{\rm eff}$ is the relevant temporal-window width.
Using the approximate common-distance estimate from Section~\ref{sec:SI22}, we adopt $D_{\rm min}=134.6$~Mpc and $D_{\rm max}=277.9$~Mpc. Restricting this shell to the $N_{\rm 3D}=19{,}368$ distance-compatible overlap pixels gives a sky fraction $f_\Omega\simeq0.025$ and $V_{\rm eff}=f_\Omega\frac{4\pi}{3}\left(D_{\rm max}^3-D_{\rm min}^3\right)\simeq1.99\times10^6~{\rm Mpc^3}$. 

The archival matching uses a broad 10--30~d rest-frame peak-delay window to remain inclusive across SE-SN diversity. For the rate calculation, however, $\Delta t_{\rm eff}$ represents the width of the relevant rise-time distribution rather than the mean explosion-to-peak delay. Type~IIb SNe span approximately 16--26~d in the rest frame\maincite{2015A&A...574A..60T}; we therefore adopt $\Delta t_{\rm eff}=10$~d. Using local volumetric rates $R_{\rm SE}=2.2^{+0.8}_{-0.6}\times10^{-5}~{\rm Mpc^{-3}\,yr^{-1}}$ and $R_{\rm IIb}=0.8^{+0.6}_{-0.4}\times10^{-5}~{\rm Mpc^{-3}\,yr^{-1}}$\citemeth{2025A&A...703A..34P}, the Poisson probability $P(\geq1)=1-\exp(-N_{\rm exp})$ is approximately 0.69 for the full SE population and 0.35 for Type~IIb SNe.

These values are approximate and do not include trials factors, the detailed ZTF/BTS detection and classification selection function, or the rarity of the specific fast and broad-lined properties of SN~2019rsq. In particular, $P_{\rm IIb}\simeq0.35$ refers to the full Type~IIb population rather than an SN~2019rsq-like subclass. We do not assign a formal probability to the narrower phenomenological class because its volumetric rate and selection function are not established. Increasing the effective temporal width would only increase the expected number of unrelated SNe and make the optical coincidence less constraining (\suppref{supp:stats}). Thus, the presence of a generic SE or Type~IIb SN is not itself the principal statistical anomaly; the unusual element is the clustered GW pair, with the supernova supplying the spatial, distance, temporal and stellar collapse anchor for a common physical interpretation.

\section{Gravitational wave event robustness checks}
\label{sec:SI3}

We assess whether S190910d and S190910h remain signal-like under independent re-analysis and whether their inferred source properties remain compatible with a common-origin interpretation involving SN~2019rsq. We test the two public alerts with complementary analyses: offline matched-filter recovery, intrinsic parameter inference for S190910d, SN-conditioned inference for S190910h and a geometric amplitude test for S190910d. Unless stated otherwise, parameter-estimation masses are source-frame quantities and search-template masses are detector-frame quantities.

\subsection{Public-alert context}

S190910d was reported by the LIGO Scientific Collaboration\citemeth{2009RPPh...72g6901A} and Virgo Collaboration\citemeth{2012JInst...7.3012A} on 2019 September 10 at 01:26:19.243 UTC through the General Coordinates Network (GCN)\citemeth{2019GCN.25695....1L}. It was identified in low latency by the SPIIR\maincite{spiir_7} and MBTAOnline\citemeth{2016CQGra..33q5012A} pipelines with a false-alarm rate of $3.7\times10^{-9}$~Hz, approximately one false alarm per eight years, and a 98\% NSBH classification with ${\rm HasNS}>99\%$. Approximately 7.1~hr later, S190910h was identified by GstLAL\citemeth{2017PhRvD..95d2001M} and reported at 08:29:58.544 UTC\citemeth{2019GCN.25707....1L}. Its false-alarm rate was $3.6\times10^{-8}$~Hz, approximately one per ten months, with a 61\% BNS classification, ${\rm HasNS}>99\%$ and ${\rm HasRemnant}>99\%$.

Both candidates triggered substantial electromagnetic follow-up but were treated as independent alerts; their probability maxima differ, and SN~2019rsq was discovered only several weeks later. The triplet therefore emerged only retrospectively. 
Both S190910d and S190910h were disseminated as public GW candidates during O3a and were not retracted. Neither was recovered above the offline catalogue threshold in GWTC-2, while both are listed among the O3a public alerts in GWTC-3\citemeth{2021PhRvX..11b1053A,2023PhRvX..13d1039A}. The contemporaneous follow-up is summarized in \suppref{supp:gwdetails}.

\subsection{Matched-filter recovery}
\label{sec:gw_search}

We re-analysed both candidates with the SPIIR matched-filter framework\maincite{spiir_7}, using updated data conditioning and a local background from the approximately 7.7-day segment immediately preceding the events. S190910d is recovered with network SNR of $\rho_{\rm net}=8.6$ and a false-alarm rate of $4.4~{\rm yr}^{-1}$ by strongly asymmetric templates with detector-frame chirp mass near $9\,M_\odot$. S190910h is recovered in L1 with SNR 8.2, a false-alarm rate of $14.1~{\rm yr}^{-1}$ and $\mathcal{M}^{\rm det}\simeq1.53\,M_\odot$.

S190910d is independently recovered by the IAS O3a analysis as GW190910\_012619, with network SNR 8.24 and $p_{\rm astro}=0.58$\maincite{Olsen2022IAS,IASO3aData}; the IAS inference also favours a strongly asymmetric source-frame mass region of order $34+3\,M_\odot$. For both candidates, the matched-filter time series follows the expected template-autocorrelation morphology and each ranks above the loudest trigger in the corresponding local-background segment (Extended Data Fig.~\ref{fig:gwrobust}). These diagnostics support treating both alerts as signal-like candidates but do not establish either as a confident GW detection. Detector-specific morphology, background ranking and search-template realizations are given in \suppref{supp:gwdetails}.

\subsection{Source-parameter inference and common-source consistency}
\label{sec:gw_source_inference}

For S190910d, we performed dedicated precessing parameter-estimation analyses with the IMRPhenomXPHM waveform model\citemeth{Pratten2021IMRPhenomXPHM}. Because the preferred solution is highly asymmetric, we tested explicitly whether the inferred mass ratio is driven by the lower prior boundary by repeating the analysis with $q\geq0.05$ and $q\geq0.01$, while otherwise retaining the same setup. The resulting posteriors are nearly identical (Extended Data Fig.~\ref{fig:gw_parameter_inference}), with median mass ratios $q=0.0858$ and $q=0.0856$ and no accumulation at either lower boundary. The source-frame component masses are correspondingly stable, with representative values near $m_1^{\rm src}\simeq34\,M_\odot$ and $m_2^{\rm src}\simeq2.9\,M_\odot$. The strongly asymmetric solution is therefore not a prior-boundary artefact and overlaps the mass region favoured by the IAS analysis. Spin information is less robust and is treated separately in \suppref{supp:gwdetails}.

For S190910h, we directly tested the common-source hypothesis by conditioning the parameter estimation on SN~2019rsq. The sky position was fixed to that of the SN and the luminosity distance assigned a finite prior of 110--160~Mpc centred on the redshift-inferred distance of 140.6~Mpc. The conditioned posterior (Extended Data Fig.~\ref{fig:gw_parameter_inference}) remains concentrated in a low-mass source-frame region near $m_1^{\rm src}\simeq2.5\,M_\odot$ and $m_2^{\rm src}\simeq1.3\,M_\odot$, and its chirp-mass scale remains consistent with the independently recovered blind matched-filter solution. Thus, conditioning on the SN position and distance preserves the low-mass character inferred from the search.

Applying the same fully conditioned strategy to S190910d produces a substantially less informative posterior. We therefore do not use that posterior as quantitative evidence for the association. Instead, the intrinsic mass scale is taken from the prior-robust precessing analysis above, while compatibility with the SN distance is tested independently through the observed GW amplitude. The asymmetric treatment of the two candidates is therefore data-driven: SN-conditioning preserves an informative low-mass solution for S190910h, whereas for S190910d the robust intrinsic mass inference and the more degenerate extrinsic amplitude consistency are tested separately.

\subsection{Geometric amplitude consistency of S190910d}
\label{sec:S190910d_projection}

We tested whether the modest observed SNR of S190910d is compatible with a source at the SN position and distance despite the strong amplitude--orientation degeneracy of a near-threshold signal. We adopted the representative source-frame configuration $34+3\,M_\odot$ favoured by the precessing and IAS analyses, fixed the sky position and $D_L=140.6$~Mpc to SN~2019rsq, and converted the masses to the detector frame for waveform generation.

Using the Advanced LIGO--Virgo detector responses and noise power spectral densities at the event time, we generate frequency-domain waveforms with the aligned-spin IMRPhenomD approximant\citemeth{Khan2016IMRPhenomD}. To isolate geometric projection, all parameters except binary orientation are fixed, and we Monte Carlo sample isotropically distributed inclinations and polarization angles. Approximately 6.6\% of sampled orientations yield $\rho_{\rm net}\leq8.6$ (Extended Data Fig.~\ref{fig:gw_orientation}). The low-SNR tail is concentrated toward relatively unfavourable, more edge-on projections, showing that geometric suppression alone can reproduce an observed amplitude as low as S190910d for a representative source at the SN distance. This is a geometric consistency test rather than a full parameter-estimation analysis: the 6.6\% fraction is neither an association probability nor a measurement of the binary inclination. The waveform inputs, orientation sampling and limitations of the calculation are given in \suppref{supp:gwdetails}.

The two events also sample substantially different detector geometries. Their separation is 7.0609~hr, corresponding to a sidereal rotation of approximately $106.2^\circ$ relative to the SN direction. At L1, the polarization-response vector $(F_+,F_\times)$ changes from approximately $(-0.221,-0.643)$ at S190910d to $(-0.669,+0.345)$ at S190910h, a rotation of $98.4^\circ$ in the response plane while the total antenna magnitude changes by only about 11\%. This does not change the physical binary inclination and is not an inclination measurement; it shows that even signals from the same sky position, and potentially from binaries sharing a similar disk orientation, can be projected very differently onto the detector network.

\section{Supernova observations and physical modelling}
\label{sec:SI4}

\subsection{Discovery and photometry}

SN~2019rsq was discovered by the Pan-STARRS survey (PS1) as PS19fsm in the $w$ band on 2019 October 1 UT\citemeth{2016arXiv161205560C,2019TNSTR1982....1C} and was subsequently detected by ZTF\citemeth{2019PASP..131a8002B,2019PASP..131g8001G} and the Asteroid Terrestrial-impact Last Alert System (ATLAS)\citemeth{2018PASP..130f4505T,2020PASP..132h5002S,2021TNSAN...7....1S}. The transient is located at $\alpha=03^{\rm h}14^{\rm m}37.73^{\rm s}$, $\delta=-07^\circ00'44.77''$ (J2000), approximately 9.9~kpc in projection from the centre of its host galaxy. Narrow host-galaxy emission lines give $z=0.0312126$, corresponding to $D_L\simeq140.6$~Mpc under the adopted cosmology. SN~2019rsq was initially classified as a broad-lined Type~Ic event\citemeth{2019TNSCR2026....1D} and later reclassified as Type~IIb as hydrogen and helium features emerged\citemeth{2024TNSCR3148....1P}.

Follow-up photometry was obtained as part of public ZTF survey observations\citemeth{2019PASP..131f8003B} with the ZTF camera\citemeth{2020PASP..132c8001D} on the Palomar 48-inch telescope (P48) in the $g$ and $r$ bands. The standard ZTF data processing and archive are described by Masci et al.\citemeth{2019PASP..131a8003M}. The P48 images were retrieved from the IPAC archive\footnote{https://irsa.ipac.caltech.edu/} and independently reduced using the \textsc{ECsnoopy} pipeline\footnote{https://sngroup.oapd.inaf.it/ecsnoopy.html}. Archival ZTF images were used as subtraction templates and transient fluxes were measured on the difference images using point spread function (PSF) fitting photometry. Science, reference and difference images were inspected epoch by epoch; representative and complete subtraction sequences are shown in Extended Data Fig.~\ref{fig:ztf_subtraction}. Additional pipeline and quality-control details are given in \suppref{supp:optical_reduction}.

For all quantitative analyses, measurements with ${\rm S/N}\geq3$ are treated as detections. Lower-significance measurements are treated as non-detections and represented by $5\sigma$ upper limits derived from the local background noise; they are excluded from the fitted light curves. We supplement the ZTF measurements with public ATLAS forced photometry in the $c$ and $o$ bands\citemeth{2018PASP..130f4505T,2020PASP..132h5002S,2021TNSAN...7....1S} obtained through the ATLAS forced-photometry service\footnote{https://fallingstar-data.com/forcedphot/}, and with PS1 $w$- and $i$-band measurements\citemeth{2016arXiv161205560C,2020ApJS..251....7F}. All photometry is reported in the AB system. Galactic foreground reddening is corrected using $E(B-V)=0.007$~mag\citemeth{2011ApJ...737..103S} and $R_V=3.1$\citemeth{1989ApJ...345..245C}. The complete photometric dataset, including upper limits, will be released through WISeREP as described in the Data Availability statement.

\subsection{Spectroscopic evolution and classification}

We analysed five spectra at rest-frame phases $-0.4$, $-0.1$, $+1.5$, $+4.4$ and $+21.4$~d relative to $g$-band maximum. Lick/KAST\citemeth{2019TNSCR2866....1H}, P200/DBSP\citemeth{2019TNSCR2026....1D} and Keck/LRIS\citemeth{2024TNSCR3148....1P} spectra are available through WISeREP\footnote{https://www.wiserep.org/}; two P60/SEDM spectra will be deposited before publication.

The sequence evolves from a broad-lined, classification-ambiguous appearance to a clear Type~IIb spectrum (Extended Data Fig.~\ref{fig:spec1}). The early $\sim6100$~\AA\ absorption can be high-velocity H$\alpha$ ($\sim2.2\times10^4~{\rm km\,s^{-1}}$) or Si~II~$\lambda6355$; the absence of clear H$\beta$ prevents a unique early identification. Later hydrogen and helium features secure the Type~IIb classification. Independent NGSF fitting\citemeth{2022TNSAN.191....1G} likewise admits both Type~IIb and Ic-BL templates near maximum but increasingly favours Type~IIb at later phases (Extended Data Fig.~\ref{fig:spec}).
For external phenomenological context, Extended Data Fig.~\ref{fig:spec1} also includes two spectra of SN~2025ulz and the currently available late-time spectrum of SN~2025adtq. SN~2025ulz shows similarly broad early-time features followed by Type~IIb-like spectral evolution. 

No statistically significant Na~I~D absorption is detected at the host redshift\citemeth{2012MNRAS.426.1465P}. Together with the continuum shapes and the absence of a systematic red colour offset relative to other Type~IIb SNe, this indicates that substantial additional host extinction is not required. We therefore apply only the Galactic foreground correction in the primary optical analysis.

\subsection{Physical light curve modelling}
\label{sec:lc_model}

We used two complementary approaches to test whether the optical evolution of SN~2019rsq can be reproduced by physically reasonable Type~IIb explosions and whether an explosion near the GW activity is compatible with the observed light curve.

First, we used \textsc{TransFit}\citemeth{2025ApJ...992...20L}, a radioactive-diffusion model for homologously expanding ejecta. We varied $M_{\rm ej}$, $v_{\rm ej}$, $M_{^{56}{\rm Ni}}$, optical opacity $\kappa$ and $t_{\rm shift}$, exploring approximately $10^4$ realizations around the nominal solution and evaluating them with $\chi^2$ (Extended Data Fig.~\ref{fig:transfit_parameter_space}). This is a local parameter-space exploration, not an MCMC posterior or set of posterior credible intervals.

The available photometry does not contain an identifiable shock-breakout or early shock-cooling component and therefore does not independently determine the explosion epoch. We instead use the modelling as a conditional consistency test: solutions are required to reproduce the observed photometry while placing the explosion within approximately three days of S190910d. Acceptable solutions occur on both sides of the GW trigger, including solutions in which the explosion precedes S190910d. The 100 best-fitting accepted realizations are shown in Fig.~\ref{fig:lc2019rsq}. Thus, the photometry does not require the explosion to occur exactly at the GW time, but permits stellar collapse within the same few-day interval.

As an independent test, we used \textsc{SNEC}\citemeth{Morozova_2015} with rotating binary-evolution structures computed with \textsc{MESA}\citemeth{Long_2022,Paxton_2011,Paxton_2013,Paxton_2015,Paxton_2018,Paxton_2019,Jermyn_2023}. We varied the mass cut/ejecta mass, kinetic energy, $^{56}$Ni mass and residual-envelope radius and compared synthetic $g$- and $r$-band light curves with the data. Implementation details are given in \suppref{supp:optical_reduction}.

The representative model shown in Fig.~\ref{fig:lc2019rsq} originates from a rotating stellar-evolution model with zero-age main-sequence (ZAMS) mass $M_{\rm ZAMS}=17\,M_\odot$ and a pre-SN mass $\simeq5\,M_\odot$. An imposed $2\,M_\odot$ mass cut gives $M_{\rm ej}\simeq3\,M_\odot$, with $E_{\rm K}\simeq10^{51}$~erg and $M_{^{56}{\rm Ni}}=0.08\,M_\odot$. This model reproduces the overall luminosity scale and temporal evolution when placed near S190910d, but is an illustrative ejecta configuration rather than a unique determination of the progenitor's initial or final mass. The independent TransFit solutions favour the same qualitative low-ejecta-mass regime, with a relatively stable radioactive-nickel scale of order $0.1\,M_\odot$.

Taken together, the diffusion-based and radiation-hydrodynamic calculations show that SN~2019rsq can be reproduced with physically plausible Type~IIb ejecta and that an explosion coincident with, or within a few days of, S190910d is not excluded. Details of the TransFit exploration and SNEC setup are given in \suppref{supp:optical_reduction}.

\subsection{Ejecta mass and total collapsing mass}

The few-solar-mass ejecta favoured by the rapid optical evolution do not imply a comparably low total collapsing mass. In black-hole-forming collapse, material can be partitioned among the central black hole, accretion disk, fallback and unbound ejecta. General-relativistic calculations of rapidly rotating 35--45\,$M_\odot$ progenitors produce massive BH--torus systems with ejecta of order a few solar masses\maincite{Fujibayashi2024}, while progenitor-dependent calculations span a broad ejecta-mass range for rapidly rotating stars\maincite{CrosatoMenegazzi2025}. These models are not fits to SN~2019rsq; they establish that its inferred ejecta scale is physically compatible with a much larger bound mass.

\subsection{Progenitor stripping and angular-momentum consistency}

The energetic, initially broad-lined appearance of SN~2019rsq is also compatible with its later Type~IIb classification. Energetic or broad-lined Type~IIb explosions are not unprecedented. SN~2003bg exhibited broad, high-velocity spectra and was modelled with
$M_{\rm ej}\sim4\,M_\odot$ and $E_{\rm k}\sim5\times10^{51}$~erg, with an inferred progenitor mass of approximately $20$--$25\,M_\odot$\maincite{Mazzali2009}. Similarly, detailed modelling of the transitional Type~Ib/IIb SN~2020cpg favoured $M_{\rm ej}\sim7\,M_\odot$, $E_{\rm k}\sim6\times10^{51}$~erg and a progenitor of approximately $25$--$30\,M_\odot$\maincite{Teffs2022}. These events demonstrate that Type~IIb phenomenology can extend into an energetic, broad-lined regime associated with comparatively massive stripped progenitors.

Still more massive Type~IIb progenitors are not excluded. Pre-explosion imaging of SN~2008ax admitted a single-star solution with an initial mass of approximately $28\,M_\odot$, although a lower-mass interacting-binary origin remained viable\citemeth{Crockett2008SN2008ax}. Stellar-evolution calculations likewise show explicitly that, for the adopted wind prescription, a $32.5\,M_\odot$ single-star model can retain approximately $0.5\,M_\odot$ of hydrogen at explosion and produce a Type~IIb outcome\citemeth{Claeys2011IIb}. These examples do not determine the progenitor mass of SN~2019rsq, but they show that a Type~IIb classification is compatible with a comparatively massive stripped progenitor, supporting the large stellar mass reservoir required by our current fragmenting-collapsar interpretation.

Radiative-transfer calculations show that small residual hydrogen masses can still produce detectable H$\alpha$ and Type~IIb-like spectral morphology\citemeth{Dessart2011,Hachinger2012}, while binary-progenitor calculations and pre-explosion constraints likewise support the survival of modest hydrogen envelopes in strongly stripped Type~Ib/IIb progenitors\citemeth{GilkisArcavi2022}. The Type~IIb classification therefore constrains the outer composition rather than requiring a low-mass central remnant or low-energy collapse.

An extended collapsar disk additionally requires sufficient specific angular momentum, independently of the ejecta-mass constraint. Rapidly rotating, strongly stripped stellar models can retain disk-forming angular momentum to collapse\maincite{Renzo2026}. The illustrative MESA progenitor used for the light-curve calculation also contains high-$j$ material, providing an internal disk-formation consistency check, but it is not used to infer the fragmentation radius. The corresponding $R_{\rm circ}$ calculation is given in \suppref{supp:optical_reduction}; the characteristic $100$--$300\,R_g$ scale in Section~\ref{sec:SI6} instead comes from dedicated collapsar-disk calculations\maincite{2007ApJ...658.1173P,2024ApJ...971L..34M,2025ApJ...991L..22C}.

\subsection{Comparison with other SE supernovae}
\label{sec:lc_compare}

We compared SN~2019rsq with benchmark Type~IIb SNe, rapidly evolving Type~IIb events and energetic or transitional SE explosions; the full comparison set and source references are given in \suppref{supp:optical_reduction}. SN~2025ulz and SN~2025adtq are included only as post-O4b phenomenological comparisons.

SN~2019rsq lies toward the rapidly evolving end of the Type~IIb population (Extended Data Fig.~\ref{fig:lc-compare}), overlapping the fast, FBOT-like Type~IIb events identified in wide-field surveys\maincite{ho2023}. Adopting S190910d as the collapse epoch gives a rest-frame rise of approximately 24~days, within the observed Type~IIb distribution\maincite{2015A&A...574A..60T}; this is a consistency check, not an independent explosion-time measurement. Its colour evolution remains broadly within the Type~IIb locus (Extended Data Fig.~\ref{fig:color}), supporting the classification and low host extinction. Detailed comparison choices are given in \suppref{supp:optical_reduction}.

\section{Host galaxy properties}
\label{sec:SI5}

SN~2019rsq is associated with SDSS~J031437.71$-$070100.9\footnote{https://skyserver.sdss.org/dr12/en/tools/explore/Summary.aspx?id=1237649962991419522} at a spectroscopic redshift $z=0.0312126$\citemeth{2015ApJS..219...12A}. The host has a visibly disturbed morphology with extended tidal features, and the SN is projected onto a faint elongated structure consistent with a tidal tail extending from the main galaxy (Fig.~\ref{fig:localization}), suggesting an interacting or recently disturbed system.

We measured matched-aperture GALEX, Pan-STARRS1 and unWISE photometry with \textsc{HOSTPHOT}\citemeth{Muller-Bravo2022} and fitted the spectral energy distribution with \textsc{Prospector}\citemeth{2021ApJS..254...22J}. The fit gives $\log_{10}(M_\star/M_\odot)=10.05^{+0.08}_{-0.09}$, ${\rm SFR}=1.12\pm0.38~M_\odot\,{\rm yr^{-1}}$, and $\log_{10}({\rm sSFR/yr^{-1}})=-10.00^{+0.12}_{-0.17}$; the fitting assumptions are given in \suppref{supp:optical_reduction}.

The SN~2019rsq and SN~2025ulz hosts have comparable stellar masses to within approximately 0.1~dex, whereas the SN~2019rsq host has a higher SFR and sSFR by approximately 0.7--0.8~dex\maincite{2025ApJ...995L..27G}. Together with the disturbed morphology and projected SN location, these properties are consistent with recent star formation and a young massive-star progenitor population. They do not distinguish among SE progenitor channels or provide independent evidence for a specific compact object merger scenario.

\section{Disk fragmentation as a plausible framework}
\label{sec:SI6}

In the disk fragmentation scenario, a massive rotating star collapses to form a central compact object (typically a black hole) surrounded by a massive gaseous accretion disk, likely within minutes after the core collapse. As the disk cools through neutrinos, it can become gravitationally unstable and potentially fragment into one or more neutron stars\maincite{2024ApJ...971L..34M,2025ApJ...991L..22C}. 

Neutron stars form preferentially in the outer regions of the accretion disk where it becomes gravitationally unstable, at characteristic radii of order\maincite{2024ApJ...971L..34M}, $R_{\rm form} \sim 100-300\,R_{\rm g}$, where \(R_{\rm g} \equiv GM_{\bullet}/c^{2}\), \(M_{\bullet}\) is the mass of the central compact object, \(G\) is the gravitational constant, and \(c\) is the speed of light. The characteristic mass of the disk-formed neutron stars is set by the local Jeans mass and can span \(\sim 0.1-1\,M_{\odot}\), with larger values being more typical for higher \(M_{\bullet}\) \maincite{2024ApJ...971L..34M,2025ApJ...991L..22C}.  
Subsequent accretion and fragment--fragment mergers can further increase the masses of the disk-born compact objects. 
The stringent angular momentum requirements on the progenitor star to produce a large disk of radius $> R_{\rm form}$ likely necessitate strong binary interaction (e.g., tidal spin-up or even a stellar merger) shortly before collapse, naturally associating such events with stripped-envelope explosions and compact progenitors, consistent with the narrow, fast-evolving Type~IIb light curve observed for SN~2019rsq.  

There are two potential sources of GW emission in this environment, both expected to be approximately time-coincident (within hours to days or less) of the core collapse. First, disk-formed neutron stars may pair into binaries and merge through GW emission, producing one or more low-mass compact-binary merger signals\maincite{2024ApJ...971L..34M}. Second, either first-generation fragments or the remnants of earlier fragment mergers may subsequently inspiral and merge with the central compact object (typically a black hole) formed at collapse, generating additional GW events \maincite{2007ApJ...658.1173P}\citemeth{2026MNRAS.545f1835L}.

Regarding the first channel, and considering for simplicity a binary composed of two neutron stars of equal mass \(M_{\rm NS}\), the GW-driven inspiral time is given by\citemeth{1964PhDT........51P},
$\tau_{\rm NSNS} = \frac{5}{512}\,\frac{c^{5} a_0^{4}}{G^{3} M_{\rm NS}^{3}} \approx 1.3~{\rm hr}\;\left(\frac{a_0}{R_{\rm H}/3}\right)^{4}\left(\frac{R_{\rm form}}{300R_{\rm g}}\right)^{4}
\left(\frac{M_{\bullet}}{35\,M_{\odot}}\right)^{8/3}
\left(\frac{M_{\rm NS}}{2\,M_{\odot}}\right)^{-5/3}$, where in the second equality we have normalized the initial semi-major axis $a_0$ of the binary to a fraction of its maximum value set by the Hill radius, $R_{\rm H} \equiv R_{\rm form}
\left(\frac{M_{\rm NS}}{3M_{\bullet}}\right)^{1/3}$, near the formation location in the disk. Owing to the steep dependence on both the initial separation and the formation radius, modest variations in these quantities readily extend the GW inspiral timescale from approximately an hour to several hours or days. For example, at \(R_{\rm form}\sim300R_{\rm g}\), increasing the initial separation from \(R_{\rm H}/3\) to approximately \(R_{\rm H}/2\) lengthens the inspiral time to \(\sim6\)--7~hr. A compact binary formed in the outer disk can therefore merge on a timescale comparable to the observed \(\simeq7.1\)~hr interval between S190910d and S190910h.

Regarding the second scenario above, the characteristic GW inspiral time for a compact object of mass
\(M_{\rm NS}\) orbiting a central object of mass \(M_{\bullet} \gg M_{\rm NS}\) on a circular orbit starting from a semi-major axis \(a_0 \sim R_{\rm form}\) is 
\[
\tau_{\rm NSBH}
= \frac{5}{256}\,
\frac{c^{5} a_0^{4}}{G^{3} M_{\bullet}^{2} M_{\rm NS}}
\simeq 1.1~{\rm hr}\,
\left(\frac{R_{\rm form}}{100\,R_{\rm g}}\right)^{4}
\left(\frac{M_{\bullet}}{35\,M_{\odot}}\right)^{2}
\left(\frac{M_{\rm NS}}{3\,M_{\odot}}\right)^{-1},
\]
where the final scaling uses \(R_{\rm form}\) expressed in units of \(R_{\rm g}=GM_{\bullet}/c^2\). The resulting direct inspiral time is naturally of order hours to days, depending sensitively on the circularization radius and compact-object masses.  

A closely related realization has recently been proposed for GW190814, whose strongly asymmetric source-frame component masses of approximately $23\,M_\odot$ and $2.6\,M_\odot$ were interpreted as the merger of a disk-born compact object with the central black hole of a collapsar system\maincite{Baibhav2026}. This provides a useful precedent for the several-tens-plus-few-solar-mass source-frame configuration favoured here for S190910d. In the same framework, interactions among multiple disk fragments can broaden the subsequent merger-time distribution from prompt hour-to-day inspirals to delays of weeks or months.

These direct GW-driven inspiral timescales therefore provide a natural way to obtain multiple compact-object mergers within hours to days of the same collapse, comparable to the observed \(\simeq7\) hr separation between S190910d and S190910h. They should not, however, be interpreted as a hard maximum delay for all subsequent compact-object interactions. In a disk containing several fragments, close encounters and few-body scattering can redistribute orbital energy and angular momentum, placing some surviving fragments or merger remnants onto wider or more eccentric orbits and extending the delay to weeks or months\maincite{Baibhav2026}. For prompt hour-to-day mergers, the associated kilonova-like emission would occur while the SN ejecta are still optically thick and before the supernova becomes optically prominent, giving rise to an event recently labeled a ``super-kilonova''\maincite{2025ApJ...995L..59K}. More delayed interactions from the scattering channel could instead occur during the optically bright SN phase or at later times, depending on the post-scattering orbit\maincite{Baibhav2026}.

The first GW candidate, S190910d, favours a strongly asymmetric black-hole--compact-object configuration with representative source-frame component masses near $m_1^{\rm src}\simeq34\,M_\odot$ and $m_2^{\rm src}\simeq3\,M_\odot$, while the second, S190910h, is consistent with a substantially lower-mass BNS-like system (Section~\ref{sec:gw_source_inference}; Extended Data Fig.~\ref{fig:gw_parameter_inference}).
A possible interpretation is that the stellar core collapsed to form a massive black hole surrounded by a fragmenting accretion disk. In this picture, the \(\sim34\,M_\odot\) primary in S190910d can be naturally associated with the central black hole formed during collapse, while the \(\sim3\,M_\odot\) secondary may represent an accretion-grown disk fragment or the remnant of an earlier fragment--fragment merger. Such a compact object could form, or be driven onto, an orbit with a characteristic radius of order \(R_{\rm form}\sim100\)--\(200R_g\) and subsequently merge with the central black hole. For the representative masses adopted above, radii of \(\sim130\)--\(170R_g\) correspond to direct GW-driven inspiral times of approximately a few to \(10\) hr.
Elsewhere in the disk, two lower-mass fragments could form at larger characteristic radii, of order \(R_{\rm form}\sim300R_g\), become gravitationally bound, and merge as a compact binary. For binaries with initial separations approaching a substantial fraction of their Hill radius, for example \(a_0\sim R_H/2\), the corresponding inspiral time is of order several hours and can readily reach \(\sim7\) hr or longer. Modest variations in formation radius and initial separation can therefore produce a second low-mass merger several hours after the first, naturally encompassing the observed \(7.1\) hr separation between S190910d and S190910h. The two GW events need not represent successive stages of the same binary, but may instead arise from distinct compact-object interactions within the same fragmenting disk.

In some realizations of this scenario, the remnant of S190910h may later merge with the central black hole, potentially producing an additional GW signal. The delay of such an event need not be comparable to the direct inspiral time: if few-body scattering first places the remnant onto a wider orbit, the merger can be delayed to much later times. Detectability would also depend sensitively on mass ratio, inclination, and detector sensitivity at the time of the merger.
Additional considerations regarding fragment growth, merger ordering, hierarchical evolution and delayed interactions are discussed in \suppref{supp:dynamics}.

\section{Short-timescale detector coverage following the GW pair}
\label{sec:SI7}

Additional compact object interactions are possible in a multi-fragment disk, but their delays are not uniquely predicted. We therefore examined detector coverage during the week following S190910d only to test the prompt end of the allowed delay distribution.

We evaluated the normalized antenna-response proxy $\sqrt{F_+^2+F_\times^2}$ toward SN~2019rsq for H1, L1 and V1, together with the HL and HLV network responses and detector operating state (Extended Data Fig.~\ref{fig:detector_coverage}). The directional response varies substantially on hour-to-day timescales, with extended intervals of single-detector operation or incomplete coincidence. S190910h, S190912c and S190912d are marked only as temporal reference points.

This is a qualitative coverage diagnostic, not a detection-efficiency calculation. A prompt merger could have occurred during reduced directional sensitivity or incomplete network operation, whereas interactions delayed to weeks or months lie outside the interval considered. The absence of another public trigger in the first week therefore places only a limited constraint on additional prompt activity and no general constraint on delayed interactions. The latter motivate the auxiliary high-energy search in \suppref{supp:grb190915a}.

\section{Radio constraints on relativistic outflows}
\label{sec:radio}

To test for a radio-bright relativistic outflow, we examined public Australian SKA Pathfinder (ASKAP\footnote{http://data.csiro.au/})\citemeth{askap} and Very Large Array Sky Survey (VLASS\footnote{https://vlass-dl.nrao.edu/vlass/quicklook/})\citemeth{vlass} imaging. ASKAP spans approximately 0.86--1.66~GHz and VLASS provides coverage near 3~GHz. For each interferometric epoch we measured the local Stokes-$I$ image rms at the SN position. No post-explosion point source is detected above $5\sigma$, so the ASKAP/VLASS constraints are empirical $5\sigma$ limits, $F_{\nu,{\rm lim}}=5\sigma_{\rm rms}$ (Extended Data Fig.~\ref{fig:radio}); pre-explosion measurements are shown only for context.

We compare these limits with published radio measurements of the Type~IIb SNe SN~2011dh and SN~2003bg, the long-lived afterglow of GRB~030329\citemeth{vanderHorst2008,Mesler2013}, and SN~2025ulz\maincite{Tanner2026}. Because the comparison data were obtained at different frequencies, we do not apply a spectral rescaling; Extended Data Fig.~\ref{fig:radio} is intended to compare luminosity scales rather than construct a homogeneous single-frequency light curve.

We additionally obtained a 1800-s FAST L-band observation on 2026 August 4 beginning at 22:44:37.773 UTC (MJD~61256.9477), with a usable band of approximately 1.05--1.45~GHz. The radiometer equation gives a nominal thermal rms of $1.80~\mu{\rm Jy}$ and nominal $5\sigma$ sensitivity of $9.0~\mu{\rm Jy}$ at 1.25~GHz; the inputs and uncertainty propagation are given in \suppref{supp:radio}. This value is a nominal single-dish thermal sensitivity, not an empirical host-subtracted continuum upper limit, and is therefore shown separately from the ASKAP/VLASS measurements. The machine-readable radio measurements and observational metadata will be released through the analysis repository described in the Data Availability statement.

The ASKAP/VLASS non-detections do not exclude radio emission at luminosities typical of ordinary or radio-bright Type~IIb SNe, but they disfavour a long-lived, radio-bright outflow comparable to GRB~030329 at the distance of SN~2019rsq. They do not exclude weaker, off-axis, baryon-loaded or choked jets, or explosions in sufficiently low-density circumstellar environments\citemeth{Soderberg2004,Bietenholz2014,Bromberg2011}.

\clearpage
\bibliographystylemeth{naturemag}
\bibliographymeth{references}

\end{methods}

\begin{addendum}

\item[Acknowledgements] We thank Akash Anumarlapudi, Igor Andreoni and David L. Kaplan for advice and contributions related to the radio data and analysis, and Junjie Zhao for assistance with the gravitational-wave analysis. We thank Linqing Wen, Stephen Smartt, Matt Nicholl, Michael Coughlin and Thomas Dent for discussions and comments on the manuscript.

We thank the LVK Collaboration EM Follow-up, Low-Latency, Burst, and Multi-Messenger groups for questions and constructive discussions during presentations of this work as part of the LIGO Publications and Presentations (P\&P) review process. These discussions helped us clarify the physical interpretation and improve the overall self-consistency of the work.

This work made use of the data from FAST (Five-hundred-meter Aperture Spherical radio Telescope)(\url{https://cstr.cn/31116.02.FAST}). FAST is a Chinese national mega-science facility, operated by National Astronomical Observatories, Chinese Academy of Sciences.

Part of this research made use of the OzSTAR high-performance computing facility hosted by Swinburne University of Technology. OzSTAR is supported in part through the Australian Government’s National Collaborative Research Infrastructure Strategy (NCRIS) for astronomy, together with contributions from the Victorian Higher Education State Investment Fund (VHESIF) of the Victorian Government.

The Australian SKA Pathfinder is part of the Australia Telescope National Facility which is managed by CSIRO. Operation of ASKAP is funded by the Australian Government with support from the National Collaborative Research Infrastructure Strategy. ASKAP uses the resources of the Pawsey Supercomputing Centre. Establishment of ASKAP, the Murchison Radio-astronomy Observatory and the Pawsey Supercomputing Centre are initiatives of the Australian Government, with support from the Government of Western Australia and the Science and Industry Endowment Fund. We acknowledge the Wajarri Yamatji people as the traditional owners of the Observatory site.

This work relied on resources made available through the LIGO Laboratory, a major research facility funded by the U.S. National Science Foundation. We also acknowledge the computational infrastructure provided by the LIGO Laboratory with support from NSF Grants PHY-0757058 and PHY-0823459.

Pan-STARRS is primarily funded to search for near-Earth asteroids through NASA grants NNX08AR22G and NNX14AM74G. The Pan-STARRS science products were made possible through the contributions of the University of Hawai’i Institute for Astronomy, the Queen’s University Belfast and the University of Oxford.

\item[Funding] S.Y. acknowledges funding from the National Natural Science Foundation of China under grant Nos. 12303046 and 12433001, the Startup Research Fund of Henan Academy of Sciences No. 242041217, and the Joint Fund of Henan Province Science and Technology R\&D Program No. 235200810057. T.-W.C. acknowledges financial support from the Yushan Fellow Program of the Ministry of Education, Taiwan (MOE-111-YSFMS-0008-001-P1), and from the National Science and Technology Council, Taiwan (NSTC 114-2112-M-008-021-MY3). B.D.M. acknowledges support from the Simons Foundation (727700); the Flatiron Institute is supported by the Simons Foundation. M.B. acknowledges the Department of Physics and Earth Science of the University of Ferrara for financial support through the FIRD 2025 grant. Z.-Q.Y. is supported by the National Natural Science Foundation of China (Grant No. 12305059). The ASPIRE GW program under grant JPMJAP2320 also provided support for this project. 

\item[Contributions] S.Y. conceived and led the project, coordinated the observations and analyses, performed the archival and statistical analyses, interpreted the results and wrote the original manuscript. T.-W.C. contributed to the early development of the project, scientific interpretation and manuscript revision. W.G. led the gravitational-wave data analysis, parameter estimation and robustness tests. B.D.M. developed the theoretical interpretation of the system and contributed extensively to the physical modelling and manuscript development. Y.Z. carried out the supernova light-curve, analytic and hydrodynamic modelling and associated literature analyses, with contributions and supervision from H.G., L.-D.L. and Y.-W.Y. A.A. contributed to scientific discussions and interpretation. J.B. and J.Y. contributed the FAST observations and data analysis. T.d.B., C.-C.L., G.S.H.P. and R.J.W. provided Pan-STARRS data and survey support. M.B. contributed to the physical and theoretical modelling and interpretation. E.C. contributed to the coincidence rate analysis. S.C. contributed to the gravitational-wave signal analysis. M.M.K. provided ZTF data and SEDM spectroscopy and contributed extensively to the scientific interpretation and revision of the manuscript. D.K. contributed to the Type~IIb supernova analysis and scientific interpretation. C.-H.L. performed the host-galaxy analysis. J.S. contributed to the supernova data analysis and interpretation. Z.-Q.Y. contributed to the early development of the project and interpretation of the gravitational-wave results. X.-J.Z. contributed to the gravitational-wave and radio analyses and independently checked key results. Z.-H.Z. provided project supervision, funding acquisition and critical review of the manuscript. All authors discussed the results and reviewed the manuscript.

\item[Competing Interests] The authors declare no competing interests.

\item[Correspondence] Correspondence and requests for materials should be addressed to Sheng Yang~(email: sheng.yang@hnas.ac.cn) or Zong-Hong Zhu~(zhuzh@bnu.edu.cn).

\item[Data Availability]

The optical spectra of SN~2019rsq used in this study are available through the WISeREP archive (\url{https://www.wiserep.org/object/14508}); the two additional P60/SEDM spectra will be deposited in the same archive before publication. The complete photometric dataset used in this work, including the independently remeasured ZTF photometry, ATLAS forced photometry, Pan-STARRS measurements, non-detections and upper limits, will also be made publicly available through WISeREP.

The public GW alert localization products and metadata were obtained from GraceDB, queried on 2025 October 31. The ZTF Bright Transient Survey catalogue was queried on the same date. The ASKAP and VLASS imaging used in this work is publicly available through their respective archives.

The GW and supernova analysis samples, the machine-readable radio measurements and limits, the FAST observational metadata and nominal thermal sensitivity, and the numerical data required to reproduce the analyses and figures will be made publicly available through the accompanying analysis repository (\url{https://github.com/saberyoung/sn2019rsq-gw-analysis}). A fixed version associated with this Article will be archived in a persistent DOI-bearing repository before publication. Data underlying the analyses are available to editors and referees upon request during peer review.

\item[Code Availability]

Custom code used to construct the GW and supernova analysis samples, perform the archival GW--SN cross-match, calculate the Monte Carlo coincidence probabilities and geometric projection tests, carry out the analyses described in this work, and generate the figures will be made publicly available at \url{https://github.com/saberyoung/sn2019rsq-gw-analysis}. The exact version used for this study, together with the corresponding analysis inputs and environment information required for reproducibility, will be archived in a persistent DOI-bearing repository before publication. The code is available to editors and referees upon request during peer review.

\end{addendum}


\clearpage

\section*{Extended Data}
\setcounter{figure}{0}
\renewcommand{\figurename}{Extended Data Fig.}

\begin{figure*}[t]
  \centering
  \begin{subfigure}[t]{0.32\textwidth}
    \centering
    \includegraphics[width=\linewidth]{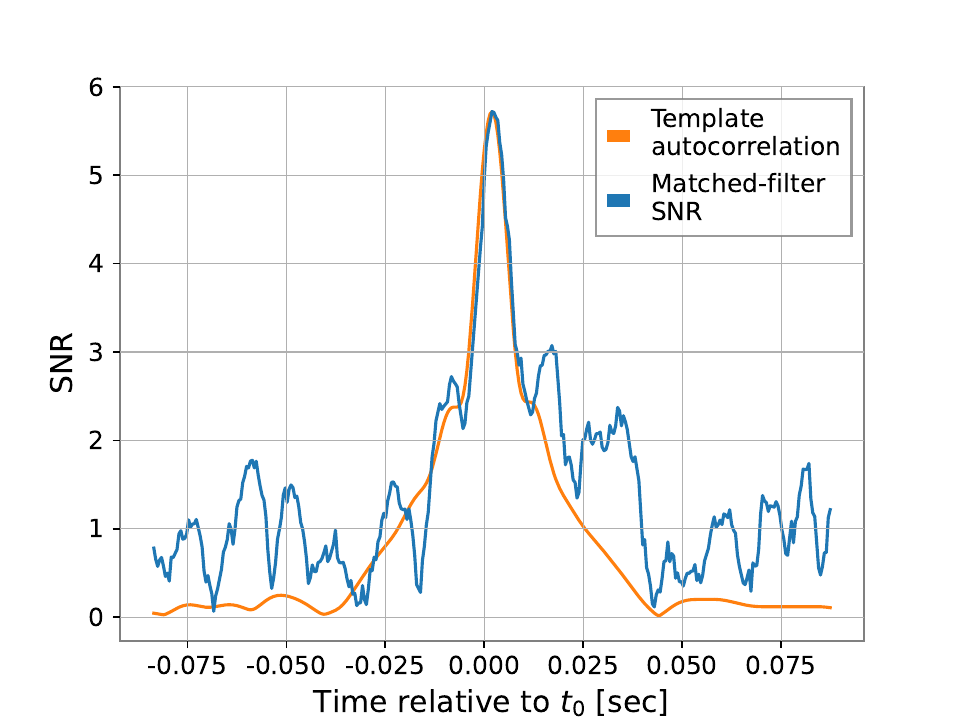}
    \caption{S190910d: H1}
    \label{fig:S190910d_snr_h1}
  \end{subfigure}\hfill
  \begin{subfigure}[t]{0.32\textwidth}
    \centering
    \includegraphics[width=\linewidth]{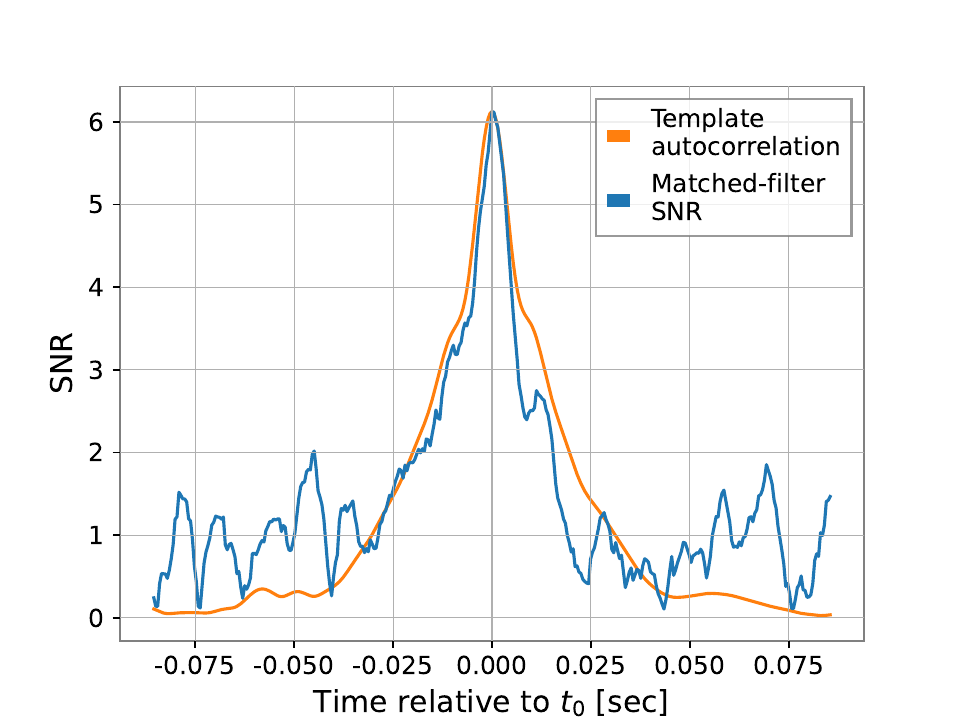}
    \caption{S190910d: L1}
    \label{fig:S190910d_snr_l1}
  \end{subfigure}\hfill
  \begin{subfigure}[t]{0.32\textwidth}
    \centering
    \includegraphics[width=\linewidth]{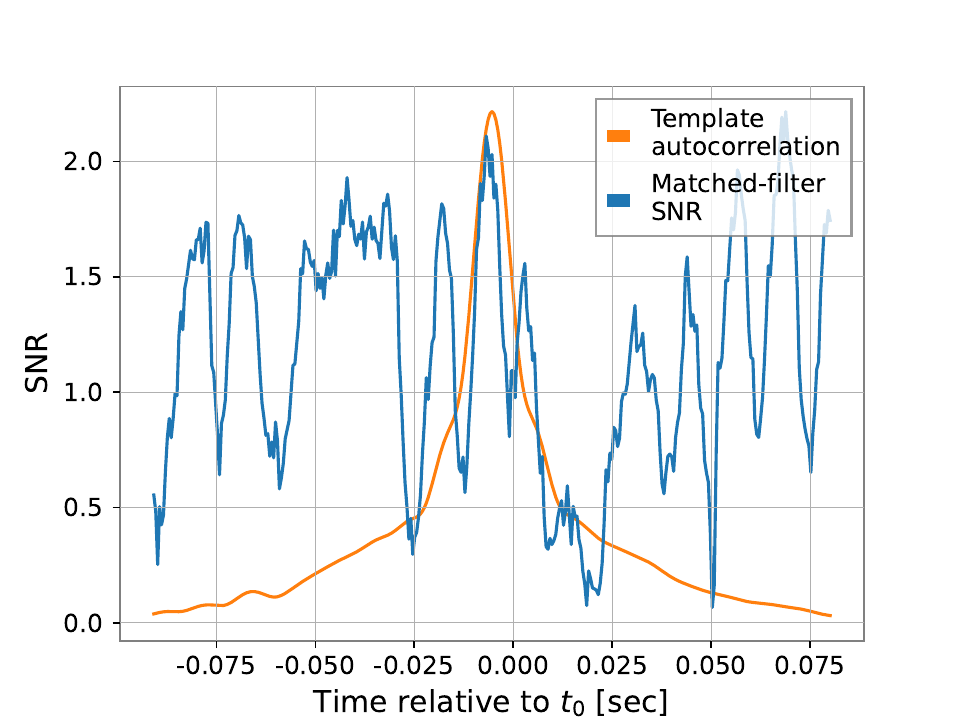}
    \caption{S190910d: V1}
    \label{fig:S190910d_snr_v1}
  \end{subfigure}
  \vspace{0.6em}
  \begin{subfigure}[t]{0.32\textwidth}
    \centering
    \includegraphics[width=\linewidth]{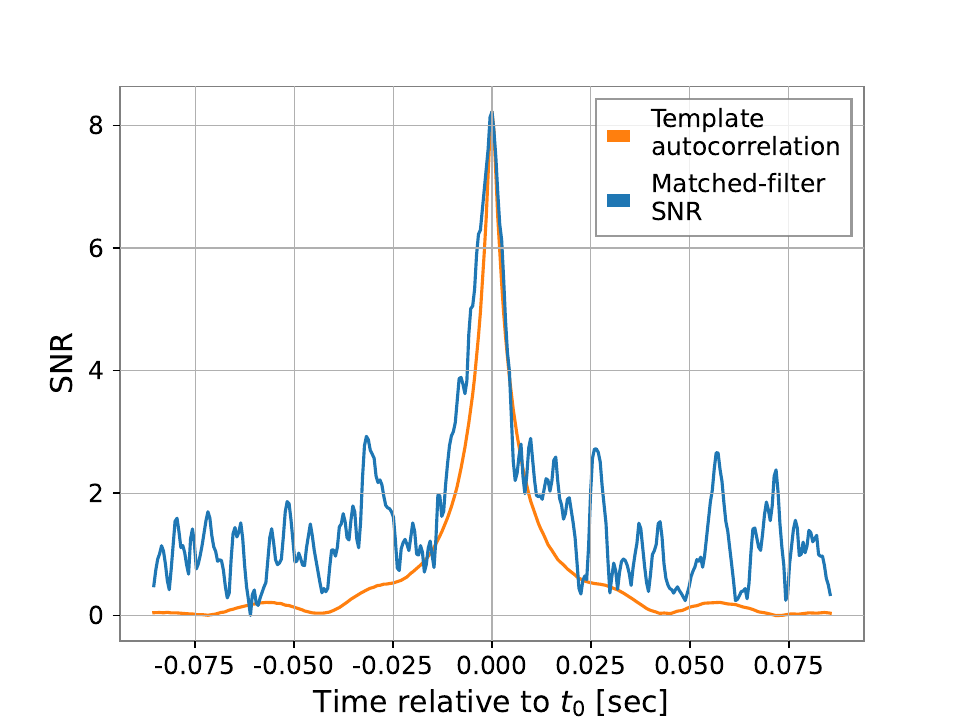}
    \caption{S190910h: L1}
    \label{fig:S190910h_snr_l1}
  \end{subfigure}\hfill
  \begin{subfigure}[t]{0.32\textwidth}
    \centering
    \includegraphics[width=\linewidth]{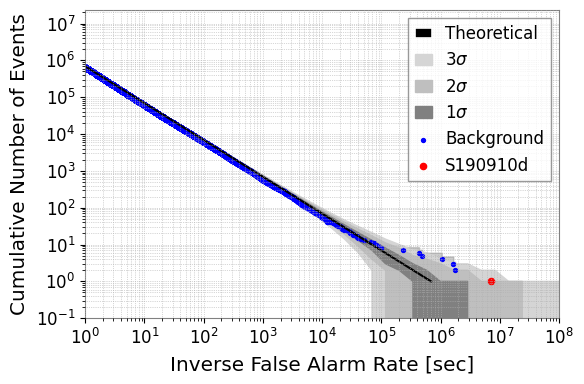}
    \caption{S190910d background}
    \label{fig:S190910d_bg}
  \end{subfigure}\hfill
  \begin{subfigure}[t]{0.32\textwidth}
    \centering
    \includegraphics[width=\linewidth]{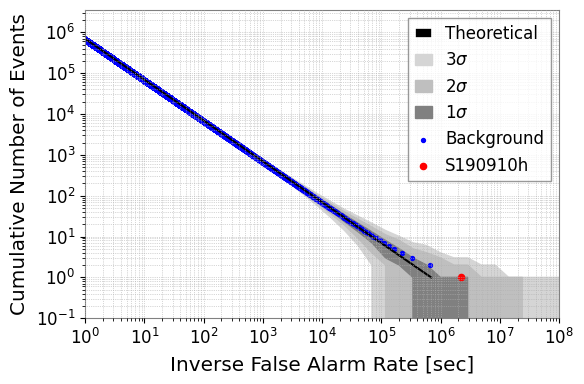}
    \caption{S190910h background}
    \label{fig:S190910h_bg}
  \end{subfigure}
  \caption{\textbf{Matched-filter morphology and local-background ranking of S190910d and S190910h.}
  \textbf{a--d,} Absolute matched-filter signal-to-noise ratio, $|\rho(t)|$, compared with the autocorrelation of the best-matching template, aligned to the merger time $t_0$. The template autocorrelation is normalized to unit peak and rescaled to the peak amplitude of $|\rho(t)|$, providing the expected time-domain envelope for a signal consistent with the recovered template. S190910d shows a clear template-consistent morphology in H1 and L1, with weaker correspondence in V1 owing to the lower sensitivity of that detector at the event time. S190910h was observed in L1 only, where the matched-filter time series closely follows the expected template envelope.
  \textbf{e,f,} Candidate ranking relative to coincident triggers in the approximately 7.7-day data segment preceding each event. The cumulative number of triggers is shown as a function of inverse false-alarm rate (IFAR); blue points denote the empirical background, the black curve shows the theoretical expectation, and grey bands indicate the corresponding $1\sigma$, $2\sigma$, and $3\sigma$ fluctuations. The candidate events are shown in red, with FARs of $4.4\,{\rm yr}^{-1}$ for S190910d and $14.1\,{\rm yr}^{-1}$ for S190910h. Both rank above the loudest background trigger in the analysed segment but remain of marginal blind-search significance. Together, the morphology and background diagnostics support treating the two events as signal-like candidates without reclassifying either as a confident gravitational-wave detection.}
  \label{fig:gwrobust}
\end{figure*}

\begin{figure*}[t]
    \centering
    \includegraphics[width=0.45\textwidth]{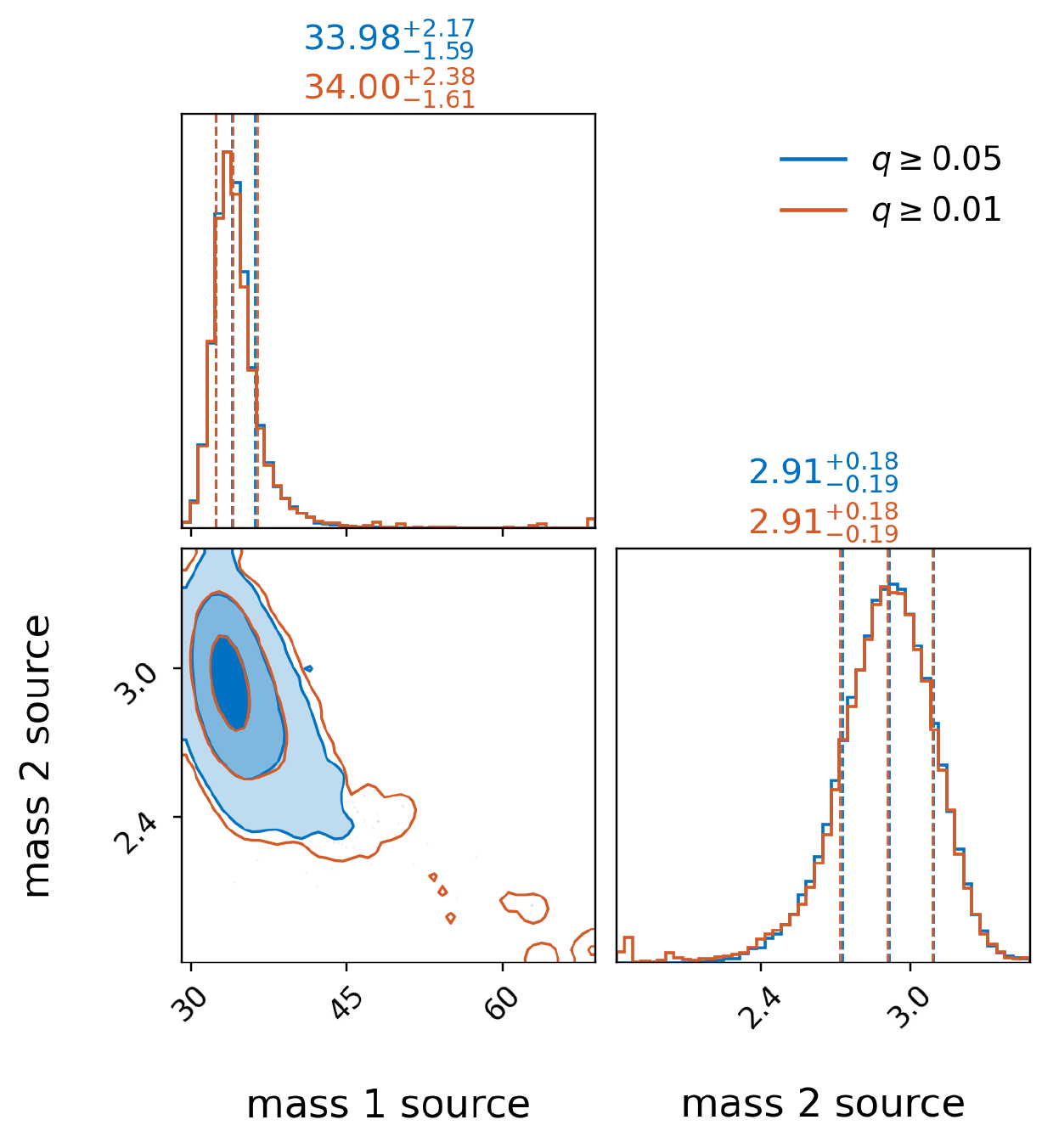}
    \hfill
    \includegraphics[width=0.48\textwidth]{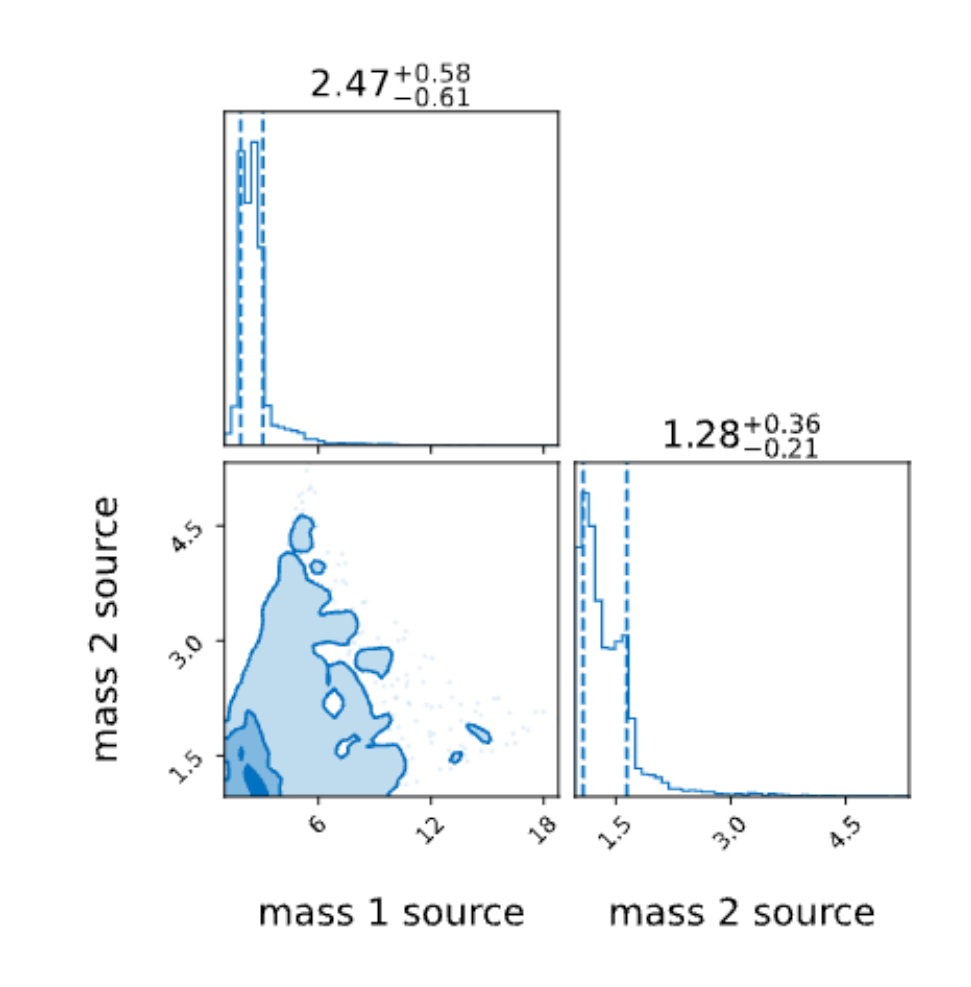}
    \caption{\textbf{Source-frame component-mass inference for S190910d and S190910h.}
    \textbf{Left,} Source-frame component-mass posteriors for S190910d from precessing parameter-estimation analyses with lower mass-ratio bounds of $q\geq0.05$ and $q\geq0.01$. The two posterior distributions are nearly identical, with median mass ratios $q=0.0858$ and $q=0.0856$, respectively, and neither accumulates at the lower prior boundary. The inferred component-mass region is therefore insensitive to extending the prior towards more asymmetric systems and favours a strongly asymmetric configuration with representative source-frame masses near $m_1^{\rm src}\simeq34\,M_\odot$ and $m_2^{\rm src}\simeq2.9\,M_\odot$. This mass region is independently consistent with that favoured by the IAS analysis of the same trigger.
    \textbf{Right,} Source-frame component-mass posterior for S190910h when the sky position is fixed to SN~2019rsq and the luminosity distance is assigned a finite $110$--$160$~Mpc prior centred on the supernova distance. The conditioned inference remains concentrated in a low-mass region near $m_1^{\rm src}\simeq2.5\,M_\odot$ and $m_2^{\rm src}\simeq1.3\,M_\odot$. The two panels serve different consistency tests: the S190910d analysis establishes a prior-robust intrinsic mass scale, whereas the S190910h analysis tests whether its low-mass solution is preserved under the SN~2019rsq common-source hypothesis.}
    \label{fig:gw_parameter_inference}
\end{figure*}

\begin{figure*}[t]
    \centering
    \includegraphics[width=0.48\textwidth]{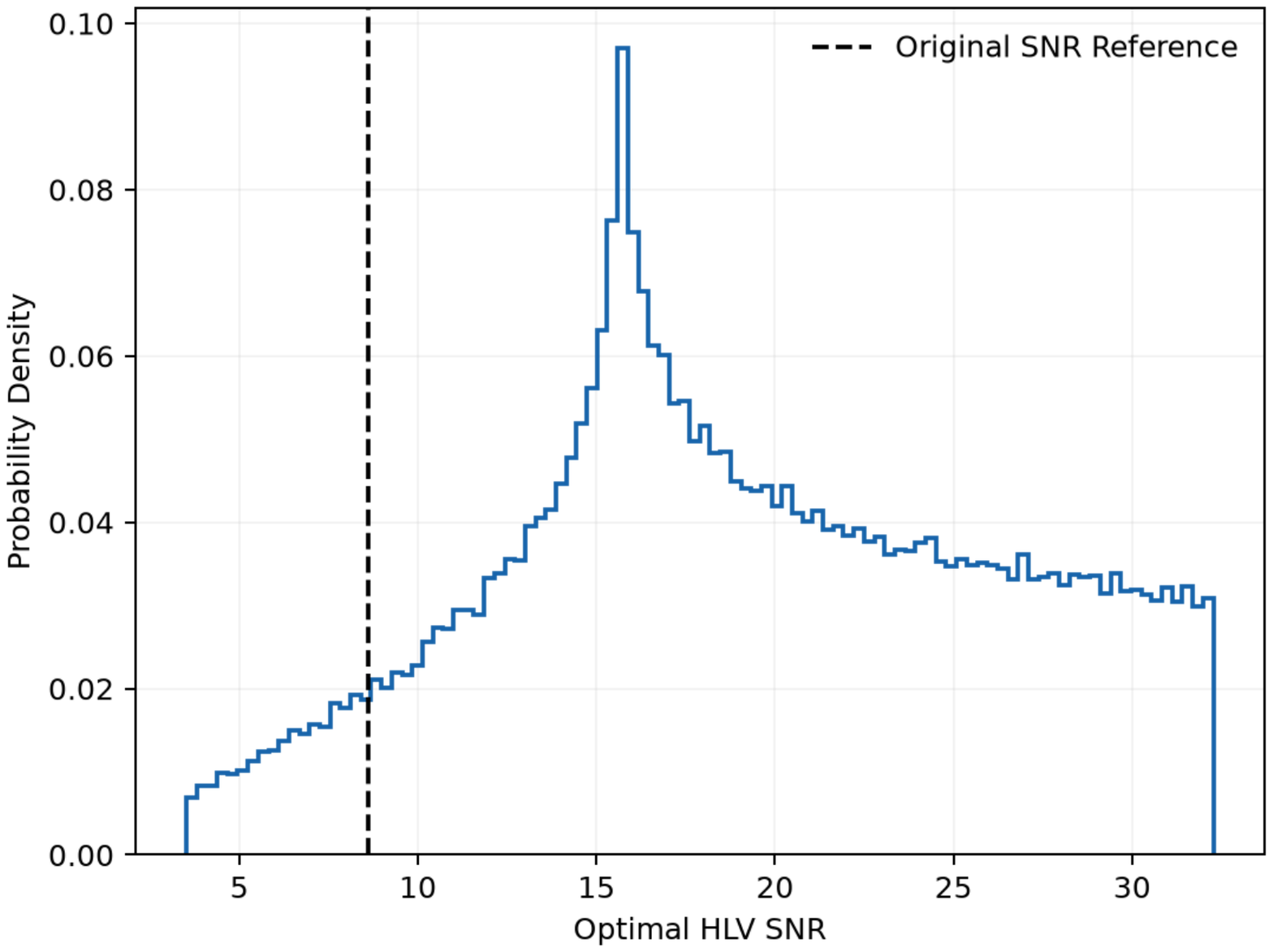}
    \hfill
    \includegraphics[width=0.48\textwidth]{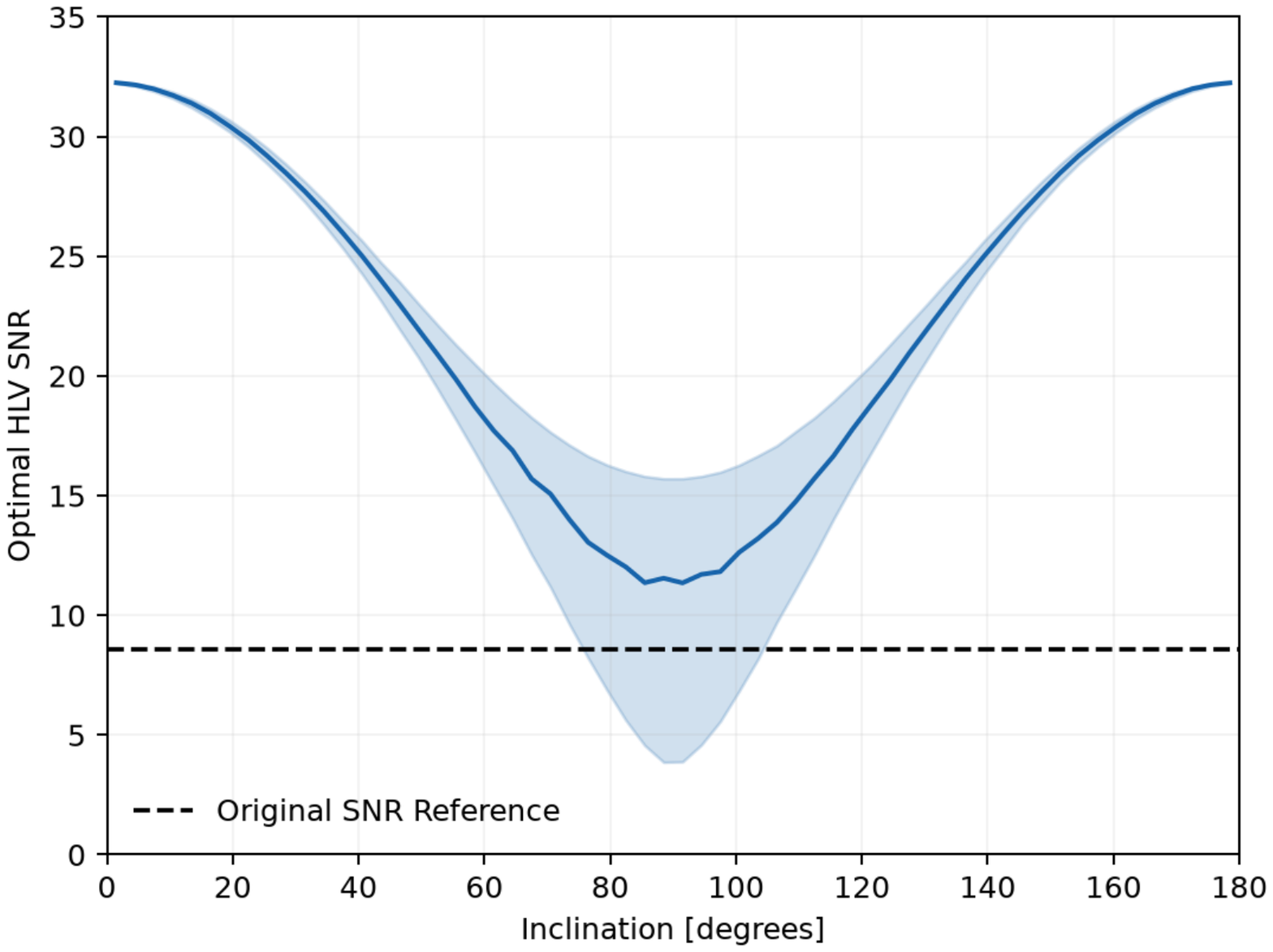}
    \caption{\textbf{Geometric amplitude consistency of S190910d with the distance of SN~2019rsq.}
    \textbf{Left,} Distribution of the optimal HLV network signal-to-noise ratio obtained by sampling isotropically distributed binary orientations for a representative S190910d-like source placed at the sky position and luminosity distance of SN~2019rsq, $D_L=140.6$~Mpc. The adopted source-frame component masses, $m_1^{\rm src}=34\,M_\odot$ and $m_2^{\rm src}=2.9\,M_\odot$, are converted to their corresponding detector-frame values at $z\simeq0.031$ for waveform generation. The vertical dashed line marks the observed reference value, $\rho_{\rm net}\simeq8.6$. Approximately $6.6\%$ of isotropically distributed orientations yield $\rho_{\rm net}\leq8.6$, demonstrating that geometric projection alone can reproduce an observed amplitude as low as that of S190910d for a source at the supernova distance.
    \textbf{Right,} Network SNR as a function of binary inclination. The solid curve shows the median network response, while the shaded region denotes the 5th--95th percentile range generated by variation in polarization angle. The response decreases strongly towards edge-on configurations, where projection and polarization effects can produce SNRs comparable to or below the observed value. This calculation is a geometric consistency test of the SN-distance hypothesis; it is neither a measurement of the binary inclination nor a probability that S190910d is physically associated with SN~2019rsq.}
    \label{fig:gw_orientation}
\end{figure*}

\begin{figure*}[t]
    \centering
    \begin{subfigure}[t]{0.72\textwidth}
        \centering
        \includegraphics[width=\linewidth]{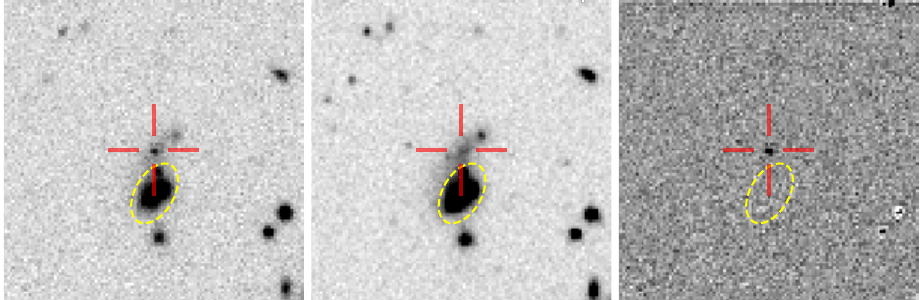}
        \caption{Representative science, reference and difference images.}
        \label{fig:subtraction_triplet}
    \end{subfigure}
    \vspace{0.8em}
    \begin{subfigure}[t]{0.48\textwidth}
        \centering
        \includegraphics[width=\linewidth]{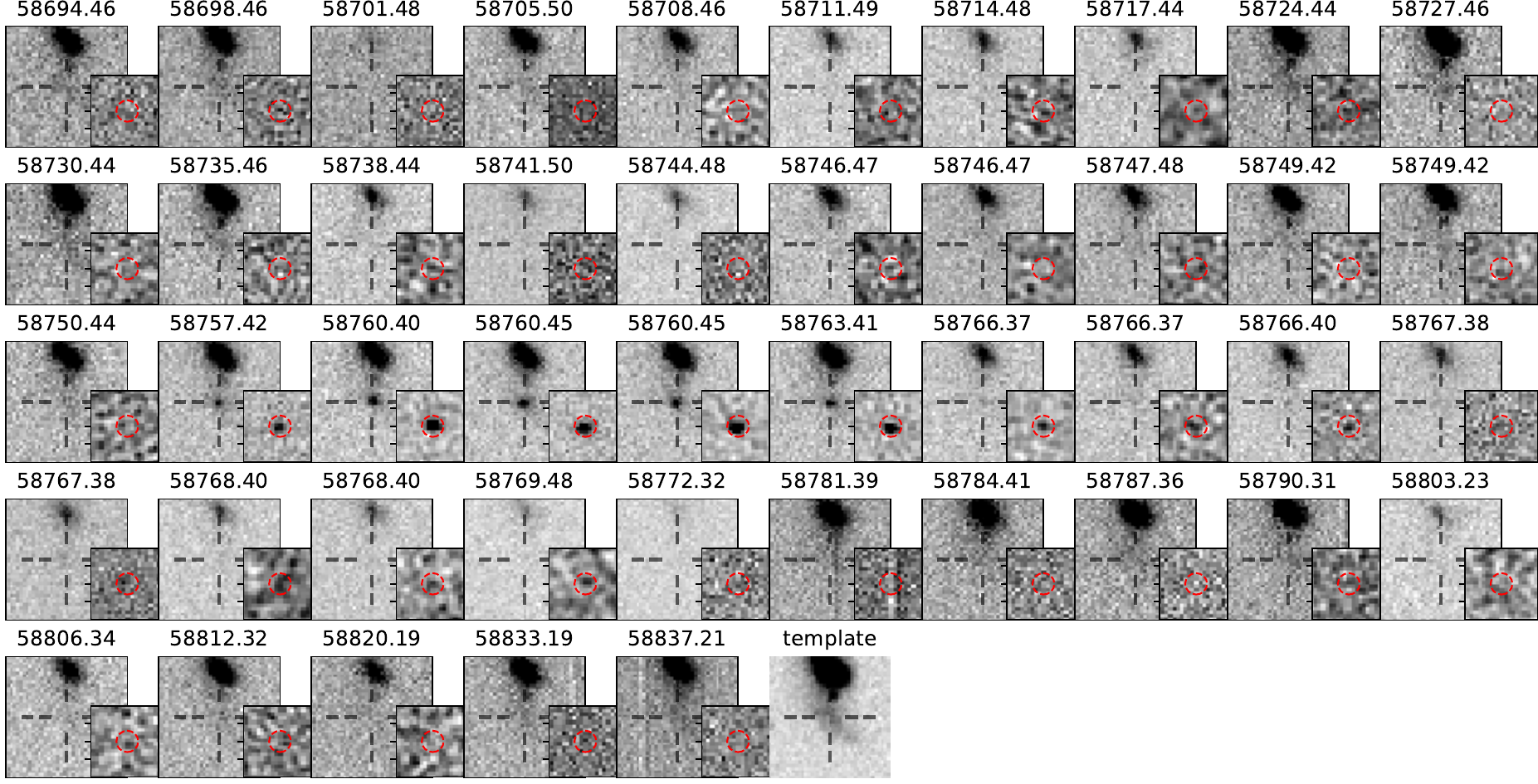}
        \caption{ZTF $g$-band subtraction sequence.}
        \label{fig:subtraction_g}
    \end{subfigure}
    \vspace{0.8em}
    \begin{subfigure}[t]{0.48\textwidth}
        \centering
        \includegraphics[width=\linewidth]{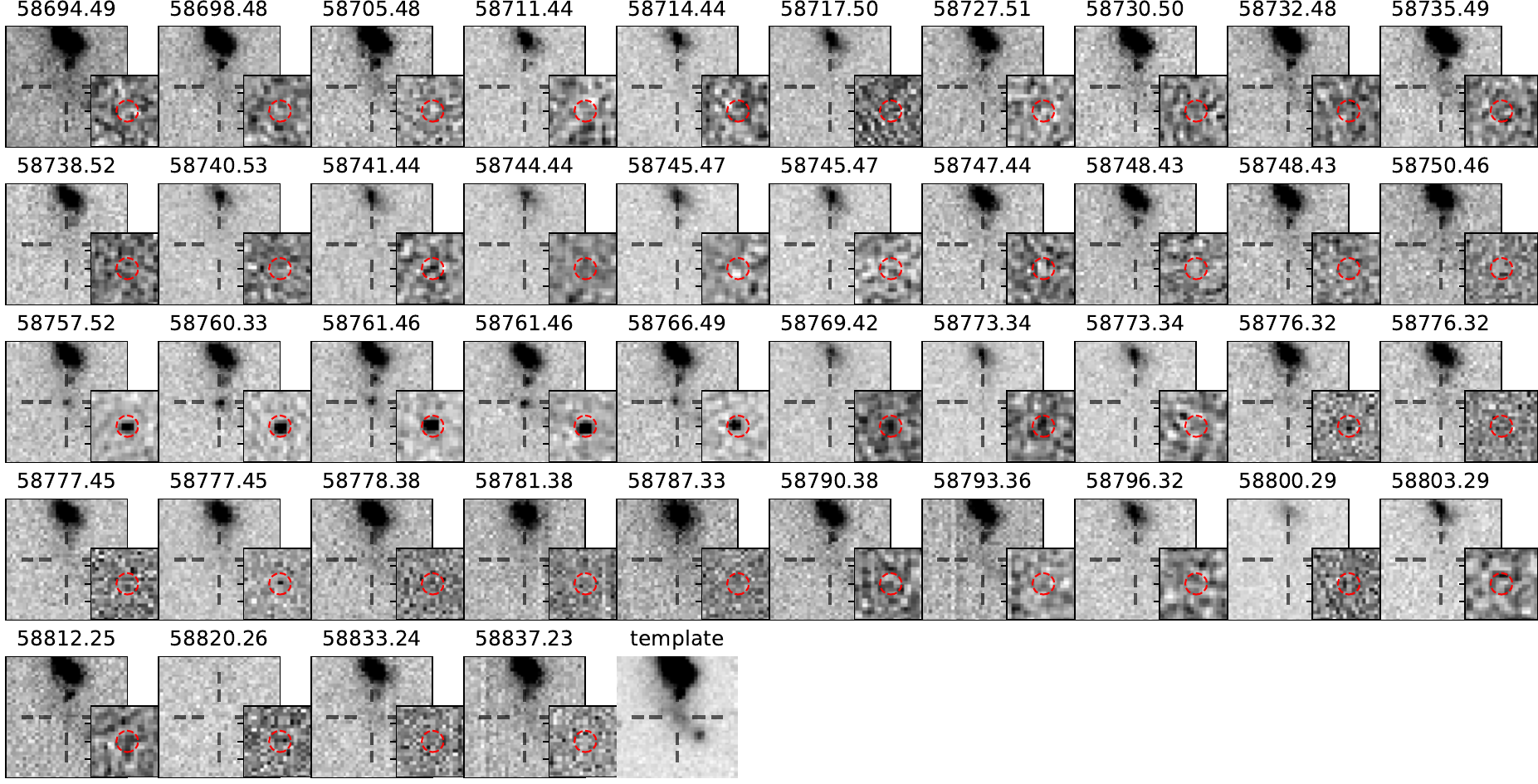}
        \caption{ZTF $r$-band subtraction sequence.}
        \label{fig:subtraction_r}
    \end{subfigure}
    \caption{\textbf{ZTF image subtraction and photometric quality control for SN~2019rsq.}
    \textbf{a,} Representative ZTF $r$-band science, reference and difference images centred on SN~2019rsq for the epoch MJD~58757.52. The reference image was constructed from pre-explosion ZTF observations, and the difference image was produced with the \textsc{ECSNOOPY} subtraction pipeline. The SN position is marked in the cutouts, and the host-galaxy aperture is indicated for spatial reference.
    \textbf{b,c,} Complete $g$- and $r$-band image-subtraction sequences used for the independent photometric remeasurement. The Modified Julian Date of each science exposure is shown above the corresponding cutout, and the final cutout in each sequence is the pre-explosion reference image. The sequences were inspected epoch by epoch to identify subtraction artefacts, host-galaxy residuals and variations in the local background at the transient position. Only measurements with ${\rm S/N}\geq3$ are treated as detections in the quantitative light curves and model fits; lower-significance measurements are treated as non-detections and converted to $5\sigma$ upper limits from the local background noise.}

    \label{fig:ztf_subtraction}
\end{figure*}

\begin{figure}[t]
    \centering
    \includegraphics[width=0.88\columnwidth]{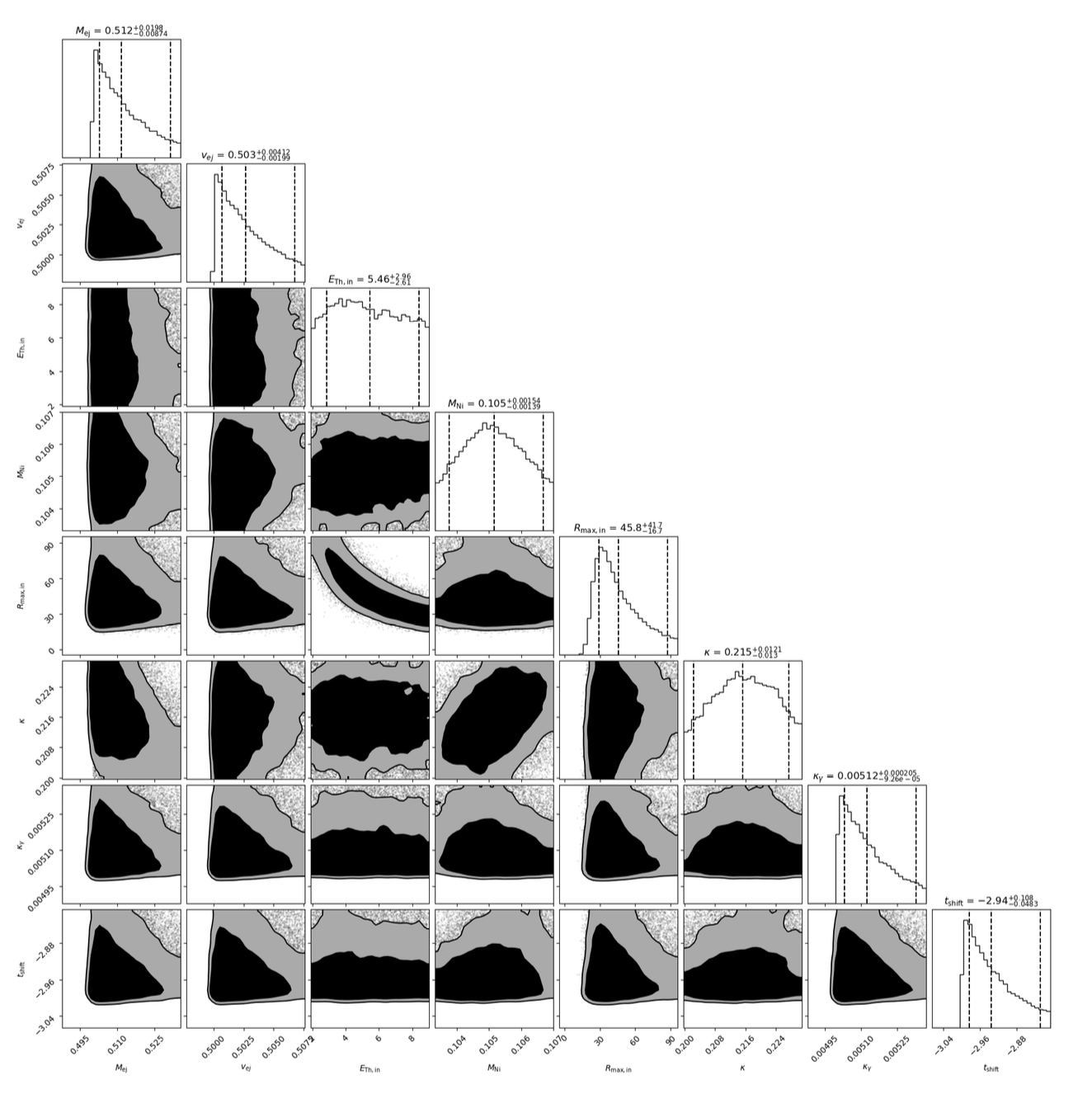}
    \caption{\textbf{Parameter-space exploration for the TransFit modelling of SN~2019rsq.}
    Distributions and covariances of the accepted light-curve model parameters obtained with the \texttt{TransFit} analysis. Starting from the nominal best-fitting solution, approximately $10^4$ parameter realizations were explored and evaluated against the observed photometry using the adopted $\chi^2$ statistic. The diagonal panels show the one-dimensional distributions of the accepted model realizations, while the off-diagonal panels illustrate correlations and degeneracies between parameters. Vertical dashed lines mark the median and central 68\% range of the accepted realizations. The broad and correlated parameter distributions reflect the limited early-time photometric coverage and the absence of a directly observed shock-breakout or shock-cooling phase. The analysis is therefore used to identify physically allowed light-curve solutions and to test whether an explosion occurring near the epoch of S190910d is compatible with the observations, rather than to provide a unique determination of the explosion epoch or progenitor parameters.}
    \label{fig:transfit_parameter_space}
\end{figure}

\begin{figure}[ht!]
  \centering
  \includegraphics[width=0.85\columnwidth]{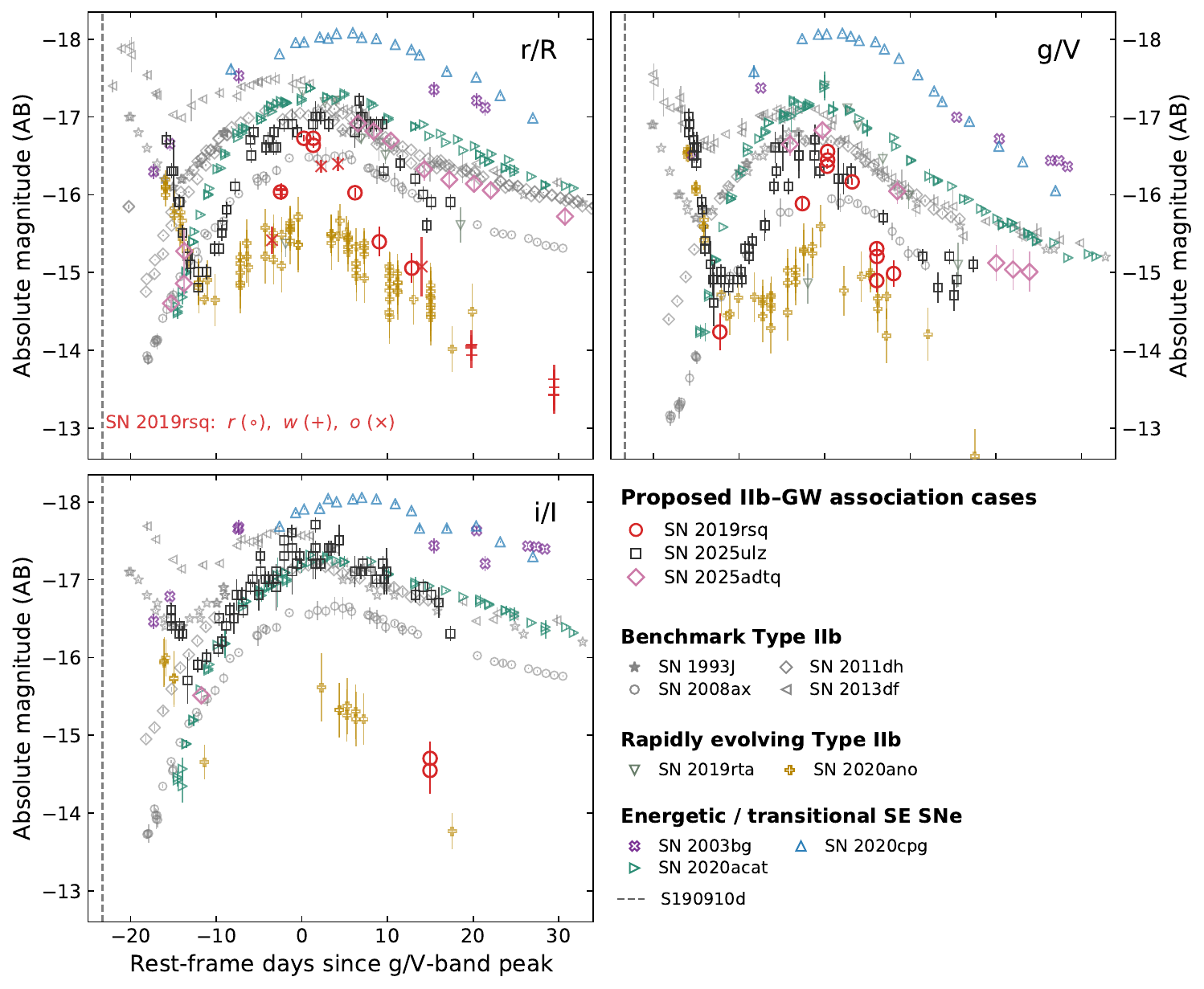}
  \caption{\textbf{Rest-frame optical light curve comparison of SN~2019rsq with Type~IIb and related stripped-envelope supernovae.} Absolute magnitude light curves in the $r/R$ (upper left), $g/V$ (upper right), and $i/I$ (lower left) bands are shown as a function of rest-frame phase relative to the respective $g/V$-band maximum. SN~2019rsq is highlighted with red open circles; its available $w$- and $o$-band measurements are additionally shown in the $r/R$ panel with $+$ and $\times$ symbols, respectively. The two other proposed Type~IIb--GW association cases, SN~2025ulz and SN~2025adtq, are shown with black open squares and magenta open diamonds. For context, the comparison sample is divided into benchmark Type~IIb SNe (SN~1993J, SN~2008ax, SN~2011dh, and SN~2013df), rapidly evolving Type~IIb SNe (SN~2019rta and SN~2020ano), and energetic or transitional stripped envelope SNe (SN~2003bg, SN~2020acat, and SN~2020cpg), as indicated in the legend. The vertical dashed line marks the epoch of the GW candidate S190910d relative to the optical maximum of SN~2019rsq.}
  \label{fig:lc-compare}
\end{figure}

\begin{figure}[ht!]
  \centering
  \includegraphics[width=0.85\columnwidth]{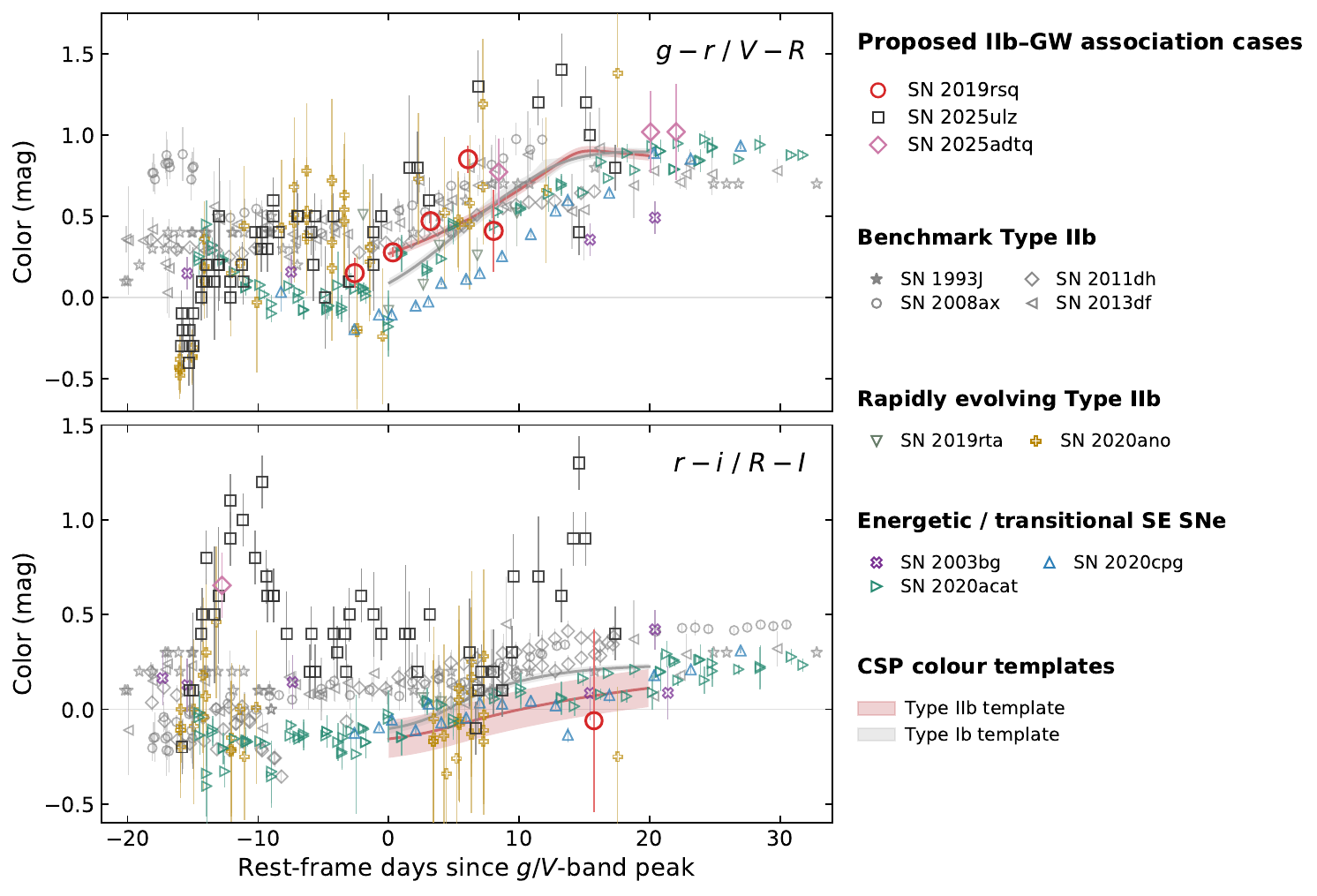}
  \caption{\textbf{Rest-frame optical colour evolution of SN~2019rsq compared with Type~IIb and related stripped-envelope supernovae.} The upper and lower panels show the evolution of $g-r/V-R$ and $r-i/R-I$, respectively, as a function of rest-frame phase relative to the respective $g/V$-band maximum. SN~2019rsq is highlighted with red open circles, while SN~2025ulz and SN~2025adtq are shown as black open squares and magenta open diamonds, respectively. The same benchmark, rapidly evolving, and energetic/transitional comparison samples as in Fig.~\ref{fig:lc-compare} are shown using matching colours and symbols. Colour measurements are constructed from temporally matched photometry in the relevant filter pairs; the comparatively sparse sampling of SN~2019rsq reflects the limited availability of sufficiently near-simultaneous multi-band observations. The pink and grey curves and shaded regions show the empirical Carnegie Supernova Project colour templates for Type~IIb and Type~Ib supernovae, respectively. SN~2019rsq remains broadly consistent with the observed Type~IIb colour locus over the epochs sampled, although the absence of sufficiently early colour measurements prevents a direct test of whether it underwent the pronounced early colour excursion observed in SN~2025ulz.}
  \label{fig:color}
\end{figure}

\begin{figure}[ht!]
  \centering
  \includegraphics[width=0.85\columnwidth]{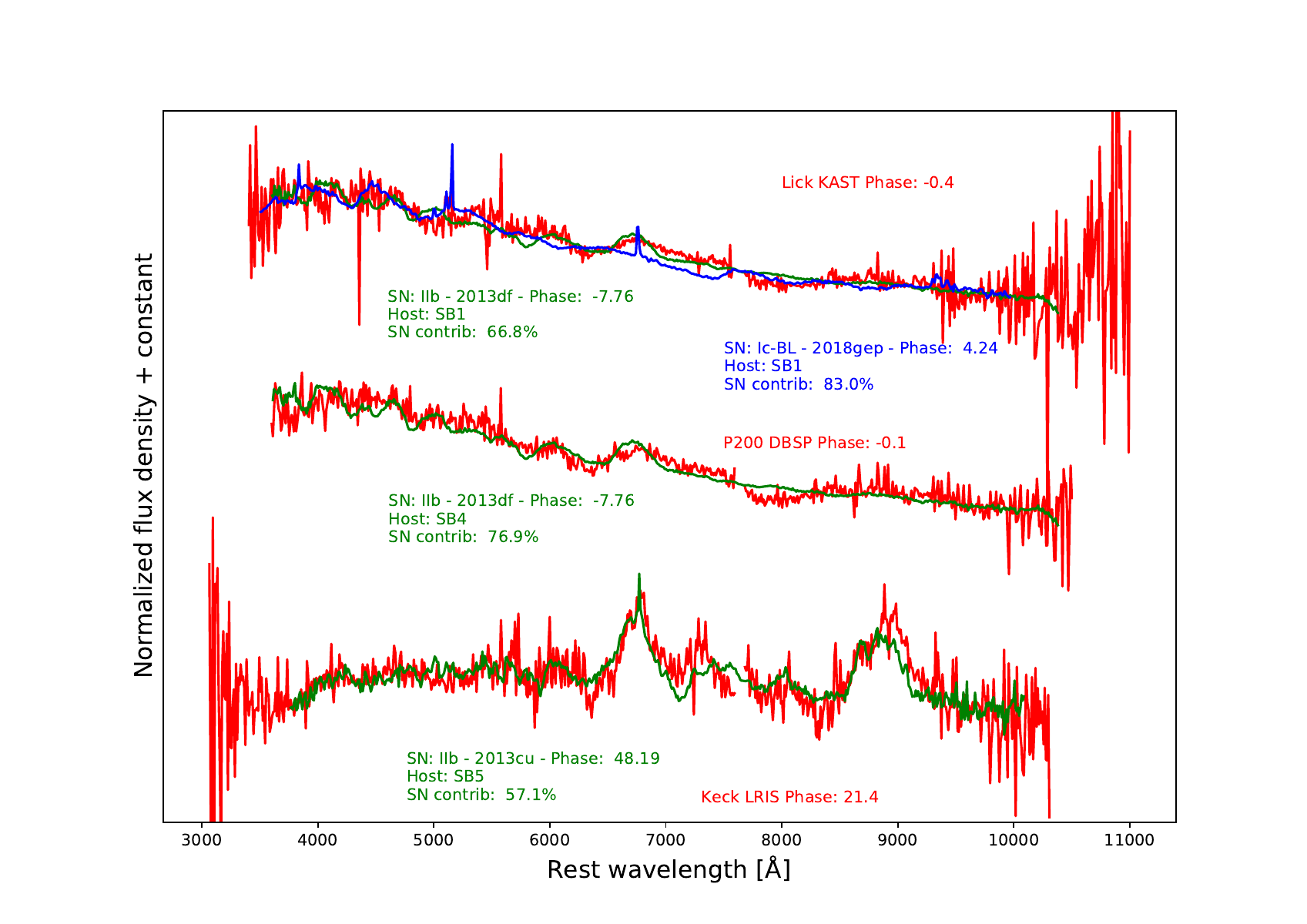}
  \caption{\textbf{NGSF spectral fitting and the evolving classification of SN~2019rsq.} Independent spectral classification of SN~2019rsq using the NGSF tool. Observed spectra (red) obtained near maximum light with Lick/KAST and P200/DBSP and at a later epoch with Keck/LRIS are compared with linear combinations of SN and host galaxy templates. For the earliest Lick/KAST spectrum, representative acceptable matches include both a Type~IIb template (SN~2013df; green) and a broad-lined Type~Ic template (SN~2018gep; blue), illustrating the classification ambiguity at early phases when the hydrogen features are weak. The P200/DBSP spectrum is well represented by a Type~IIb template, and by $\sim21$ rest-frame days after maximum the Keck/LRIS spectrum is clearly consistent with a Type~IIb classification, with more developed hydrogen and helium features. The early ambiguity is consistent with the historical classification sequence in which SN~2019rsq was initially reported as SN~Ic-BL before subsequent spectroscopy established its Type~IIb nature.}
  \label{fig:spec}
\end{figure}

\begin{figure}[ht!]
  \centering
  \includegraphics[width=0.8\columnwidth]{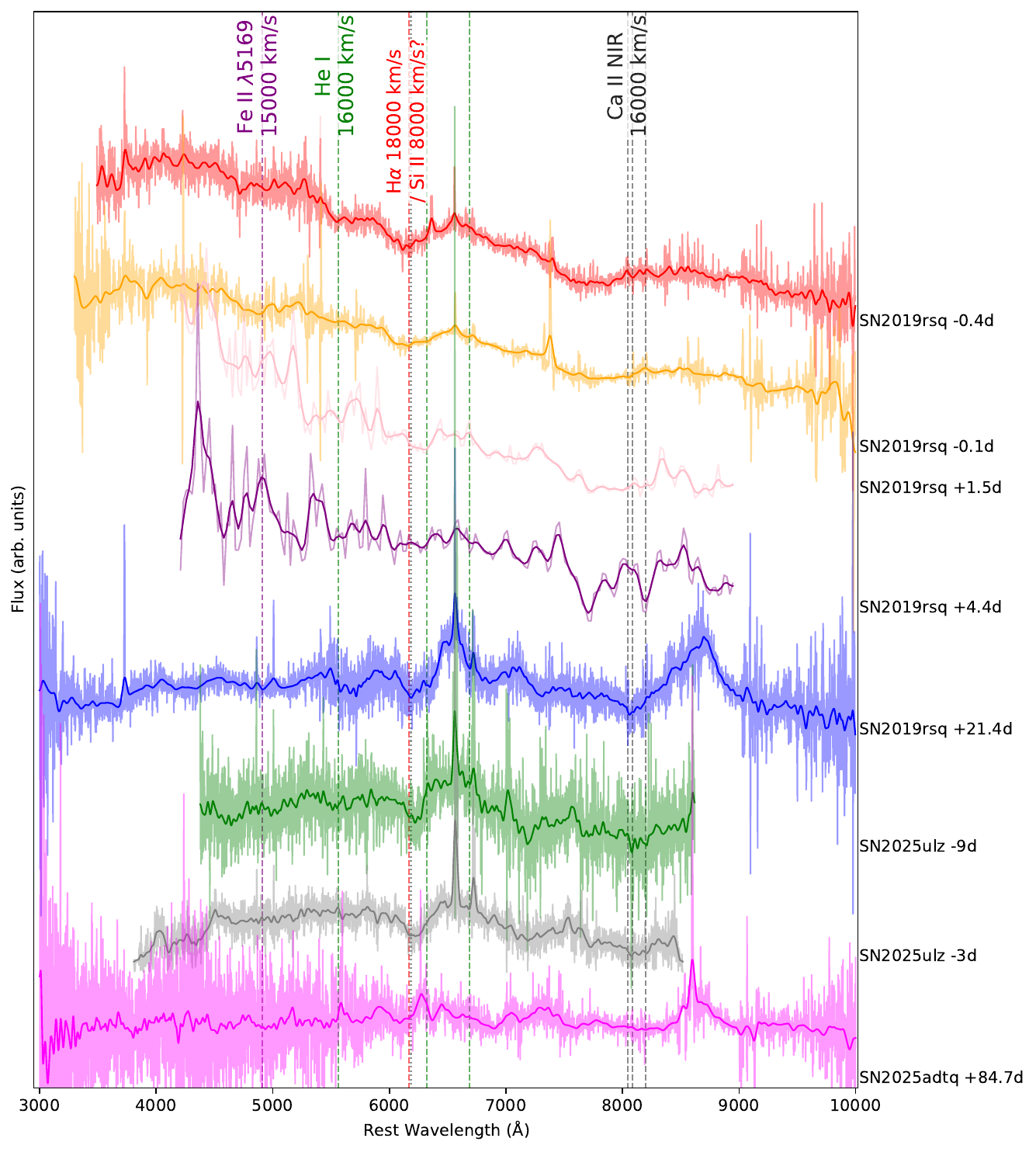}
  \caption{\textbf{Optical spectral evolution of SN~2019rsq and comparison with the other proposed Type~IIb--GW association cases.} Rest-frame optical spectra of SN~2019rsq obtained at five epochs are shown together with two spectra of SN~2025ulz and one late-time  spectrum of SN~2025adtq. All phases are given in rest-frame days relative to the adopted $g/V$-band maximum of each object, and the spectra are arbitrarily scaled and vertically offset for clarity. The two P60/SEDM spectra of SN~2019rsq provide low-resolution coverage during the early post-peak phase, bridging the higher-quality P200/DBSP and Lick/KAST spectra near maximum and the later Keck/LRIS spectrum. Dashed lines indicate illustrative line-identification guides for SN~2019rsq, including Fe~II~$\lambda5169$, He~I, H$\alpha$, and the Ca~II near-infrared triplet; the feature near $\sim 6200$~\AA\ may alternatively be associated with Si~II~$\lambda6355$ at lower expansion velocity. The spectral sequence of SN~2019rsq shows a progression from an early broad-lined appearance toward a more typical Type~IIb-like spectrum at later phases. SN~2025ulz exhibits similarly broad early-time features, whereas the currently public spectrum of SN~2025adtq samples only a much later epoch and serves primarily as a late-time Type~IIb reference.}
  \label{fig:spec1}
\end{figure}

\begin{figure}[ht!]
  \centering
  \includegraphics[width=0.95\columnwidth]{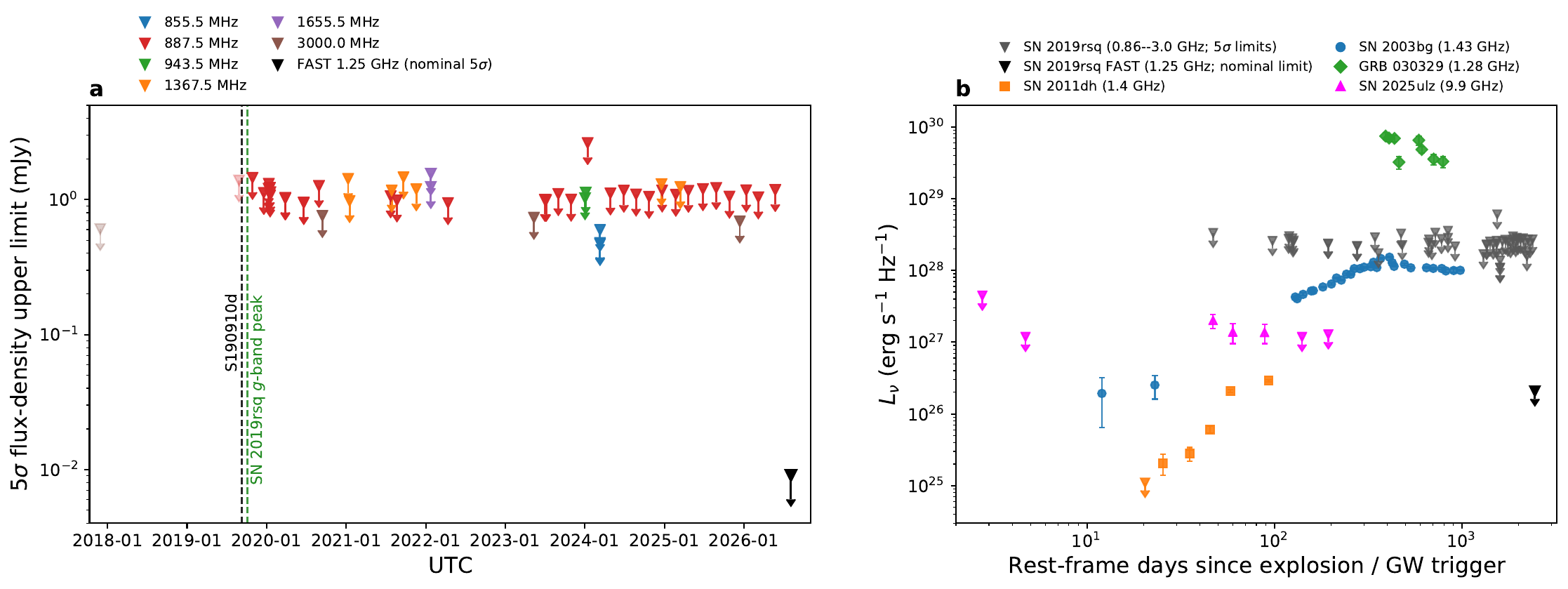}
  \caption{\textbf{Radio constraints on SN~2019rsq and comparison with radio-emitting Type~IIb supernovae and GRB afterglows.} 
  \textbf{a,} Multi-epoch radio constraints at the position of SN~2019rsq. Downward triangles show $5\sigma$ flux-density upper limits derived from the local image rms of archival ASKAP and VLASS observations spanning approximately 0.86--3.0~GHz. Measurements obtained before S190910d are shown with reduced opacity and are included only for temporal context. The black and green vertical dashed lines mark the epochs of S190910d and the $g$-band maximum of SN~2019rsq, respectively. The late-time FAST L-band observation is shown separately in black. Unlike the interferometric ASKAP/VLASS measurements, the FAST value represents the nominal $5\sigma$ thermal sensitivity, $9.0\,\mu{\rm Jy}$ at a nominal central frequency of 1.25~GHz, inferred from the radiometer equation; it is not an empirical host-subtracted flux-density upper limit.
  \textbf{b,} Spectral-luminosity constraints on SN~2019rsq compared with selected radio transients. The ASKAP/VLASS limits and nominal FAST sensitivity are shown together with published radio measurements of the Type~IIb SNe SN~2011dh and SN~2003bg, the long-lived radio afterglow of GRB~030329, and the proposed SN--GW association SN~2025ulz. Representative observing frequencies are indicated in the legend. No spectral rescaling is applied between frequencies, so the comparison is intended to provide luminosity context rather than a strictly frequency-matched radio light curve. Times are shown in the source rest frame relative to the adopted explosion or GW-trigger epoch of each transient. The ASKAP/VLASS limits do not exclude radio emission at luminosities typical of ordinary or radio-bright Type~IIb SNe, whereas a long-lived radio-bright afterglow comparable to GRB~030329 would have been readily detectable.} \label{fig:radio}
\end{figure}

\begin{figure}[ht!]
  \centering
  \includegraphics[width=0.95\columnwidth]{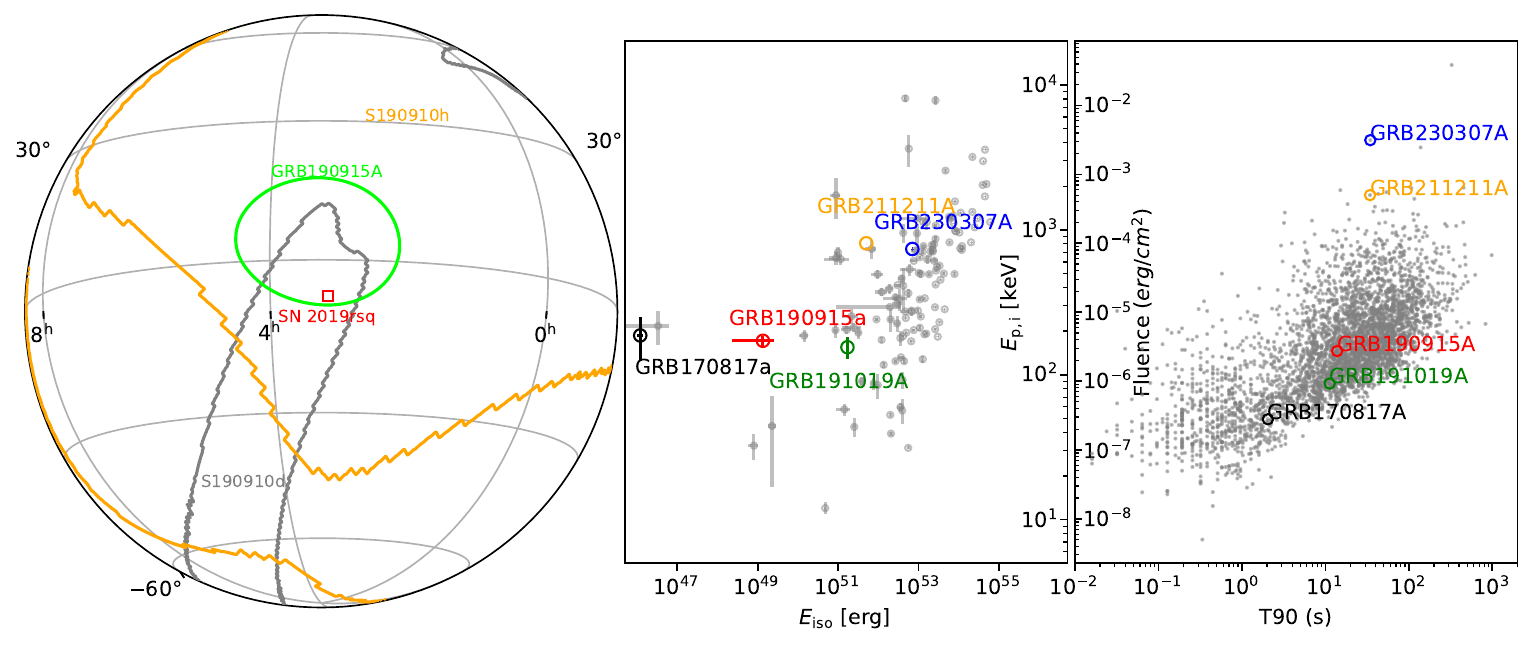}
  \caption{\textbf{GRB~190915A as an auxiliary high-energy coincidence in the SN~2019rsq/GW field.} 
  \textbf{a,} Sky localizations of S190910d (grey) and S190910h (orange), shown as 90\% credible regions, together with the position of SN~2019rsq (red square) and the localization of GRB~190915A (green). GRB~190915A occurred approximately five days after the two GW candidates and its reported localization is compatible with the broader sky region containing the GW localizations and SN~2019rsq. 
  \textbf{b,} Rest-frame spectral peak energy, $E_{\rm p,i}$, versus isotropic-equivalent gamma-ray energy, $E_{\rm iso}$, for GRB~190915A compared with the long-GRB population and selected comparison events. Because the redshift of GRB~190915A is unknown, its location in this plane is calculated under the hypothesis that it lies at the distance of SN~2019rsq, $D_L=140.6$~Mpc. Under this assumption, $E_{\rm iso}\simeq6.3\times10^{48}$~erg and the burst occupies the low-energy regime of the long-GRB population without being an obvious outlier. 
  \textbf{c,} Observed gamma-ray fluence versus $T_{90}$ for GRB~190915A and the same comparison population. GRB~190915A has $T_{90}=13.82\pm0.57$~s and a time-integrated fluence of $(2.73\pm0.03)\times10^{-6}$~erg~cm$^{-2}$, consistent with the broad distribution of long-duration GRBs. GRB~190915A is not required by the common-origin interpretation developed in this work; if physically associated, its approximately five-day delay would instead favour delayed or reactivated high-energy activity rather than prompt emission associated with either S190910d or S190910h.}\label{fig:grb190915a}
\end{figure}

\begin{figure}[ht!]
  \centering
  \includegraphics[width=0.55\columnwidth]{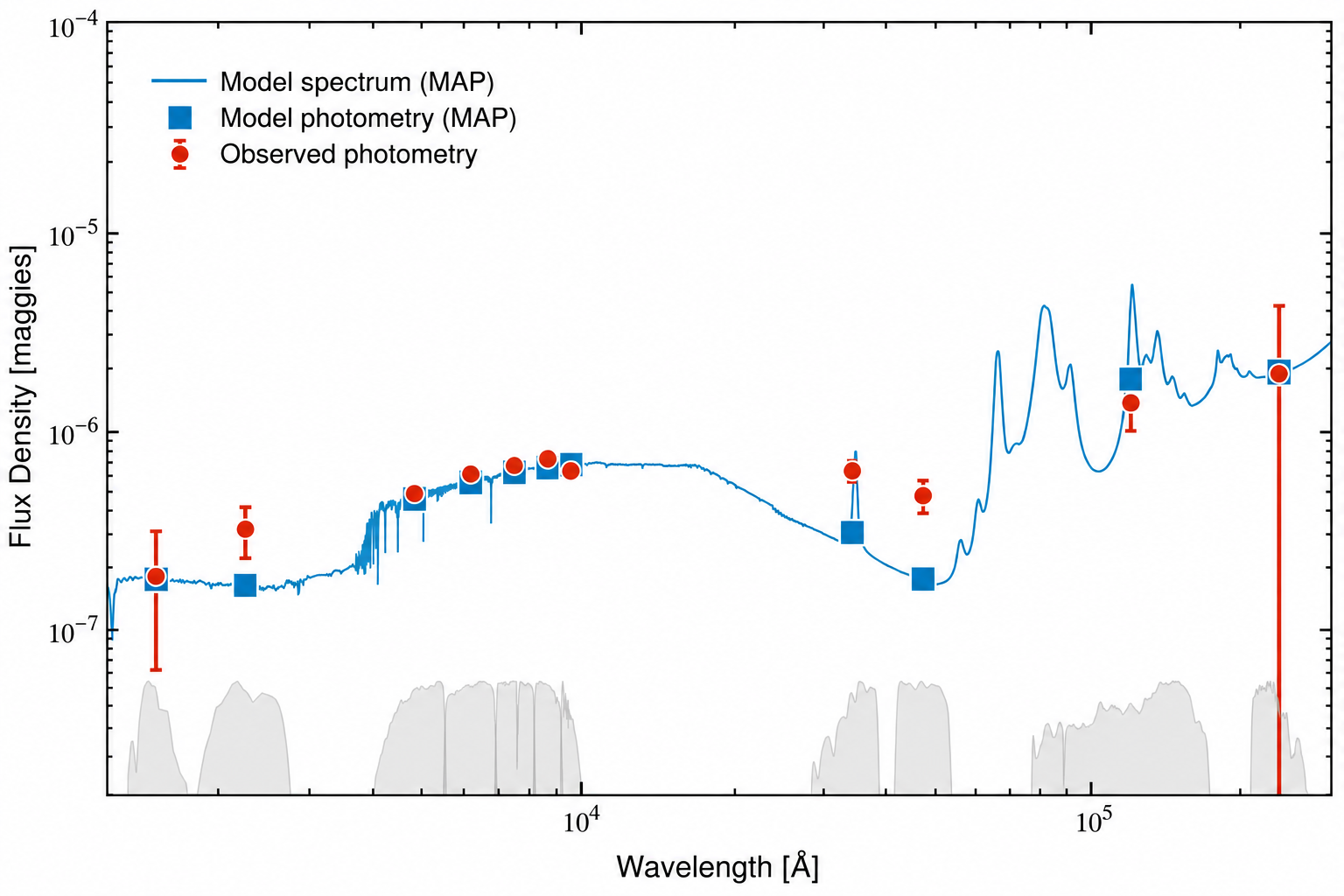}
  \includegraphics[width=0.4\columnwidth]{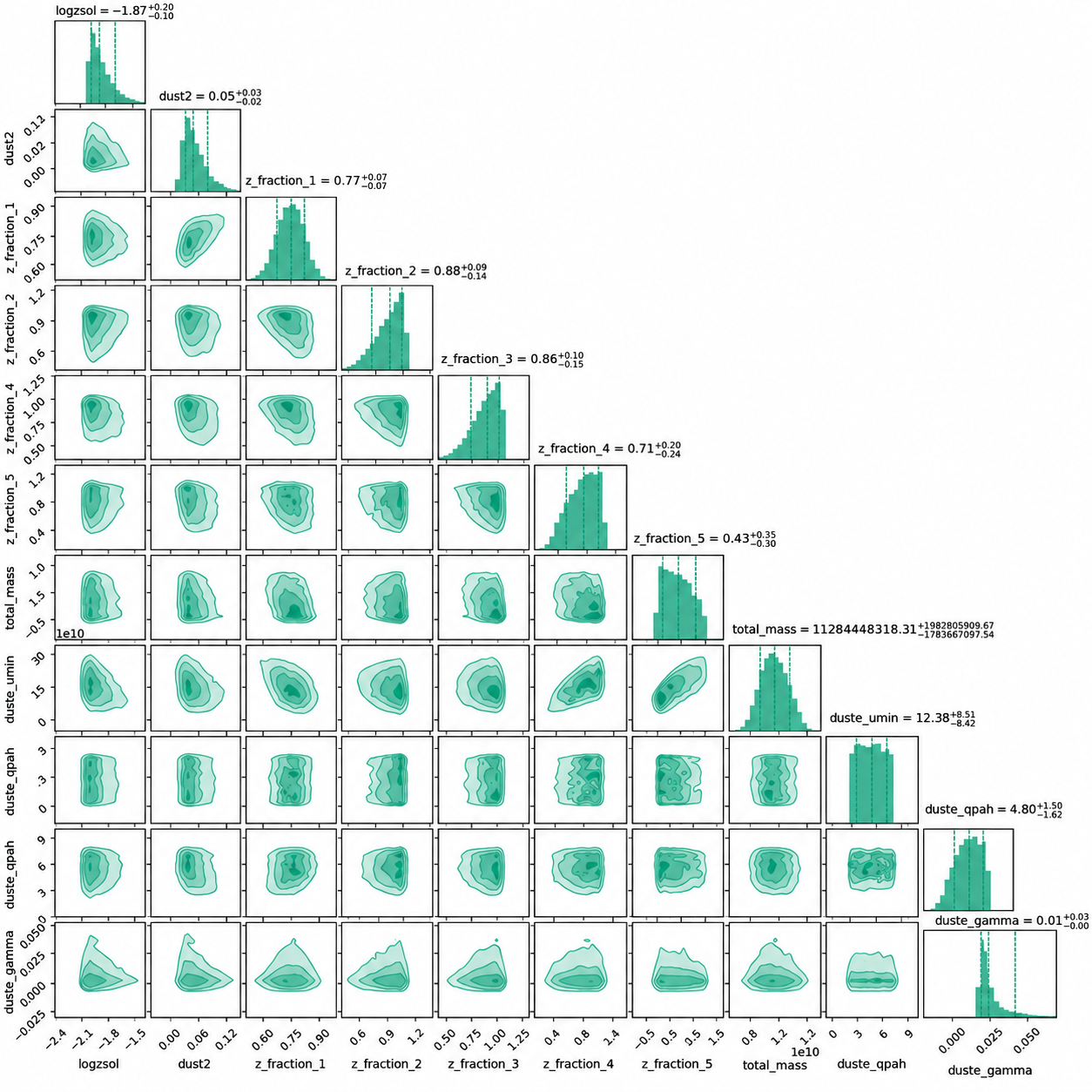}
  \caption{\textbf{Broadband SED modelling of the host galaxy of SN~2019rsq.}
  \textbf{Left,} Observed broadband photometry of the host galaxy SDSS~J031437.71$-$070100.9 from GALEX, Pan-STARRS1 and unWISE, together with the best-fitting \textsc{Prospector} model. Red points show the measured host-galaxy photometry, the solid blue curve shows the model spectrum corresponding to the maximum-a-posteriori solution, and open blue squares show the corresponding synthetic photometry. Grey curves indicate the transmission profiles of the filters used in the fit.
  \textbf{Right,} Posterior distributions from the \textsc{Prospector} broadband SED analysis. Diagonal panels show the marginalized one-dimensional distributions and off-diagonal panels show parameter covariances; contours enclose the 68\%, 95\%, and 99.7\% credible regions. The inferred global stellar mass is $\log_{10}(M_\star/M_\odot)=10.05^{+0.08}_{-0.09}$, the star-formation rate is ${\rm SFR}=1.12^{+0.38}_{-0.38}\,M_\odot\,{\rm yr}^{-1}$, and the corresponding specific star-formation rate is $\log_{10}({\rm sSFR/yr^{-1}})=-10.00^{+0.12}_{-0.17}$. Together with the disturbed morphology and tidal features visible in the host environment, these global properties are consistent with an actively star-forming system capable of hosting a young massive-star population, but do not by themselves distinguish among supernova progenitor channels or provide evidence for a specific compact-object merger scenario.}\label{fig:host_sed}
\end{figure}

\begin{figure}[ht!]
  \centering
  \includegraphics[width=0.85\columnwidth]{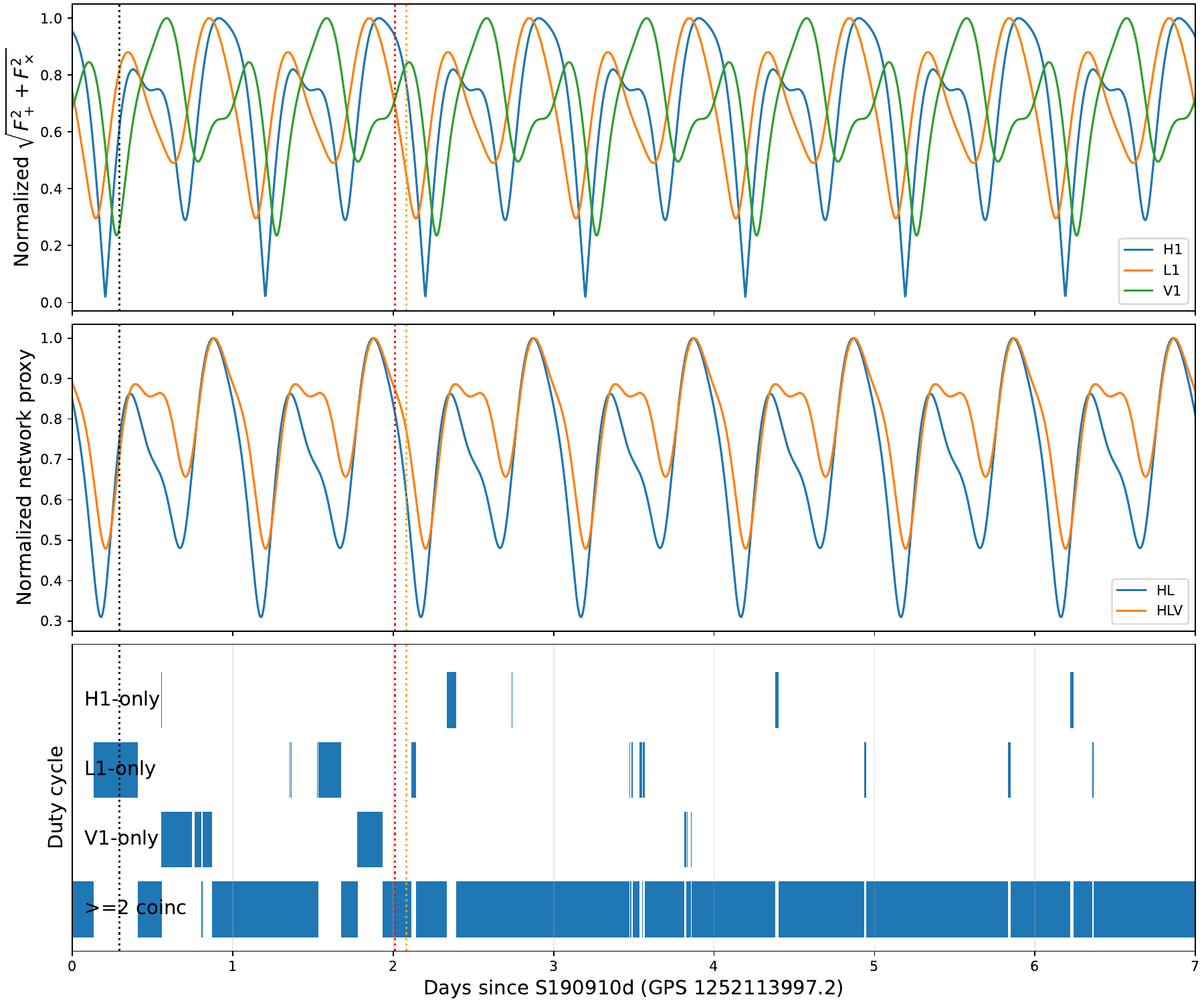}
  \caption{\textbf{Time-dependent detector response and network coverage toward SN~2019rsq following S190910d.}
  \textbf{Top,} Normalized antenna-response proxy, $\sqrt{F_{+}^{2}+F_{\times}^{2}}$, for the individual H1, L1 and V1 detectors, evaluated at the sky position of SN~2019rsq as a function of time during the week following S190910d.
  \textbf{Middle,} Corresponding normalized network-response proxy for the H1--L1 (HL) and H1--L1--V1 (HLV) detector combinations.
  \textbf{Bottom,} Detector operating state over the same interval, distinguishing periods of H1-only, L1-only and V1-only operation from intervals with two or more detectors in coincidence. S190910d defines the zero point of the time axis; vertical dotted lines mark S190910h, S190912c and S190912d for temporal reference.
  The directional antenna response varies substantially on hour-to-day timescales because of Earth's rotation, while the network also undergoes intervals of incomplete multi-detector coverage. Consequently, the sensitivity to an additional compact-binary merger from the direction of SN~2019rsq was not uniform during the first week. These quantities provide a qualitative assessment of short-timescale observational coverage rather than a full detection-efficiency calculation. The absence of another public GW trigger during this interval therefore places only a limited constraint on an additional prompt merger and does not test interactions dynamically delayed to the week-to-month timescales permitted by multi-fragment scattering.} \label{fig:detector_coverage}
\end{figure}

\clearpage

\begin{supplement}

\suppnote{Archival sample completeness and external checks}{supp:samples}

\paragraph{Broader transient-side completeness check.}
The formal coincidence analysis uses the spectroscopically classified ZTF BTS SE sample defined in Methods. As an independent completeness check, we repeated the archival cross-match using the broader TNS database. Because homogeneous peak epochs are unavailable for the full TNS sample, the reported discovery date was used as an approximate temporal proxy where necessary, and spatially and temporally plausible matches were inspected individually. This broader search did not reveal an additional GW--transient configuration comparable to SN~2019rsq/S190910d--S190910h. The TNS selection function is heterogeneous and discovery date is not equivalent to the explosion epoch or main optical peak; the search is therefore used only as a qualitative completeness check and does not enter the formal coincidence-rate calculation.

\paragraph{Post-cutoff comparison systems.}
The formal archival search and all primary coincidence probabilities are restricted to the predefined O3--O4b interval ending on 2025 January 28. Consequently, S250818k/SN~2025ulz and S251112cm/SN~2025adtq do not enter the primary GW or SN samples, Monte Carlo calculations, or quoted primary probabilities. As a retrospective external check, we applied the same basic spatial and temporal matching logic to these later systems; both satisfy the basic GW--SN criteria and are retained only as external phenomenological comparisons. Their emergence after the analysis cutoff is therefore informative as an external sanity check but is not folded back into the significance calculation.

\suppnote{Robustness of the GW-pair coincidence analysis}{supp:stats}

\paragraph{Additional close-in-time GW pairs.}
Among the 10 selected GW pairs separated by less than 5~days, only three have non-zero 3D overlap in the public localization products. In addition to S190910d--S190910h ($N_{\rm 3D}=19{,}368$, $\Delta t=7.1$~hr), these are S241230ev--S250101k ($N_{\rm 3D}=5{,}872$, $\Delta t\simeq1.07$~d) and S230731an--S230802aq ($N_{\rm 3D}=356$, $\Delta t\simeq1.57$~d). The latter two pairs are BBH-dominated in their public low-latency classifications and enter the selected sample through non-zero NSBH probabilities.

S190930s--S190930t provide a useful counterexample: the alerts are separated by only approximately one hour and have projected sky overlap, but no corresponding 3D overlap. Temporal proximity plus two-dimensional overlap alone therefore does not necessarily imply a similarly compact three-dimensional configuration.

\paragraph{Temporal-threshold and sample-definition robustness.}
The observed separation of S190910d and S190910h is 7.1~hr. We also evaluated thresholds of 6, 12 and 24~hr. For the selected 43-superevent sample, the 3D coincidence probabilities are $6.7\times10^{-5}$, $7.6\times10^{-5}$, $1.26\times10^{-4}$ and $2.69\times10^{-4}$ for 6, 7.1, 12 and 24~hr, respectively. Repeating the same calculation with all non-retracted public superevents in the analysis interval gives $7.56\times10^{-5}$, $9.17\times10^{-5}$, $1.56\times10^{-4}$ and $3.09\times10^{-4}$. The similarity of the results shows that the inferred rarity is not produced solely by the low-mass/mass-gap source-class selection. All values in this note are pairwise probabilities under the adopted null and should not be interpreted as a catalogue-wide look-elsewhere probability.

\paragraph{Two-dimensional robustness test.}
Because distance inference for near-threshold GW candidates can be sensitive to waveform and prior assumptions, we repeated the experiment without line-of-sight distance information. Requiring only the temporal criterion and $N_{\rm 2D}>100$ gives selected-sample probabilities of $1.18\times10^{-4}$, $1.41\times10^{-4}$, $2.62\times10^{-4}$ and $5.11\times10^{-4}$ for 6, 7.1, 12 and 24~hr. The corresponding all-alert values are $1.10\times10^{-4}$, $1.37\times10^{-4}$, $2.35\times10^{-4}$ and $4.62\times10^{-4}$. Thus, the unusual clustering is already present in time plus projected sky position, while the public distance information provides genuine additional discrimination.

\paragraph{Common-distance and SN-rate assumptions.}
At the SN coordinates, the public 3D GW localization products give radial-ansatz parameters $(\mu,\sigma)=(372.37,174.57)$~Mpc for S190910d and $(172.60,78.55)$~Mpc for S190910h. Removing the common $D_L^2$ volumetric prior and multiplying the corresponding Gaussian likelihood kernels gives an approximate common-distance likelihood centred at $206.23\pm71.63$~Mpc. Its $\pm\sigma$ interval defines the effective shell 134.6--277.9~Mpc used in the approximate volumetric calculation. The angular factor is calculated from the $N_{\rm 3D}=19{,}368$ distance-compatible pixels rather than the full 2D intersection. As in Methods, Section~\ref{sec:SI2}, this is a consistency construction rather than a formal joint parameter-estimation posterior.

The 10-day effective temporal width used in the SN-rate calculation is distinct from the broader 10--30~day rest-frame matching window. The former represents the approximate width of the relevant Type~IIb rise-time distribution; increasing it would increase the expected number of unrelated SNe and therefore make the optical coincidence less constraining. The resulting probabilities of approximately 0.69 (SE) and 0.35 (Type~IIb) should be regarded as population-level, order-of-magnitude estimates rather than a survey-selection-corrected likelihood. We deliberately do not reduce these probabilities by conditioning post hoc on the unusual fast or broad-lined properties of SN~2019rsq.

\suppnote{Contemporaneous follow-up and additional GW re-analysis details}{supp:gwdetails}

\paragraph{Contemporaneous follow-up context.}
S190910d and S190910h were treated as separate public GW alerts in real time and each prompted substantial electromagnetic follow-up, as documented in O3 follow-up studies\citemeth{Coughlin2020O3a,deJaeger2022}. The public S190910h classification included ${\rm HasRemnant}>99\%$, contributing to its interest as an electromagnetic-counterpart target at the time. ZTF reported multiple optical candidates in the two localization regions, and subsequent spectroscopy with facilities including GTC and WHT rejected several promptly reported transients as unrelated. Although the two public sky localizations overlap substantially, their highest-probability regions are different and no common-system interpretation was proposed during the contemporaneous campaigns. SN~2019rsq was discovered only several weeks later and therefore could not have informed the real-time GW follow-up. The triplet configuration becomes apparent only retrospectively. The extent of follow-up demonstrates that both alerts were scientifically relevant EM-search targets, but is not used as evidence that either trigger is astrophysical or physically associated with SN~2019rsq.

\paragraph{Local-background construction and matched-filter morphology.}
The offline SPIIR re-analysis uses updated data conditioning and a local background from the O3 segment immediately preceding the events (O3 chunk19; GPS 1251349203--1252015022), corresponding to approximately 7.7~days. For S190910d, $|\rho(t)|$ closely follows the normalized autocorrelation envelope of the best-matching template in H1 and L1; the correspondence is weaker in V1 because of its lower sensitivity and small SNR contribution. S190910d ranks above the loudest background trigger in the analysed segment but remains between the approximate $2\sigma$ and $3\sigma$ fluctuation bands. The original low-latency alert used an H1L1 coincidence, whereas the offline ranking uses an H1L1V1 background; Virgo adds little coherent support.

S190910h occurred while only L1 was operating. Its matched-filter time series similarly follows the expected template-autocorrelation morphology and ranks above the loudest local-background trigger, while remaining between the approximate $1\sigma$ and $2\sigma$ fluctuation bands. These diagnostics support a signal-like interpretation but do not establish either event as a confident detection or replace the official LVK assessment.

\paragraph{Matched-filter template realizations.}
For S190910d, one high-ranking SPIIR template has detector-frame component masses $88.2+1.96\,M_\odot$ and $\mathcal{M}^{\rm det}=8.93\,M_\odot$, while other high-ranking templates occupy substantially lower-primary-mass configurations near $34+3\,M_\odot$. For a near-threshold, highly asymmetric candidate, matched filtering constrains the chirp-mass scale more robustly than the individual component decomposition. We therefore do not interpret any single search-template component masses as parameter-estimation measurements. The independent IAS recovery and dedicated precessing parameter estimation motivate the representative source-frame $34+3\,M_\odot$ region used in the physical discussion. For S190910h, the highest-ranking template has detector-frame component masses approximately $2.93+1.11\,M_\odot$ and $\mathcal{M}^{\rm det}\simeq1.53\,M_\odot$.

\paragraph{Additional parameter-estimation results.}
For S190910d, the IMRPhenomXPHM analyses with $q\geq0.05$ and $q\geq0.01$ give median mass ratios 0.0858 and 0.0856 and do not accumulate at the lower prior boundary. The precessing analyses also favour negative effective spin, with median $\chi_{\rm eff}\simeq-0.84$ and an approximate 90\% credible interval $[-0.98,-0.56]$. We treat the spin inference more cautiously than the mass hierarchy because it is more sensitive to waveform freedom and parameter degeneracies; lower- or moderate-spin aligned configurations remain viable. The spin result is therefore not used as a discriminator between formation channels.

For S190910h, fixing the sky position to SN~2019rsq and adopting a 110--160~Mpc distance prior preserves the source-frame solution near $2.5+1.3\,M_\odot$. Applying the same fully conditioned strategy to S190910d yields a substantially less informative posterior, which is why the intrinsic and extrinsic consistency tests for S190910d are kept separate in Methods.

\paragraph{Geometric-projection calculation.}
For the S190910d amplitude test, the adopted source-frame masses $34+3\,M_\odot$ correspond to detector-frame values of approximately $35.1+3.0\,M_\odot$ at $z\simeq0.031$. The representative waveform uses $\chi_{\rm eff}=-0.87$ only to define a reproducible realization; this value is not treated as independent evidence for stellar rotation or disk fragmentation. The geocentric coalescence time is fixed to S190910d and the sky position and luminosity distance to SN~2019rsq. Frequency-domain waveforms are generated with IMRPhenomD using the detector responses and noise power spectral densities appropriate to the event time. Orientations are sampled with $\cos\iota\sim\mathcal{U}[-1,1]$ and $\psi\sim\mathcal{U}[0,\pi)$. Approximately 6.6\% of realizations yield $\rho_{\rm net}\leq8.6$. Because the waveform is non-precessing and omits higher-order modes, the fraction is interpreted only as a geometric consistency diagnostic.

\suppnote{Additional optical reduction, modelling and comparison details}{supp:optical_reduction}

\paragraph{ZTF image reduction and photometry.}
The ZTF/P48 images were processed with the \textsc{ECsnoopy} workflow. The reduction uses \textsc{Astropy}\citemeth{2013A&A...558A..33A} and \textsc{Photutils}\citemeth{2016ascl.soft09011B} for PSF modelling and photometry, \textsc{HOTPANTS}\citemeth{2015ascl.soft04004B} for PSF-matched subtraction, \textsc{SExtractor}\citemeth{1996A&AS..117..393B} for source detection, \textsc{SCAMP}\citemeth{2006ASPC..351..112B} for astrometric calibration and \textsc{SWarp}\citemeth{2010ascl.soft10068B} for resampling/coaddition. Archival ZTF observations were used as subtraction templates; the PSF was constructed from isolated field stars and refitted after local-background refinement.

\paragraph{Subtraction and low-S/N quality checks.}
Science, reference and difference-image triplets were inspected at every epoch for host residuals, background structure and subtraction artefacts. Low-significance residual flux was examined only as an internal diagnostic. All fitted light curves require ${\rm S/N}\geq3$; lower-significance measurements are treated as non-detections and represented by $5\sigma$ limits from the local background noise.

\paragraph{Spectral sequence, line ambiguity and NGSF classification.}
The five-spectrum sequence comprises near-maximum P200/DBSP and Lick/KAST observations, two lower-resolution P60/SEDM epochs, and a Keck/LRIS spectrum at approximately $+21.4$ rest-frame days. At the host redshift, the broad absorption around $\sim6100$~\AA\ can be associated either with high-velocity H$\alpha$ ($\sim2.2\times10^4~{\rm km\,s^{-1}}$) or Si~II~$\lambda6355$ at lower velocity; the absence of a comparably clear H$\beta$ feature prevents an unambiguous early identification. NGSF\citemeth{2022TNSAN.191....1G} reproduces this phase-dependent ambiguity, with the earliest spectrum admitting both Type~IIb and Ic-BL templates and later spectra increasingly favouring Type~IIb. No significant host-frame Na~I~D is detected\citemeth{2012MNRAS.426.1465P}.

\paragraph{TransFit parameter-space exploration.}
Methods describes the five varied parameters and the approximately $10^4$ local realizations used in the TransFit analysis\citemeth{2025ApJ...992...20L}. For the conditional consistency test, accepted solutions reproduce the photometry while placing the explosion within approximately three days of S190910d; the 100 best-fitting accepted curves are shown in Fig.~\ref{fig:lc2019rsq}. Acceptable realizations occur on both sides of the GW trigger. Because no shock breakout or early cooling component is observed, $t_{\rm shift}$ remains covariant with the ejecta parameters. The displayed distributions therefore summarize the explored local solution space and are not posterior credible intervals.

\paragraph{Radiation-hydrodynamic modelling.}
The independent SNEC calculations\citemeth{Morozova_2015} use rotating binary-evolution structures computed with MESA\citemeth{Long_2022,Paxton_2011,Paxton_2013,Paxton_2015,Paxton_2018,Paxton_2019,Jermyn_2023}. The grid varies the imposed mass cut/ejecta mass, explosion energy, $^{56}$Ni mass and residual-envelope structure, using a thermal-bomb prescription; $^{56}$Ni is fully mixed in the adopted grid. A modest host-extinction range is included as a nuisance degree of freedom, although the spectra and colours favour little additional extinction. The representative numerical configuration is given in Methods. The mass cut is imposed by the explosion calculation rather than predicted by stellar evolution, and degeneracies among the explosion parameters mean that the calculation constrains a viable ejecta configuration rather than a unique progenitor mass.

\paragraph{Illustrative angular-momentum consistency check.}
The mass-partition, residual-hydrogen and stellar-angular-momentum arguments are given in Methods. As an internal check, the illustrative MESA structure retains a specific-angular-momentum profile $j(m)$. Material of specific angular momentum $j$ circularizes at
\[
R_{\rm circ}=\frac{j^2}{GM_\bullet},\qquad
\frac{R_{\rm circ}}{R_g}=\left(\frac{jc}{GM_\bullet}\right)^2.
\]
For plausible post-collapse central masses, the high-$j$ layers can circularize at tens to hundreds of gravitational radii. Because $R_{\rm circ}/R_g$ depends sensitively on $M_\bullet$ and the illustrative $17\,M_\odot$ model was not jointly optimized with the GW parameters, this check does not determine the physical fragmentation radius. The $100$--$300\,R_g$ scale in Section~\ref{sec:SI6} instead comes from dedicated collapsar-disk calculations; independent stellar-evolution models support the existence of rapidly rotating, strongly stripped progenitors with substantial retained angular momentum\maincite{Renzo2026}.

\paragraph{Construction of the photometric comparison sample.}
The comparison sample combines benchmark Type~IIb SNe, rapidly evolving Type~IIb events and energetic/transitional SE explosions, together with the post-O4b SN~2025ulz and SN~2025adtq systems. For qualitative comparison, SDSS-like and Johnson--Cousins filters are treated approximately as $g/V$, $r/R$ and $i/I$. We do not transform all datasets onto a single synthetic system; the comparison is intended to establish relative luminosity and timescale trends rather than precision cross-system photometric equivalence. Rest-frame phases are measured relative to the respective $g$- or $V$-band maxima.

\paragraph{Colour-curve construction.}
Rest-frame $g-r/V-R$ and $r-i/R-I$ colours are constructed from temporally matched or sufficiently near-simultaneous measurements. Empirical Type~IIb and Type~Ib colour sequences from the Carnegie Supernova Project\citemeth{2018A&A...609A.135S} are used as population references. SN~2019rsq remains broadly within the Type~IIb distribution over the observed interval; its early-time multiband coverage is insufficient to test the stronger early colour evolution reported for SN~2025ulz.

\paragraph{Host-galaxy SED-fitting details.}
The host analysis uses matched-aperture GALEX, Pan-STARRS1 and unWISE photometry measured with \textsc{HOSTPHOT}\citemeth{Muller-Bravo2022}. The \textsc{Prospector} fit\citemeth{2021ApJS..254...22J} adopts the Prospector-$\alpha$ framework with a non-parametric star-formation history and \textsc{dynesty} sampling. The original fit used a log-uniform total stellar-mass prior over $10^8$--$10^{12}\,M_\odot$, a stellar-metallicity range $-3<\log(Z/Z_\odot)<0.19$, dust optical depth $0<\hat\tau_2<4$, and standard dust-emission priors including $0.1<U_{\min}<25$ and $0.5<Q_{\rm PAH}<7.0$. These choices are included for reproducibility; the host analysis is used only to establish a disturbed, star-forming environment and not to discriminate between compact object formation channels.

\suppnote{Additional fragment evolution and future diagnostics}{supp:dynamics}

The characteristic fragmentation radii, merger channels and inspiral timescales relevant to the proposed interpretation are described in Methods, Section~\ref{sec:SI6}. Here we emphasize two additional aspects that are not captured by the simple vacuum inspiral scalings.

First, the mass of a self-gravitating fragment at formation need not be the mass measured in a later GW merger. Gas accretion, migration, fragment--fragment mergers and close encounters can modify both the mass and orbit of a disk-born compact object before coalescence\maincite{2025ApJ...991L..22C,Baibhav2026}\citemeth{2026MNRAS.545f1835L}. We therefore use the observed component masses to identify a physically compatible evolved-fragment regime, rather than to infer unique fragment birth masses or a unique interaction history.

Second, hierarchical evolution may leave an additional observable signature in the orbital dynamics. Numerical-relativity calculations of hierarchical subsolar-mass mergers in collapsar-disk-like environments show that repeated mergers can retain measurable eccentricity into the final coalescence\citemeth{Wu2026Eccentricity}. A future detection of significant eccentricity in otherwise similar low-mass or strongly asymmetric GW events would therefore provide an independent test of this formation channel.

\suppnote{Additional radio-observation and comparison details}{supp:radio}

\paragraph{ASKAP and VLASS measurements.}
For each ASKAP and VLASS epoch, the Stokes-$I$ image was inspected at the SN position and the local rms measured in a nearby source-free region. The quoted constraints are empirical $5\sigma$ limits based on those local image rms values. ASKAP spans 855.5, 887.5, 943.5, 1367.5 and 1655.5~MHz, while VLASS provides 3-GHz measurements. Pre-explosion epochs are retained only as contextual checks on persistent or unrelated emission. The complete machine-readable measurements will be released with the analysis repository.

\paragraph{Radio spectral-luminosity comparison.}
For each transient, $L_\nu=4\pi D_L^2F_\nu$ and time is shown in the source rest frame. The comparison set comprises SN~2011dh, SN~2003bg, GRB~030329 and SN~2025ulz at representative observing frequencies of approximately 1.4, 1.43, 1.28 and 9.9~GHz, respectively. No spectral index is assumed to transform the measurements to a common frequency, avoiding a model-dependent correction. Extended Data Fig.~\ref{fig:radio} therefore compares radio luminosity scales rather than a homogeneous single-frequency light curve.

\paragraph{FAST nominal thermal sensitivity.}
For FAST, the expected thermal noise is
\[
\sigma_{\rm th}=\frac{T_{\rm sys}}{G\sqrt{n_{\rm pol}\Delta\nu t_{\rm int}}}.
\]
Using $T_{\rm sys}=29.6\pm0.9$~K, $G=13.7\pm0.3$~K~Jy$^{-1}$, $n_{\rm pol}=2$, $\Delta\nu=400$~MHz and $t_{\rm int}=1800$~s gives $\sigma_{\rm th}\simeq1.80~\mu{\rm Jy}$ and $5\sigma_{\rm th}\simeq9.0~\mu{\rm Jy}$. Treating the quoted $T_{\rm sys}$ and gain uncertainties as independent gives an uncertainty of approximately $0.3~\mu{\rm Jy}$ on the nominal $5\sigma$ sensitivity. This is not the rms of a host-subtracted continuum image: the FAST single-dish beam can include host continuum, unrelated sources and diffuse background. The FAST point is therefore displayed separately and is not used as an empirical upper limit equivalent to the ASKAP/VLASS measurements.

\suppnote{GRB~190915A as an auxiliary high-energy coincidence}{supp:grb190915a}

We searched for high-energy transients in temporal and spatial proximity to the GW/SN system. The long-duration GRB~190915A (Fermi trigger GRB190915240) was reported independently by \textit{Fermi}/GBM\citemeth{2019GCN.25750....1F} and Insight-HXMT/HE\citemeth{2019GCN.25760....1Z} on 2019 September 15 at 05:44:57.911 UTC, approximately five days after the GW pair. Its reported localization is compatible with the broader region containing the two GW localizations and SN~2019rsq.

Prompt-emission parameters were obtained from the public Fermi/GBM burst archive\footnote{https://heasarc.gsfc.nasa.gov/W3Browse/fermi/fermigbrst.html} and are consistent with independent BALROG localization/spectral analyses\citemeth{10.1093/mnras/stx2853,Berlato_2019,2019GCN.25751....1K}. The catalogue gives $T_{90}=13.82\pm0.57$~s, time-integrated fluence $S_\gamma=(2.73\pm0.03)\times10^{-6}~{\rm erg\,cm^{-2}}$, $E_{\rm peak}\simeq166$~keV and low-energy photon index $\alpha\simeq-0.66$.

If GRB~190915A is hypothetically placed at $D_L=140.6$~Mpc, its isotropic-equivalent gamma-ray energy is approximately
\[
E_{\rm iso}\simeq\frac{4\pi D_L^2S_\gamma}{1+z}\simeq6.3\times10^{48}~{\rm erg},
\]
neglecting a small low-redshift $K$-correction\citemeth{Bloom_2001,2002A&A...390...81A}. At that assumed distance, its prompt properties occupy the broad low-energy end of the long-GRB distribution and do not provide distinctive evidence for physical association.

GRB~190915A is therefore not required by the core interpretation and is not counted as part of the association. If it were related, its approximately five-day delay would be more naturally interpreted as a separate episode of delayed or reactivated central-engine activity than as prompt emission associated with either S190910d or S190910h. The projection calculation for S190910d also permits unfavourable viewing orientations, but it is not an inclination measurement and does not require any later interaction to share the same axis. The absence of a prompt classical GRB does not by itself exclude central-engine activity in a stripped-envelope explosion. SN~2009bb and SN~2012ap provide useful phenomenological precedents, exhibiting relativistic or mildly relativistic ejecta despite the absence of a detected GRB\citemeth{Soderberg2010,Chakraborti2015}; both are Type~Ic/Ic-BL events rather than direct analogues of the Type~IIb SN~2019rsq. The ASKAP/VLASS non-detections provide a more direct constraint on a powerful, long-lived GRB~030329-like radio afterglow.

\suppdiscussion{Comparison with other proposed Type~IIb--GW associations}{supp:comparison}

In addition to SN~2019rsq/S190910d--S190910h, two SE supernovae have recently been discussed in spatial and temporal coincidence with low-mass GW candidates: SN~2025ulz/S250818k and SN~2025adtq/S251112cm\maincite{2025ApJ...995L..59K,2025ApJ...995L..27G,Franz_2025,Hall2026S251112cm}. Although the systems are heterogeneous and individually inconclusive, they share several unusual phenomenological features.

First, all three involve low-significance, low-mass or NS-involved GW candidates rather than standard high-significance catalogue-like detections. SN~2019rsq is unique in being associated with two near-threshold candidates separated by only 7.1~hr and having substantial 3D localization overlap. Second, all three optical transients are Type~IIb SNe. SN~2019rsq lies toward the fast-evolving, initially broad-lined end of the Type~IIb population; SN~2025ulz likewise shows rapid evolution and broad early spectral features, while SN~2025adtq provides an additional Type~IIb/GW-candidate comparison case. Third, the putative compact object activity in each system occurs on a timescale compatible with the stellar-collapse/early-SN phase rather than with the long evolutionary delays of ordinary field compact binaries. For SN~2019rsq, the main optical peak follows S190910d by approximately 24~days, consistent with a normal Type~IIb rise time if the GW activity occurred near collapse. For SN~2025adtq, the published analysis places the estimated explosion epoch approximately two days before S251112cm; SN~2025ulz was likewise identified in temporal and spatial coincidence with S250818k.

Important differences remain. SN~2025ulz currently has the most extensive electromagnetic follow-up, including reported non-thermal radio emission consistent with fast ejecta interacting with circumstellar material and also compatible with an off-axis relativistic outflow\maincite{Tanner2026}. The interpretation nevertheless remains debated: the ENGRAVE follow-up favours an unrelated Type~IIb supernova contaminant\citemeth{Ackley2026ENGRAVE}, while the independent radio detection leaves a central-engine/off-axis-outflow interpretation viable. SN~2019rsq instead has interferometric radio limits that disfavour a powerful, long-lived GRB~030329-like afterglow. SN~2025adtq currently provides more limited electromagnetic information. None of the three systems individually establishes compact object merger activity during stellar collapse, but together they motivate targeted searches for rapidly evolving Type~IIb SNe associated with low-mass, NS-involved or temporally clustered GW activity in future observing runs.

\end{supplement}

\end{document}